\documentclass[twocolumn]{aastex631}

\shorttitle{The High-Mass Young Stellar Object X3}
\shortauthors{Pei{\ss}ker et al.}

\begin{document}

%\title{High mass star formation in the Galactic center: multiwavelength observations of the Young Stellar Object X3 close to Sgr A*}

\title{X3: a high-mass Young Stellar Object close to the supermassive black hole Sgr~A*}

\correspondingauthor{Florian Pei{\ss}ker}
\email{peissker@ph1.uni-koeln.de}

\author[0000-0002-9850-2708]{Florian Pei$\beta$ker}
\affil{I.Physikalisches Institut der Universit\"at zu K\"oln, Z\"ulpicher Str. 77, 50937 K\"oln, Germany}

\author[0000-0001-6450-1187]{Michal Zaja\v{c}ek}
\affil{Department of Theoretical Physics and Astrophysics, Faculty of Science, Masaryk University, Kotl\'a\v{r}sk\'a 2, 611 37 Brno, Czech Republic}

\author[0000-0001-7134-9005]{Nadeen B. Sabha}
\affil{Institut f\"ur Astro- und Teilchenphysik, Universit\"at Innsbruck, Technikerstr. 25, 6020 Innsbruck, Austria}
\author[0000-0001-8185-8954]{Masato Tsuboi}
\affil{Institute of Space and Astronautical Science, Japan Aerospace Exploration Agency,
3-1-1 Yoshinodai, Chuo-ku, Sagamihara, Kanagawa 252-5210, Japan}

\author[0000-0001-9069-4955]{Jihane Moultaka}
\affil{Institut de Recherche en Astrophysique et Planetologie (IRAP), Universit\'e de Toulouse, CNRS, Observatoire Midi-Pyrénées (OMP), Toulouse, France}

\author[0000-0001-5342-5713]{Lucas Labadie}
\affil{I.Physikalisches Institut der Universit\"at zu K\"oln, Z\"ulpicher Str. 77, 50937 K\"oln, Germany}

\author[0000-0001-6049-3132]{Andreas Eckart}
\affil{I.Physikalisches Institut der Universit\"at zu K\"oln, Z\"ulpicher Str. 77, 50937 K\"oln, Germany}
\affil{Max-Plank-Institut f\"ur Radioastronomie, Auf dem H\"ugel 69, 53121 Bonn, Germany}

\author[0000-0002-5760-0459]{Vladim\'ir Karas}
\affil{Astronomical Institute, Czech Academy of Sciences, Bo\v{c}n\'{i} II 1401, CZ-14100 Prague, Czech Republic}

\author[0000-0002-5859-3932]{Lukas Steiniger}
\affil{I.Physikalisches Institut der Universit\"at zu K\"oln, Z\"ulpicher Str. 77, 50937 K\"oln, Germany}

\author[0000-0001-6165-8525]{Matthias Subroweit}
\affil{I.Physikalisches Institut der Universit\"at zu K\"oln, Z\"ulpicher Str. 77, 50937 K\"oln, Germany}

\author[0000-0002-3004-6208]{Anjana Suresh}
\affil{I.Physikalisches Institut der Universit\"at zu K\"oln, Z\"ulpicher Str. 77, 50937 K\"oln, Germany}

\author[0000-0002-4902-3794]{Maria Melamed}
\affil{I.Physikalisches Institut der Universit\"at zu K\"oln, Z\"ulpicher Str. 77, 50937 K\"oln, Germany}

\author[0000-0002-8382-2020]{Yann Cl\'enet}
\affil{LESIA, Observatoire de Paris, Universit\'e PSL, CNRS, Sorbonne Universit\'e, Universit\'e de Paris, 5 place Jules Janssen, 92195 Meudon, France}

\begin{abstract}

To date, the proposed observation of Young Stellar Objects (YSOs) in the Galactic center (GC) still raises the question where and how these objects could have formed due to the violent vicinity of Sgr~A*. Here, we report the multi-wavelength detection of a highly dynamic YSO close to Sgr~A* that might be a member of the IRS13 cluster. We observe the beforehand known coreless bow-shock source X3 in the near- and mid-infrared (NIR/MIR) with SINFONI (VLT), NACO (VLT), ISAAC (VLT), VISIR (VLT), SHARP (NTT), and NIRCAM2 (KECK). In the radio domain, we use CO continuum and H30$\alpha$ ALMA observations to identify system components at different temperatures and locations concerning the central stellar source. It is suggested that these radio/submm observations in combination with the NIR Br$\gamma$ line can be associated with a protoplanetary disk of the YSO which is consistent with manifold VISIR observations that reveal complex molecules and elements such as PAH, SIV, NeII and ArIII in a dense and compact region. Based on the photometric multi-wavelength analysis, we infer the mass of $15^{+10}_{-5} M_{\odot}$ for the YSO with a related age of a few $10^4$ yr. Due to this age estimate and the required relaxation time scales for high-mass stars, this finding is an indication for ongoing star formation in the inner parsec. The proper motion and 3d distance imply a relation of X3 and IRS13. We argue that IRS13 may serve as a birthplace for young stars that are ejected due to the evaporation of the cluster. 

\end{abstract}


\section{Introduction} \label{sec:1}

The properties of star formation are well defined and require environments characterized by a sufficiently low temperature ($<$20K) and a high gas density \citep[][]{Hsieh2021}. In contrast, the temperature of gaseous-dusty filaments in the Galactic center exceeds a gas temperature of $\sim 6000\,{\rm K}$ and a dust temperature of $\sim 250$K \citep{Cotera1999, Moser2017}. Furthermore, the velocity dispersion of all known objects close to Sgr~A* outpaces typical numerical values in star formation regions by several magnitudes \citep{Larson1981, Genzel2000}. Due to the presence of the $4\,\times\,10^{6}\,M_{\odot}$ supermassive black hole Sgr~A*, tidal forces hinder gas clumping, which is necessary for the formation of stars. However, \cite{Yusef-Zadeh2013} found at a distance of about 0.6 pc from Sgr~A* SiO clumps that imply high-mass star formation. The authors suggest that the observed clumps are an indication of {\it in situ} star formation that took place during the last 10$^4$ - 10$^5$ years. In an subsequent analysis, ALMA observations showed high-velocity clumps in the {\it inner parsec} but also within the Circum-Nuclear Disk (CND) directed towards Sgr~A* \citep{Moser2017, Hsieh2021}. Some of these dense clumps meet the conditions necessary for star formation. In addition, the authors of \cite{Jalali2014} propose that the interplay between Sgr~A* and in-spiralling clumps of several ten Solar masses could trigger star formation. Similar interpretations of the observational findings can be found in the theoretical works of \cite{Nayakshin2007} and \cite{Hobbs2009}.
The in-spiralling of clumps and/or cluster was already suggested by \citet{Portegies-Zwart2003} and \citet{Maillard2004} to explain the high-mass stars of IRS 13 and IRS 16 \citep[for a review, see][]{Genzel2010}. While the IRS 16 cluster is the origin of prominent stellar winds originating at high-mass stars \citep[][]{Krabbe1991, Krabbe1995}, IRS 13 could contain dust-enshrouded YSOs \citep[][]{Eckart2004} and evolved Wolf-Rayet (WR) stars \citep[][]{Moultaka2005}. The presence of dusty objects of IRS 13 are accompanied by the finding of dusty D-sources in or close to the S-cluster \citep[][]{Eckart2013, Peissker2020b}\footnote{\cite{Ciurlo2020} denote these objects as G-sources.}. It is evident that the finding of two distinct populations with sources that share photometric properties such as the H-K and K-L colors imply a common formation history. Furthermore, it is {suggested} that more sources such as X7 \citep[][]{peissker2021} and G2/DSO \citep[][]{peissker2021c} can be found in the {\it inner parsec} \citep[][]{Yusef-Zadeh2015, Yusef-Zadeh2017}. 

At a distance of about one arcsecond from IRS 13, the previously identified bow-shock source X3 is located, which was assumed to be coreless \citep[][]{Clenet2003, muzic2010}. It is speculated that the bow shock shape of X3 is created by stellar winds originating at the IRS 16 cluster \citep[][]{Wardle1992, peissker2021}.
Here we revisit the morphological and emission properties of the bow-shock source X3 by analyzing an extensive archival data set observed between 1995 and 2020. The near- and mid-infrared observations were carried out in the H- (1.65 $\mu m$), K- (2.20 $\mu m$), L- (3.80 $\mu m$), and M-band (4.80 $\mu m$) with a total on-source integration time of several weeks. The mid-infrared domain is accompanied by narrow-filter observations between 8-20$\mu m$ using VISIR covering the N- and Q-band. In addition to radio/submm observations at 232 GHz (1292 $\mu m$) and 343 GHz (874 $\mu m$), we include 3-dimensional Integral Field Unit (IFU) data cubes observed with SINFONI. Based on the analysis presented here, we propose the detection of a high-mass YSO (HMYSO) associated with X3 {characterized by strong outflows with velocities of several hundred km/s}. This surprising and novel interpretation of the X3-system suggests an urgent need to revise our view on this one and similar objects in the Milky Way center. To justify our claim, we will investigate the spectral footprint of the X3 system and compare it to close-by early- and late-type stars. Furthermore, we present a comprehensive multi-wavelength analysis to target the complexity of the X3-system. We discuss formation scenarios consistent with theoretical models that aim to explain the origin of young stars in the {\it inner parsec}.\newline

In the following Sec. \ref{sec:analysis}, we provide an overview of the different telescopes, instruments, and methods that are used for the analysis. Hereafter, we present the { main} results in Sec. \ref{sec:results}. These results will be discussed in Sec. \ref{sec:discuss} and are { followed} by the conclusions in Sec. \ref{sec:conclusion}.

\section{Data and Tools}
\label{sec:analysis}

{In this section}, we describe the data and analysis tools that are used in this work. {Data tables} can be found in Appendix \ref{ref:data_appendix}. In Table \ref{tab:telescopes_instruments}, we list all the instruments/telescopes used with the {corresponding} wavelength domain.
\begin{table}[hbt!]
    \centering
    \begin{tabular}{|cc|}
         \hline 
         \hline
           Telescope/Instrument & Wavelength [$\mu m$] \\
         \hline
            VLT(NACO) &  1.6, 2.1, 3.8, 4.7   \\
            VLT(SINFONI) &  1.4-2.3   \\
            VLT(ISAAC) &  3.0-4.5   \\
            VLT(VISIR) &  8.0-20.0   \\
            NTT(SHARP) &  2.2   \\
            NIRCAM2(KECK) &  1.6, 2.1, 3.7 \\
            OSIRIS(KECK) &   2.1 \\
            ALMA         &  874, 1292 \\
       \hline
    \end{tabular}
    \caption{Telescopes and instruments used in this work. The listed wavelength for the ALMA observations corresponds to 232 GHz and 343 GHz.}
    \label{tab:telescopes_instruments}
\end{table}
Throughout this work, we will not distinguish between the various sub-bands such as for example the K$_S$-, K'-, and K-band since the color difference ($|K-K'|$, $|K-K_S|$ , $|K'-K_S|$) is smaller than the typical photometric uncertainty of about 0.1-0.2 mag \citep[][]{Ott1999, Eckart2004}.

\subsection{Very Large Telescope}

The Very Large Telescope (VLT) is located at Cerro de Paranal (Chile) at a height of about 2500 m. The VLT harbor four main telescopes (Unit Telescopes, abbreviated with UTs) with a individual dish size of 8.2 m and four smaller Auxiliary Telescopes (ATs) with a dish size of 1.8 m/each. While every UT is capable of Adaptive Optics (AO) to correct for the turbulent atmosphere using a Natural Guidance Star (NGS), UT4 additionally supports the use of a Laser Guide Star (LGS). Except for VISIR, the VLT instruments SINFONI, NACO, and ISAAC are already decommissioned. However, ERIS \citep[][]{Davies2018} is the successor to NACO and SINFONI that combines the capabilities of both instruments. ERIS is scheduled for first light in 2022.  

\subsubsection{SINFONI and NACO}

SINFONI \citep[][]{Eisenhauer2003, Bonnet2004} is a NIR instrument that operates between $1.1\,-\,2.45\,\mu m$ and is supported by an Integrated Field Unit (IFU). The IFU is responsible for the shape of the data, which is arranged as a 3d cube containing two spatial and one spectral dimension. This data setup allows for the analysis of individual emission lines by subtracting the underlying continuum. If the source of interest { exhibits} a Line Of Sight (LOS) velocity with $\neq\,0\,$km/s, it is identical to the Doppler shift of a line with respect to the rest wavelength. Since the K-band around $2.2\,\mu m$ is not affected by telluric emission and absorption features, the Br$\gamma$ line with a rest wavelength of $2.1661\,\mu m$ is commonly used for the analysis of stars in the K-band because of the minimized confusion.\newline
For the observations, an exposure time of 300s per single data cube is used. By combining and stacking single data cubes, we are able to artificially increase the on-source integrating time. The spatial pixel scale for the here presented SINFONI data is $0.1"$ with a Field Of View (FOV) of $3"\,\times\,3"$ per single data cube, the grating is set to the H+K band ($1.45\,\mu m\,-\,2.45\,\mu m$) with a spectral resolution $R\,=\,\frac{\lambda}{\Delta\lambda}\,=\,1500$. With a central wavelength in the H+K band of $\lambda\,=\,1.95\,\mu m$, we get the value for the smallest distinguishable wavelength of $\Delta\lambda\,=\,0.0013\,\mu m$. This, however, is an upper limit since the PSF and therefore the computed spectral resolution differs as a function of position on the detector\footnote{See the SINFONI manual, available at \url{www.eso.org}.}. Using the measured values from the instrument commissioning, we adapt the line uncertainty of $\pm\,25\,$km/s.\newline\newline
In addition to SINFONI, we use the near-infrared imager NACO \citep[][]{Lenzen2003, Rousset2003} in the H-, K-, L-, and M-band with an FOV of about $\rm 15\,arcsec\,\times\,15\,arcsec\approx\,0.6\,pc\,\times\,0.6\,pc$ that shows almost the entire NSC/inner parsec of the GC. Because of the size of the FOV, the bright supergiant IRS7 is regularly used for AO. We choose the standard randomized dither pattern inside the largest dither-box of 4 arcsec. Calibration data like sky-, flat-, and dark-frames are observed with a standard procedure and are provided by the telescope site\footnote{Sky-frames are obtained from a nearby empty region.}. The reduction of the data is done by using DPUSER (Ott, MPE Garching) { as well as} in-built scripts such as dead pixel correction. 

\subsubsection{ISAAC and VISIR}

The Infrared Spectrometer And Array Camera \citep[ISAAC, see][]{Moorwood1998} and VLT Imager and Spectrometer for mid Infrared \citep[VISIR, see][]{Lagage2004} instruments are capable to perform imaging and spectroscopic observations in the MIR { domain}. The ISAAC data which is used to create a MIR cube ($3.0\,\mu m\,-\,4.5\,\mu m$) is already discussed and analyzed in \cite{Moultaka2005} where the authors describe the reduction process in detail.
The ISAAC observations were performed in July 2003 using the long-wavelength and low resolution mode with $R=700$.
In this list, VISIR is the only instrument that is currently mounted at the VLT, namely, UT3. It is, like ISAAC, a MIR imager with spectrometer capabilities. In contrast to the continuum observations with ISAAC, several filters of VISIR were used. The selected filters for the observations that took place in 2004 are PAH1r1 ($8.19\,\mu m$), PAH1 ($8.59\,\mu m$), ArIII ($8.99\,\mu m$), SIVr1 ($10.02\,\mu m$), SIVr2 ($11.11\,\mu m$), PAH2 ($11.25\,\mu m$), NeIIr1 ($12.51\,\mu m$), NeII ($12.81\,\mu m$), NeIIr2 ($13.04\,\mu m$), Q2 ($18.72\,\mu m$), and Q3 ($19.50\,\mu m$). In order to increase the signal-to-noise ratio, we stack individual frames to create a final mosaic. 

\subsection{NTT telescope}

The ESO New Technology Telescope (NTT) uses active optics to correct for distortion effects caused by the dish ($\sim 3.6$m). It was commissioned in 1989 and hosted instruments such as the NIR Speckle-Camera SHARP. The data used in the work was published in \cite{Menten1997} and is high-pass filtered with the Lucy-Richardson algorithm \citep[][]{Lucy1974}. We {refer to \cite{peissker2020a, Peissker2022} where we describe and discuss different aspects of the high-pass filter technique.}

\subsection{Keck observatory}

The Keck observatory is located at Mount Mauna Kea. With an elevation of over 4000 m, it is one of the highest ground-based observatories. The observatory consists of two single telescopes, Keck I and Keck II with a dish size of 10 m each. Both telescopes can work as an interferometer with an increased baseline compared to the single dish setup.

\subsubsection{NIRCAM2}

NIRCAM operates in the NIR in the K$_S$-band at a central wavelength of $2.12\,\mu m$. The spatial pixel scale is $0.0099$ mas, the size of a single exposure with an FOV of about $10\,\times\,10$ arcsec is comparable to the NACO data. We use the KOA archive\footnote{\url{www.koa.ipac.caltech.edu}} to download data observed in 2019. For the observations carreid out in 2019 (PI: Tuan Do, UCLA), a LGS was used to perform the AO correction. We use the pre-calibrated, i.e., science ready data (dark-, flat-, and sky-corrected), and apply the shift and add algorithm to maximize the signal-to-noise ratio. 

\subsubsection{OSIRIS}

The OH-Suppressing Infrared Imaging Spectrograph (OSIRIS) mounted at the KeckII telescope. Comparable to SINFONI, OSIRIS provides 3d data cubes that consist of 2 spatial and 1 spectral dimension \citep[][]{Larkin2006, Mieda2014} and uses an Integrated Field Spectrograph (IFS) supported by an AO. 
For the purpose of guidance in a crowded FOV, a science camera is coupled with the instrument producing NIR continuum data. For this work, we will utilize this science camera and investigate the K-band emission of the X3 system in 2020. The presented data was observed with the Kn3 filter that represents the K-band ($2.12-2.22\,\mu m$). The corresponding FOV is $2048\,\times\,2048$ pixel and a spatial resolution of 10 mas. The science ready archival data downloaded from the Keck Online Archive (KOA) is shifted and added to suppress noise and artefacts.

\subsection{ALMA}

The Atacama Large (Sub)Millimeter Array (ALMA) is located on the Chajnantor plateau. With 66 radio telescopes with a maximum distance of about 16 km between single units, it is the largest ground-based observatory to date. The antennas cover a wavelength range of 31 to 1000 GHz. Here used data (PI Masato Tsuboi, project code: 2015.1.01080.S) was previously discussed and analyzed in \cite{Tsuboi2017, tsuboi2019, Tsuboi2020a, Tsuboi2020b} and makes use of Band 7 ($\approx\,350$ GHz). In addition, we use archival science ready data\footnote{\url{https://almascience.eso.org/aq/}} observed at 232 GHz (H30$\alpha$) by PI Lena Murchikova \citep[project code: 2016.1.00870.S, see][]{Murchikova2019}.

\subsection{Methods}

In the following, we will describe the analyzing techniques used in this work. Since we used and discussed these tools in detail in previous works \citep[e.g.,][]{peissker2020a, Peissker2020b}, we will refer the reader to the related publications.

\subsubsection{High-pass filter}

For the K-band detection presented in Fig. \ref{fig:x3_system}, we used a high-pass filter to minimize the influence of overlapping PSF. We applied the high-pass filter that can be described as an {\it image sharpener} to all H- and K-band images presented in this work (except the Br$\gamma$ line maps shown in Fig. \ref{fig:linemaps_sinfo_sinfo} {and} Fig. \ref{fig:osiris_2020}, Appendix \ref{sec:appendix_further_detections}). The motivation for this process are the extended PSF wings of bright sources in the investigated FOV. If the PSF and especially its wings become too dominant, some image details are suppressed. The process is outlined in the following:
\begin{enumerate}
    \item The input image ($I_{in}$) is smoothed with a Gaussian that matches the PSF of $I_{in}$
    \item The resulting image $I_{low}$ is a low-pass-filtered version of $I_{in}$
    \item $I_{in}$ is the result of the real image (i.e., the natural scene without the influence of a PSF) composed with an "all filter" PSF
\end{enumerate}
This above relation can be mathematically described with 
\begin{equation}
    I_{in}\,-\,I_{low}\,=\,I_{high}
    \label{eq:highpass}
\end{equation}
where $I_{high}$ is the desired high-pass filtered image of the input frame. In addition to the process described, we can again use a Gaussian and apply it to $I_{high}$ to construct a smoothed version of the final result. With this procedure, broader PSF wings are again added to the data. However, the influence of the PSF wings is significantly reduced. In \cite{Peissker2020b}, Fig. 2, we show a comparison between filtered and non filtered data to visualize the advantage of an {\it image sharpener}. Compared to the Lucy-Richardson algorithm \citep{Lucy1974}, the {\it image sharpener} is accessible and easy to use. It can be classified as a cosmetic procedure and shows similarities to Angular Differential Imaging \citep{Marois2006}. However, we add that the SHARP data of 1995 \citep[see Sec. \ref{sec:results} and][]{Menten1997} is treated with a Lucy-Richardson (LR) deconvolution algorithm. We guide the interested reader to \cite{Peissker2022} for a detailed description of the LR process. 

\subsubsection{Keplerian fit}

For the trajectory of X3, \citet{muzic2010} derived a linear solution. Since we have an access to a data baseline that is twice as long as the one analyzed by \citet{muzic2010}, we will apply a Keplerian solution to the data.
Based on the analysis in \cite{Parsa2017}, \cite{Ali2020}, and \cite{Peissker2022}, we assume a mass of $4.0\,\times\,10^{6}\,M_{\odot}$ with a distance of $8.0\,$kpc for the central potential SgrA*. This assumption is justified by the independently derived mass of $4.0\,\times\,10^{6}\,M_{\odot}$ for Sgr~A* by \cite{eht2022} and in reasonable agreement with \cite{Do2019S2}.

\subsection{Theoretical models}

To put the observational photometric results into perspective, we { apply the 3d radiative transfer MCMC} model by \citet{Robitaille2011} to model the SED for the X3 system. The code called $\rm HYPERION$ is based on a 3-d dust continuum emission model where ray-tracing is used to enhance the results. A detailed documentation of the code is freely accessible\footnote{\url{www.hyperion-rt.org}}.
HYPERION takes various structures of YSOs into account. For example, the gaseous accretion disk, bipolar cavities, and a rotationally flattened infalling dust envelope \citep[Ulrich type, see][]{Ulrich1976} can be modeled. We refer the interested reader to the publication of \citet{SiciliaAguilar2016} where the authors provide a rough overview of the composition of a YSO (see Sec. \ref{sec:discuss}). {An additional example of successful application of the code can be found in \cite{Zajacek2017} were we successfully modeled the NIR continuum emission of the infrared excess source DSO/G2 using similar gaseous-dusty structures typical of YSOs as for the X3 system.}
For the dust grains, we use a model that is supposed to represent the dust envelope of the X3-system. Our dust model is based on \cite{Draine2003} with a slope of the optical extinction curve of $\rm R_V\,=\,A_V/(A_B-A_V)\,=\,3.1$ which appears to be suitable and agrees with detailed high-resolution studies of the region by \cite{Fritz2011}.

\section{Results} \label{sec:results}
\begin{figure*}[htbp!]
	\centering
	\includegraphics[width=.7\textwidth]{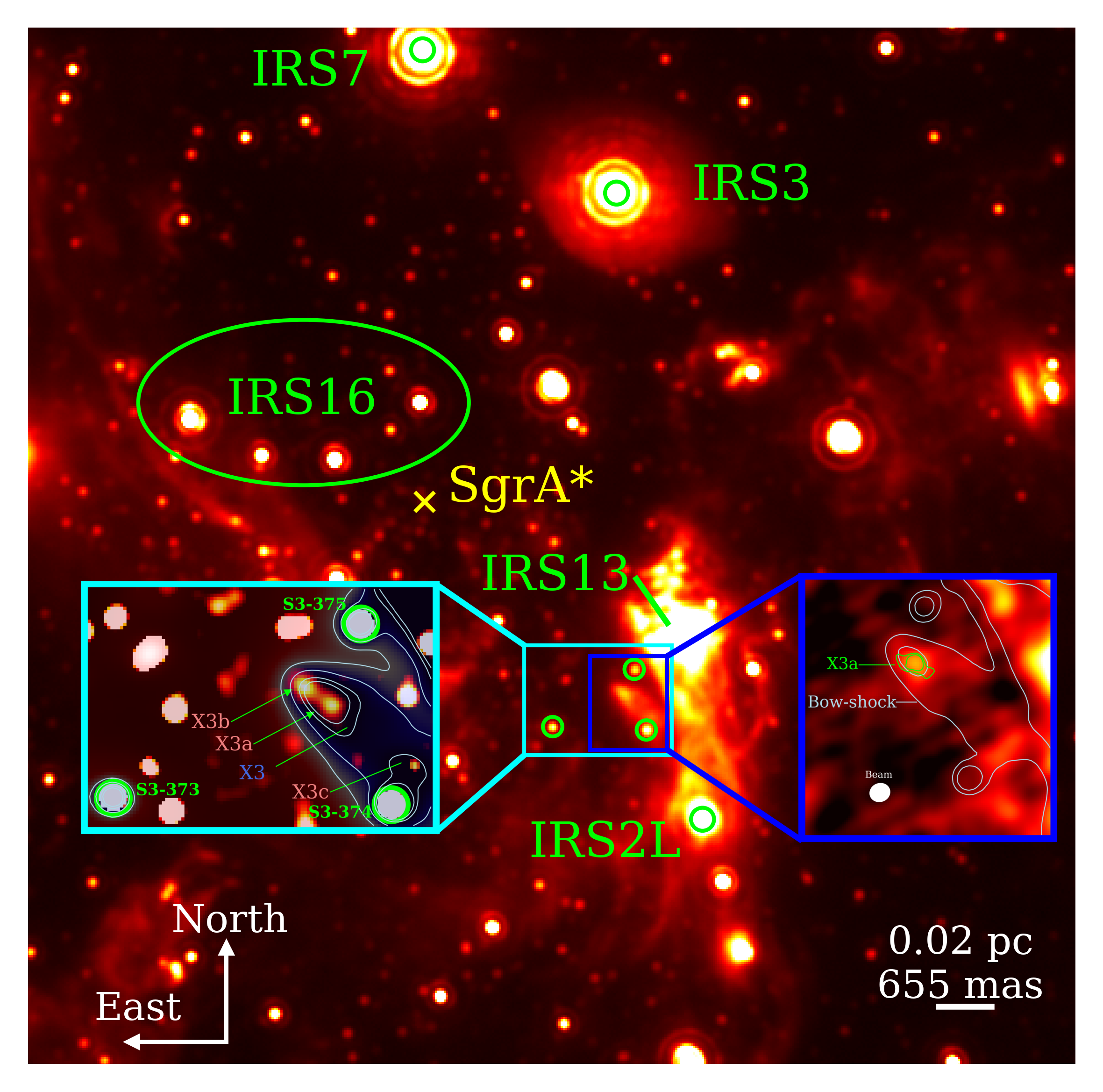}
	\caption{Finding chart for the X3 system. We show two zoomed views toward the X3 system. The cyan zoom box to the left shows a K- and L-band overlay image where blue represents the dust of the bow shock and red is associated with hot and thermal emission. Whereas X3a can be classified as a YSO, X3b and X3c are thermal blobs. The cyan box also indicates the position of the three closest and brightest stars, S3-373, S3-374, and S3-375 \citep[see][]{Gautam2019}. The blue box on the right shows the CO emission at 343 GHz {where we incorporate K-band} contours of X3a (lime colored) and the bow shock {which is primarily observed in the L-band (light blue colored)}.  
	The background image is observed with NACO in the {MIR (L-band, 3.8$\mu m$)} where we indicate prominent clusters like IRS16 or IRS13 and stars. Sgr A* is located at the position of the yellow $\times$.}
\label{fig:x3_system}
\end{figure*}

In this section, we present the results of our multiwavelength analysis. In the following section, we analyze the data that reveals the stellar counterpart of X3 with its components that are related to a YSO. Based on the presented observations and the SED model, we derive various parameters that describe the young dust-enshrouded star. { In Figure \ref{fig:x3_system}, we provide} an overview of the components that are associated with the X3 system. The three brightest and closest early-type stars to X3 are arranged as a triangle that is used throughout this manuscript for guidance to identify the investigated system in the crowded field. Please note that we { adopt} the nomenclature of \cite{Gautam2019} for S3-374.
Since the other two stars of the triangle are not named yet, we are following the established nomenclature with S3-373 and S3-375. The cyan box in Fig. \ref{fig:x3_system} shows the three {\it triangle stars} S3-373, S3-374, and S3-375. {In the following, all figures are normalized to their peak flux intensity to maintain a comparable character. Some of the closer stars might be over-saturated due to intrinsic flux density of X3.}

\subsection{Near- and Mid-Infrared detection}

During the analysis of NACO data observed in 2004, we found at the position of the dusty $L$-band bow shock source X3 three compact sources that we call X3a, X3b, and X3c (see Fig. \ref{fig:x3_system} and Table \ref{tab:components} for an overview). 
\begin{table}[hbt!]
    \centering
    \begin{tabular}{|cc|}
         \hline 
         \hline
           Name & Classification\\
           \hline
           X3   & Bow shock dust envelope  \\
           X3a  & Central stellar source \\
           X3b  & Hot thermal blob \\
           X3c  & Hot blob? \\
           \hline
    \end{tabular}
    \caption{Components of the X3 system. While we find strong evidence in our data set that justifies the listed classification of X3, X3a, and X3b, the analysis of X3c remains challenging.} 
    \label{tab:components}
\end{table}
K-band observations with the SHARP camera (NTT) of 1995 \citep[see Figure 2 in][]{Menten1997} in comparison with NACO (VLT) data of 2002 imply the absence of X3b in earlier years while X3a can be detected well above the confusion limit between 1995 and 2020. 
\begin{figure}[htbp!]
	\centering
	\includegraphics[width=0.5\textwidth]{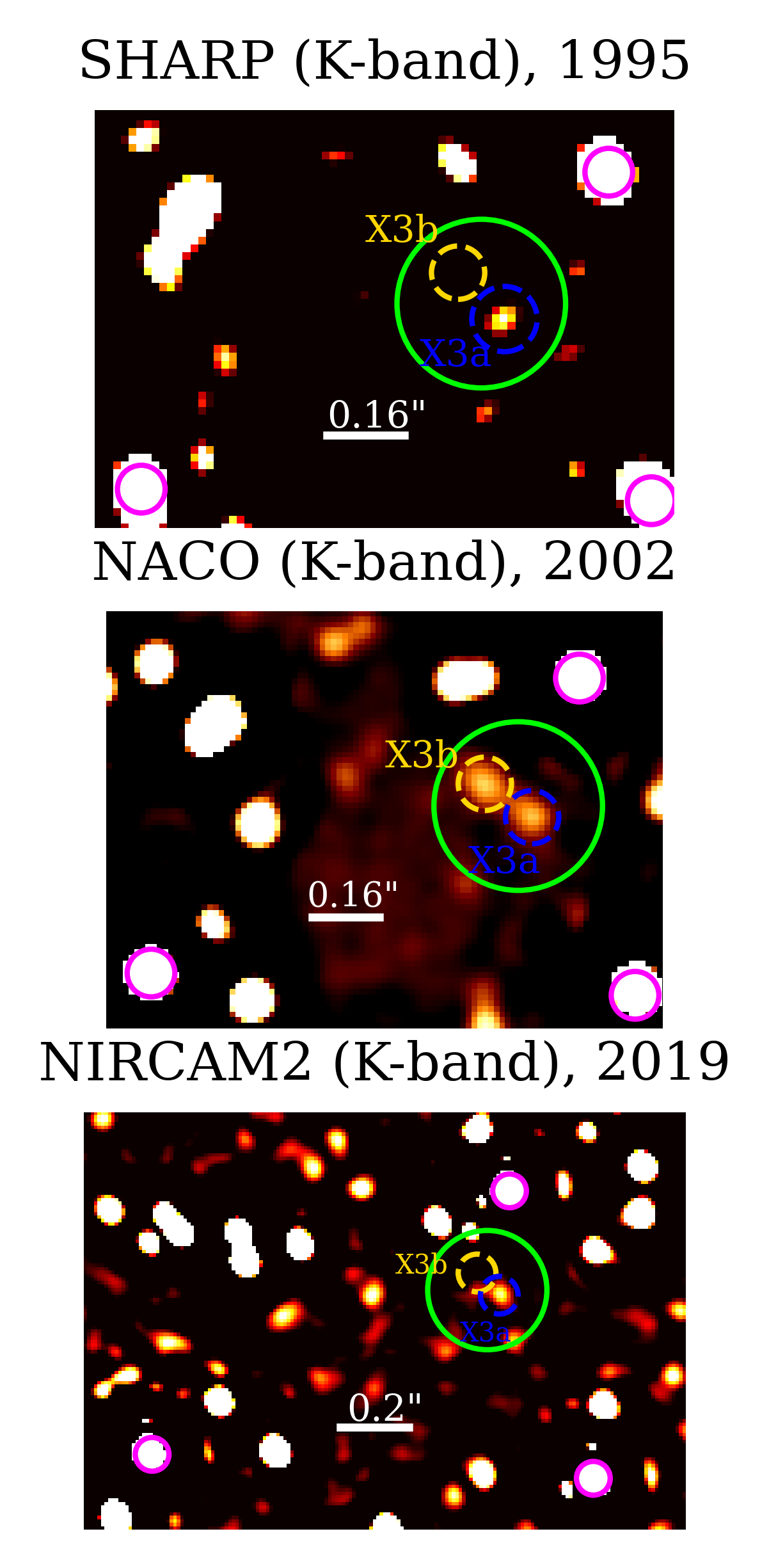}
	\caption{K-band detection of X3a in 1995, 2002, and 2019. The position of the dashed gold-colored circle marks the (expected) position of X3b. Because of the higher K-band magnitude of X3b compared to X3a in 2002, the blob should be traceable in 1995 and 2019. We indicate the position of X3a with a blue-colored dashed circle. For guidance, we marked the triangle stars S3-373, S3-374, and S3-375 with a magenta colored circle (please see Fig. \ref{fig:x3_system}). North is up, East is to the left.}
\label{fig:detection_x3a_x3b}
\end{figure}
Setting X3a as a reference source, we find that the distance of X3b decreases between 2002 and 2011 (see the K-band observations in Appendix \ref{sec:appendix_multiwavelength_detection}, Fig. \ref{fig:x3_all_k}). In the following years after 2011, the data does not show any significant $K$-band flux above the noise level for X3b. This is unexpected especially because the magnitude of X3b was higher compared to X3a in 2002. We pick three epochs to reflect this dynamical behavior of the X3 system. From the data, it is apparent that X3b exhibits no K-band emission above the noise level at the expected position neither in 1995 nor 2019 (Fig. \ref{fig:detection_x3a_x3b}).\newline
In addition to X3b, we observed a second blob in the L-band that moves along with X3a between 2002 and 2019. To exclude the chance of a sporadic fly-by, we center the available L-band NACO data on the position of the dusty envelope X3. The resulting stacked data is shown in Fig. \ref{fig:counterjet} where we also indicate the position of X3c with respect to the bow shock X3 (see also Fig. \ref{fig:x3_system}). We find a compact emission with a FHWM of about 0.1". If X3 and X3c would not be related, the flux of the blob would have been canceled out as it is eminent by the low background emission displayed in Fig. \ref{fig:counterjet}. In Sec. \ref{sec:discuss}, we estimate the statistical probability for a random fly-by event for objects close to the IRS13 cluster based on the K-band luminosity function (KLF) formulated by \cite{Paumard2006}.\newline
\begin{figure}[htbp!]
	\centering
	\includegraphics[width=0.4\textwidth]{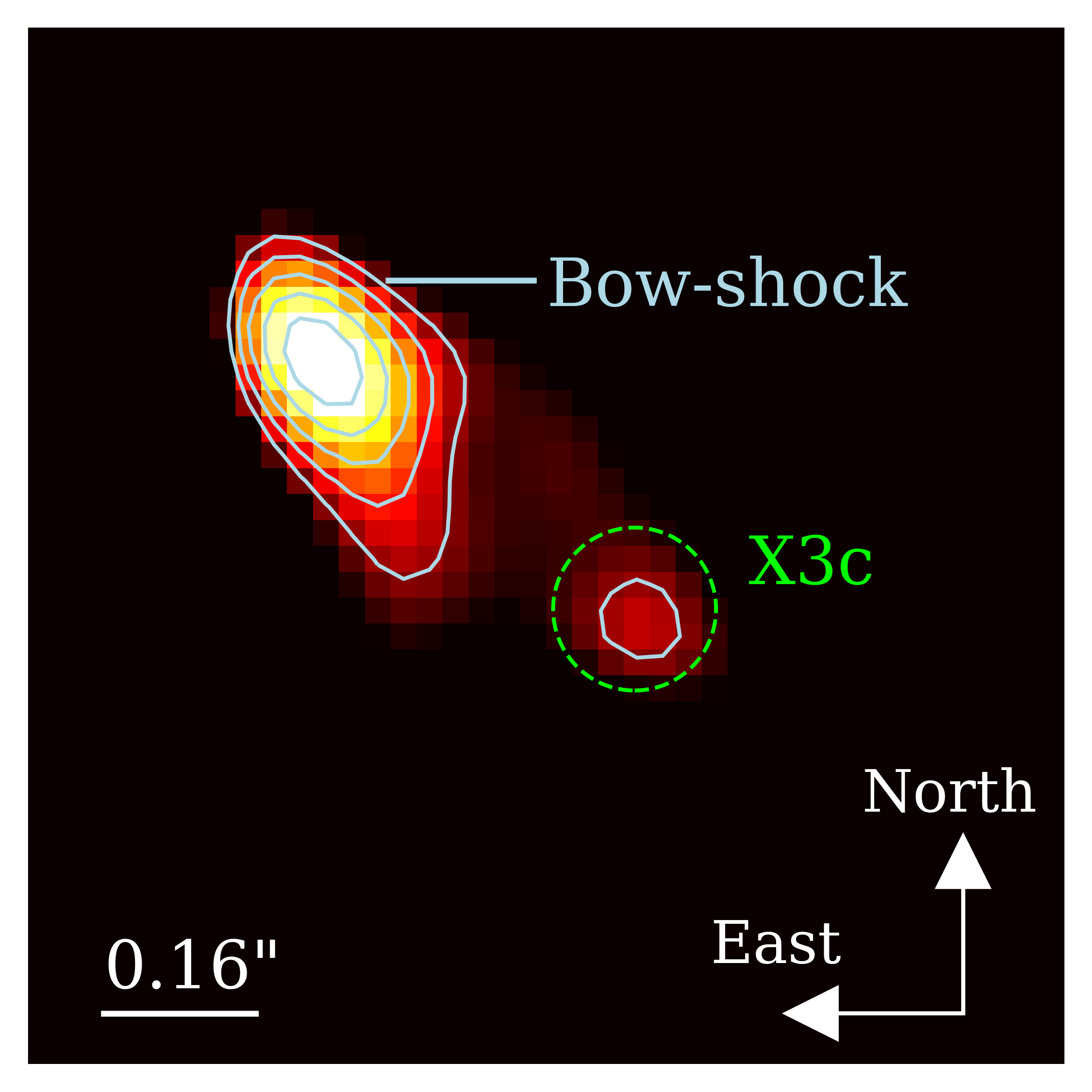}
	\caption{Dust blob X3c of the X3 system observed in the L-band with NACO. For the data presented, we stack all L-band NACO observations between 2002 and 2018 and subtract the triangle stars to emphasize the components of the system. We mark the position of X3 and the X3c blob. The contour lines refer to 20$\%$, 40$\%$, 60$\%$, 80$\%$, and 100$\%$ of the peak emission of the bow shock. Between 2002 and 2018, the distance of X3 to X3c stays constant. The measured projected length from tip to tail is about 0.8".}
\label{fig:counterjet}
\end{figure}
Furthermore, we determine the distance of X3 and X3a to Sgr~A* using the orbital elements of the well observed S2 orbit \citep[][]{Peissker2022}. From the orbital fit of S2, we calculate the position of Sgr~A* which then is used as a reference position to derive the distance of X3 and X3a. 
To account for uncertainties caused by the detector \citep{Plewa2018}, we chose $\pm\,0.5$ pixel to additionally incorporate the footprint of the variable background as well as the large distance ($>0.1$parsec) between the X3 system and Sgr~A*. With the spatial pixel scale of NACO of 0.0133 arcsec (H- and K-band) and 0.027 arcsec (L-band), this translates to an uncertainty of $\pm\,6.6$ mas and $\pm\,13.5$ mas. A short clip showing X3 and X3c is appended. 
\begin{figure}[htbp!]
	\centering
	\includegraphics[width=0.5\textwidth]{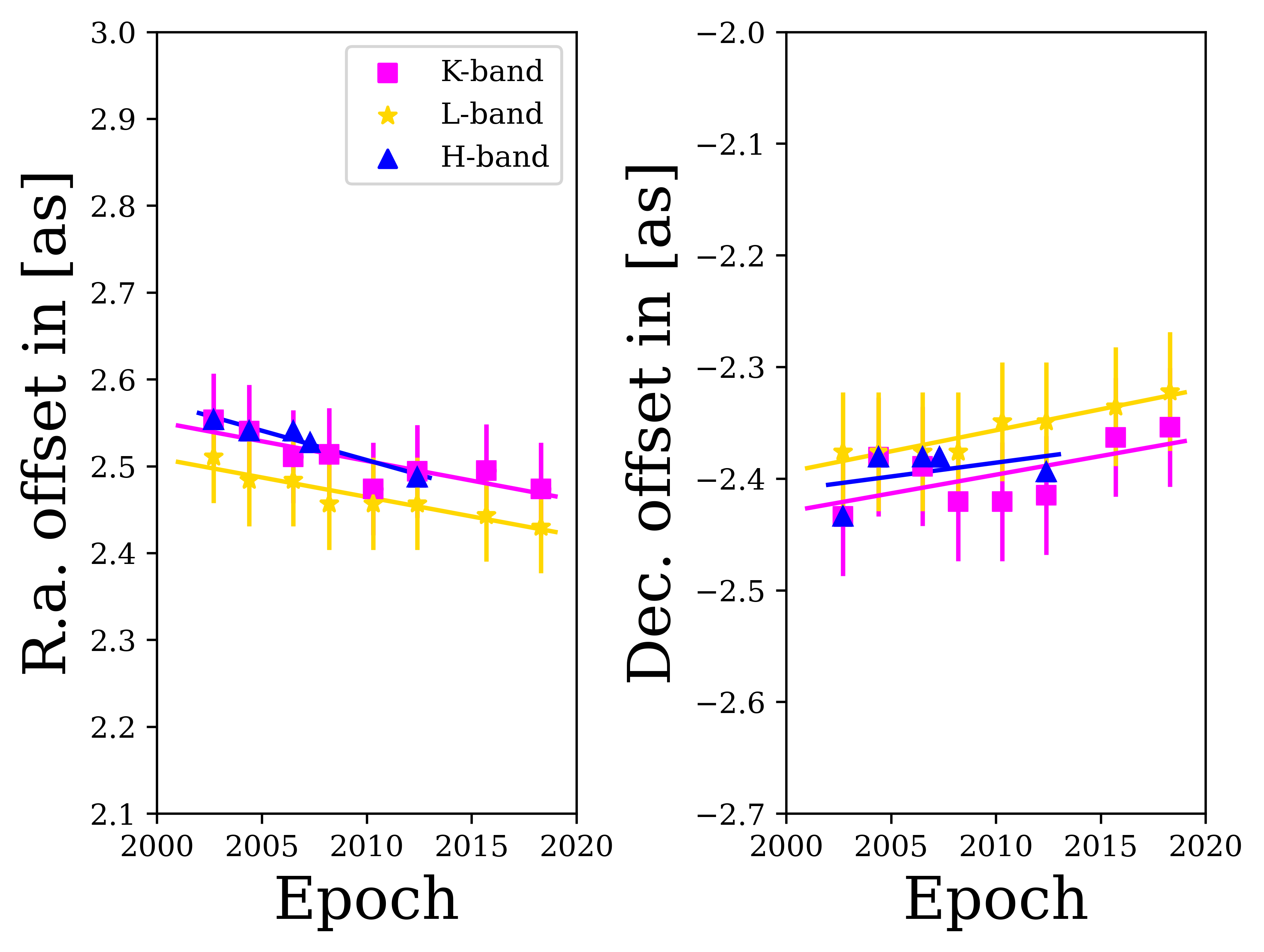}
	\caption{Proper motion of the X3-system based on NACO H-, K-, and L-band data. We use Sgr~A* as the reference position. We derive an average proper motion of the X3-system of $244\,\pm\,27$ km/s where the uncertainty is the standard deviation. Because the fitted central position of the bow shock deviates due to its dimensions from the K- and H-band, a certain offset is expected. However, the direction of the proper motion of all three objects is consistent, which is reflected by the fit.}
\label{fig:proper_motion}
\end{figure}
For the dust envelope X3 and the blob X3a, we find an averaged proper motion of $244\,\pm\,27$ km/s, as shown in Fig. \ref{fig:proper_motion}. Individual numerical values are $220\,\pm\,30$ km/s for X3a (K-band) and $229\,\pm\,30$ km/s for the bow shock X3 (L-band). For the H-band detection of X3a, we find a proper motion of $282\,\pm\,30$ km/s which is slightly higher than the L- and K-band. Given the lower data baseline and the higher noise level of the H-band data, this is expected. However, the estimation of the proper motion of X3 and X3a in the H-,K-, and L-band are in reasonable agreement considering the uncertainties of the analysis. The continuum data used for the proper motion observed with NACO is shown in Appendix \ref{sec:appendix_multiwavelength_detection}, Fig. \ref{fig:x3_all_h}, Fig. \ref{fig:x3_all_k}, and Fig. \ref{fig:x3_all_l}.\newline
As mentioned above, the distance between X3a and X3b decreases between 2002 and 2012. In addition, we find a constant distance between X3a and X3c. We reflect this manifold behavior of the two components in Fig. \ref{fig:distance_blobs} where we assumed that X3b was created around 1995. However, any other assumed epoch for the creation of X3b before 2002 does not impact the overall trend of the decreasing distance between 2002 and 2011.
\begin{figure}[htbp!]
	\centering
	\includegraphics[width=0.5\textwidth]{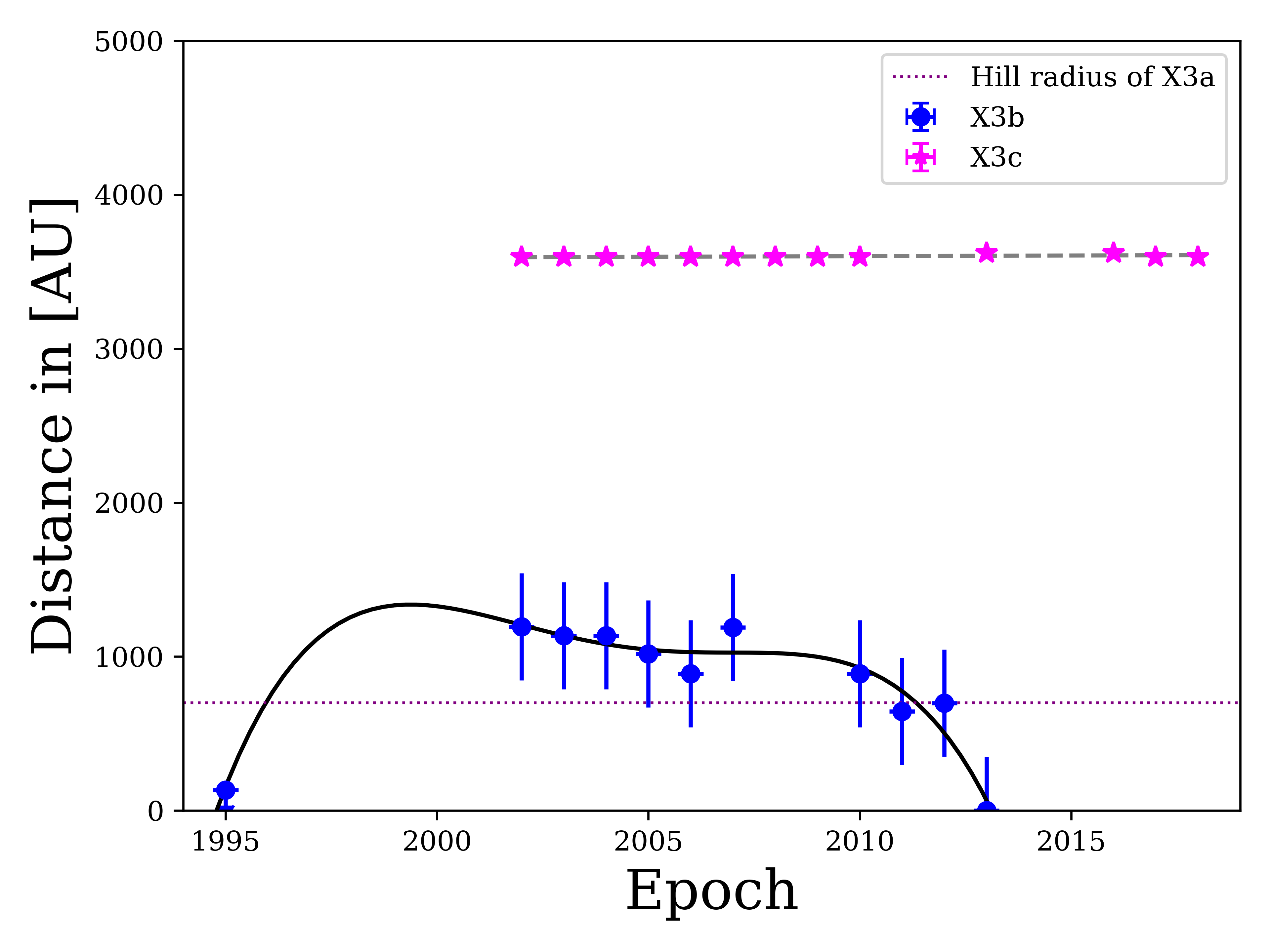}
	\caption{Distance of X3b and X3c with respect to X3a. We trace X3b without confusion until 2012 in the NACO K-band data (see Fig. \ref{fig:x3_all_k}). {The data implies that the thermal blob X3b got created or ejected between 1995 and 2002. Between 2002 and 2007, the distance of X3b to X3a stayed constant with about 1000 AU.} After 2012, we do not find any significant K-band emission above the detection limit at the expected position of X3b. In addition, we trace X3c without confusion along with X3 in the L-band between 2002 and 2019. All data points show the related standard deviation of the fit (black and grey dashed line). The {purple dotted} line indicates the Hill radius where objects are expected to be significantly influenced by the gravitational field of X3a (see Sec. \ref{sec:discuss}).}
\label{fig:distance_blobs}
\end{figure}
From the data displayed in Fig. \ref{fig:counterjet} and Fig. \ref{fig:distance_blobs}, it is evident that the proper motion of X3c matches the estimated $244\,\pm\,27$ km/s for X3a and the bow shock X3. Consequently, we identify without confusion all three components at their expected positions in 2019 (Fig. \ref{fig:2019_keck}).
\begin{figure}[htbp]
	\centering
	\includegraphics[width=0.4\textwidth]{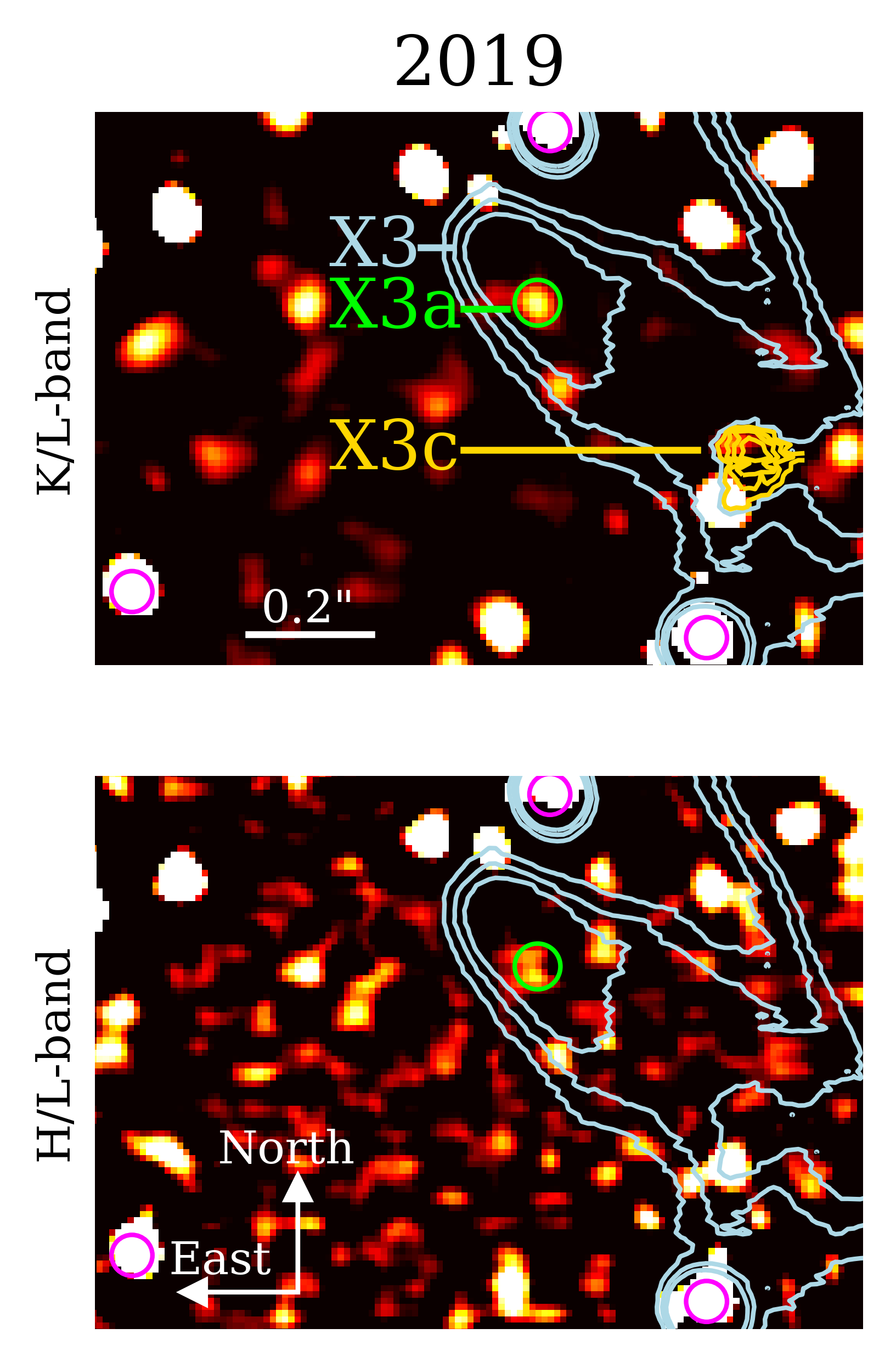}
	\caption{Multi-wavelength detection of the X3-system in the H-, K-, and L-band with NIRC2 (KECK) in 2019. The upper panel shows a combination of the K-band (background image) and the L-band ({light-blue} contours). The lower panel shows the H-band emission whereas the lightblue-colored contours are again indicating the L-band emission of the bow shock X3. The green circle is placed at the identical position showing the H- and K-band emission X3a. The blob X3c is hinted by the golden contours above S3-374 which is together with S3-373 and S3-375 indicated by a magenta colored circle (see Fig. \ref{fig:x3_system}). No emission of X3b above the noise level is detected at the expected position. We refer the interested reader to the evolution of X3a and X3b shown in Fig. \ref{fig:x3_all_k}, Appendix \ref{sec:appendix_multiwavelength_detection}. Please also consider the setup of the X3-system in 2002 as observed with NACO (Fig. \ref{fig:detection_x3a_x3b}).}
\label{fig:2019_keck}
\end{figure}
%Especially the NACO observation of 2002 shown in Fig. \ref{fig:detection_x3a_x3b} and the NIRCAM2 detection 2019 as presented in Fig. \ref{fig:2019_keck} provides a good impression of the evolution of the X3 system in the K-band.
%In the K-band, we find a matching trajectory of X3a with the dusty envelope X3 between 2002 and 2019 implying that both objects are related.  

\subsection{Line emission of the X3-system}

\begin{figure}[htbp!]
	\centering
	\includegraphics[width=.5\textwidth]{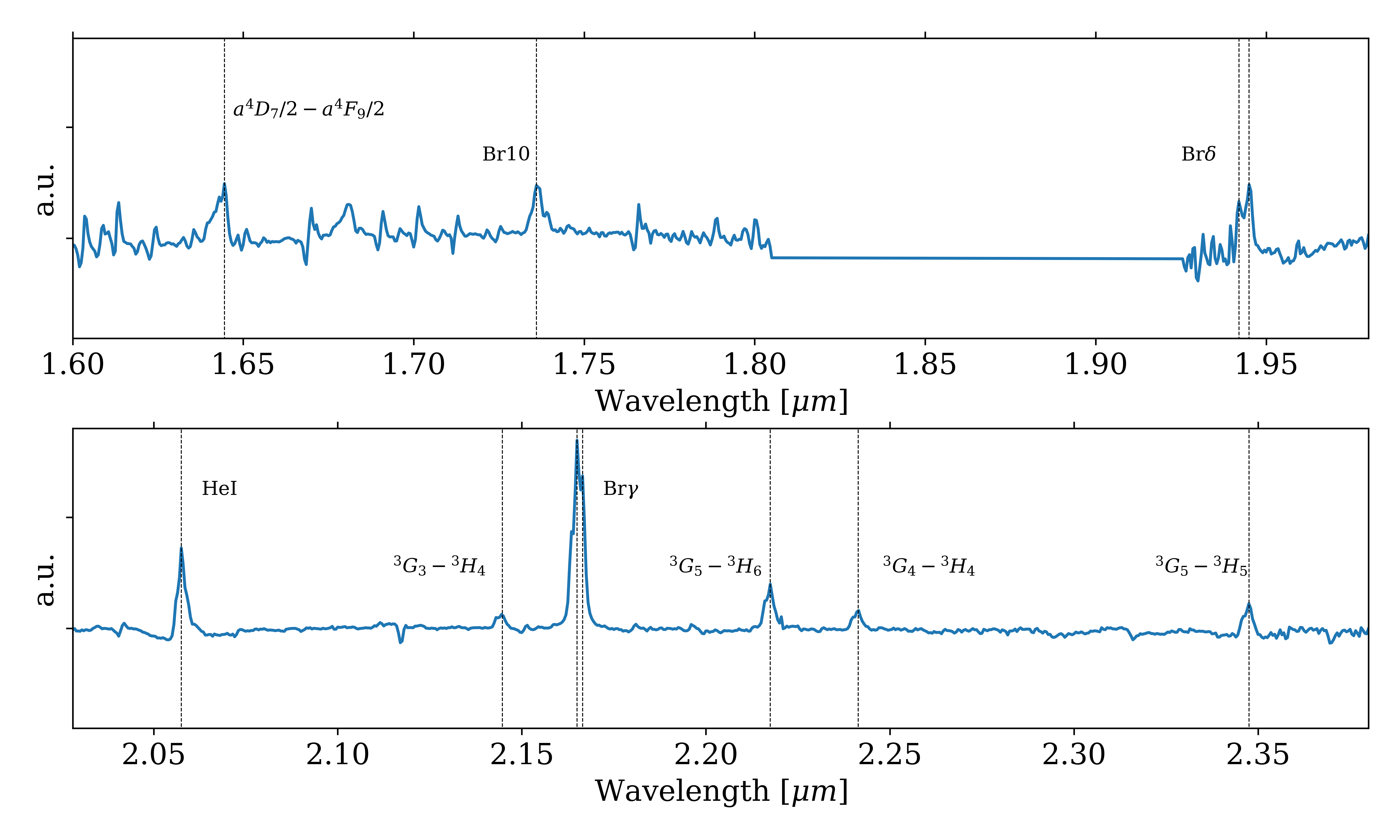}
	\caption{Spectrum (H+K band) of the X3-system extracted from the SINFONI data observed in 2014. Because of different line strengths, we subdivide the spectrum into two subplots (top and bottom). The emission shows tracers such as a double-peaked Br$\gamma$ line around 2.1661$\mu m$, and a P-Cygni profile close to the blueshifted HeI line at 2.0575$\mu m$. We find no { significant} NIR CO tracers which excludes { an evolved} late-type nature of X3a. We will discuss the missing CO absorption lines in detail in Sec. \ref{sec:discuss}.}
\label{fig:x3_spec_sinfo}
\end{figure}

\begin{figure}[htbp!]
	\centering
	\includegraphics[width=0.5\textwidth]{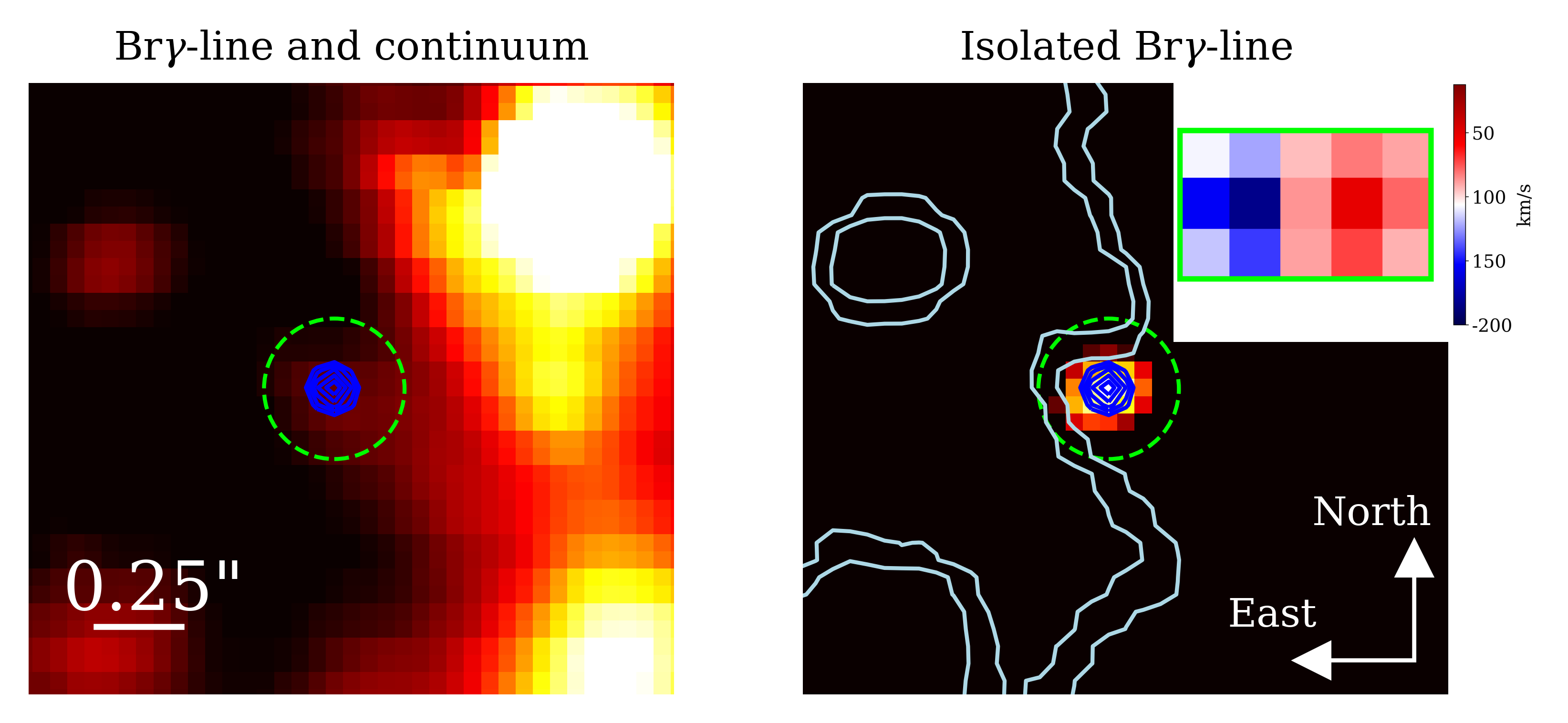}
	\caption{Combined continuum and Br$\gamma$ line emission observed with SINFONI. The left plot shows the Br$\gamma$ emission in combination with the continuum (indicated by the surrounding gas and stellar emission). The right image shows a 3 pixel smoothed isolated Br$\gamma$ line where we subtracted the continuum. For the isolated Br$\gamma$ line on the right, we integrate over the full width of the double peaked emission shown in Fig. \ref{fig:x3_spec_sinfo}. {For the blue contour lines, we adapt the normalized emission of the shown isolated Br$\gamma$-line corresponding to 50$\%$, 60$\%$, 70$\%$, 80$\%$, and 90$\%$ of the peak intensity.}
	In the upper right, we show a position-position-velocity diagram at the position of the double peak Br$\gamma$ line. The velocity range is about $\Delta v\,\simeq \,200$km/s. Additional material shows the velocity gradient and the related offset of the blue- and red-shifted Br$\gamma$ line. A short clip showing the double peak Br$\gamma$ line is appended.}
\label{fig:linemaps_sinfo_sinfo}
\end{figure}

\begin{figure}[htbp!]          %\hline
          %Pf$\delta @$3.2969 $\mu$m & n=9-5             & 3.3018 $\pm$ 0.0002   & 445 \\
	\centering
	\includegraphics[width=.4\textwidth]{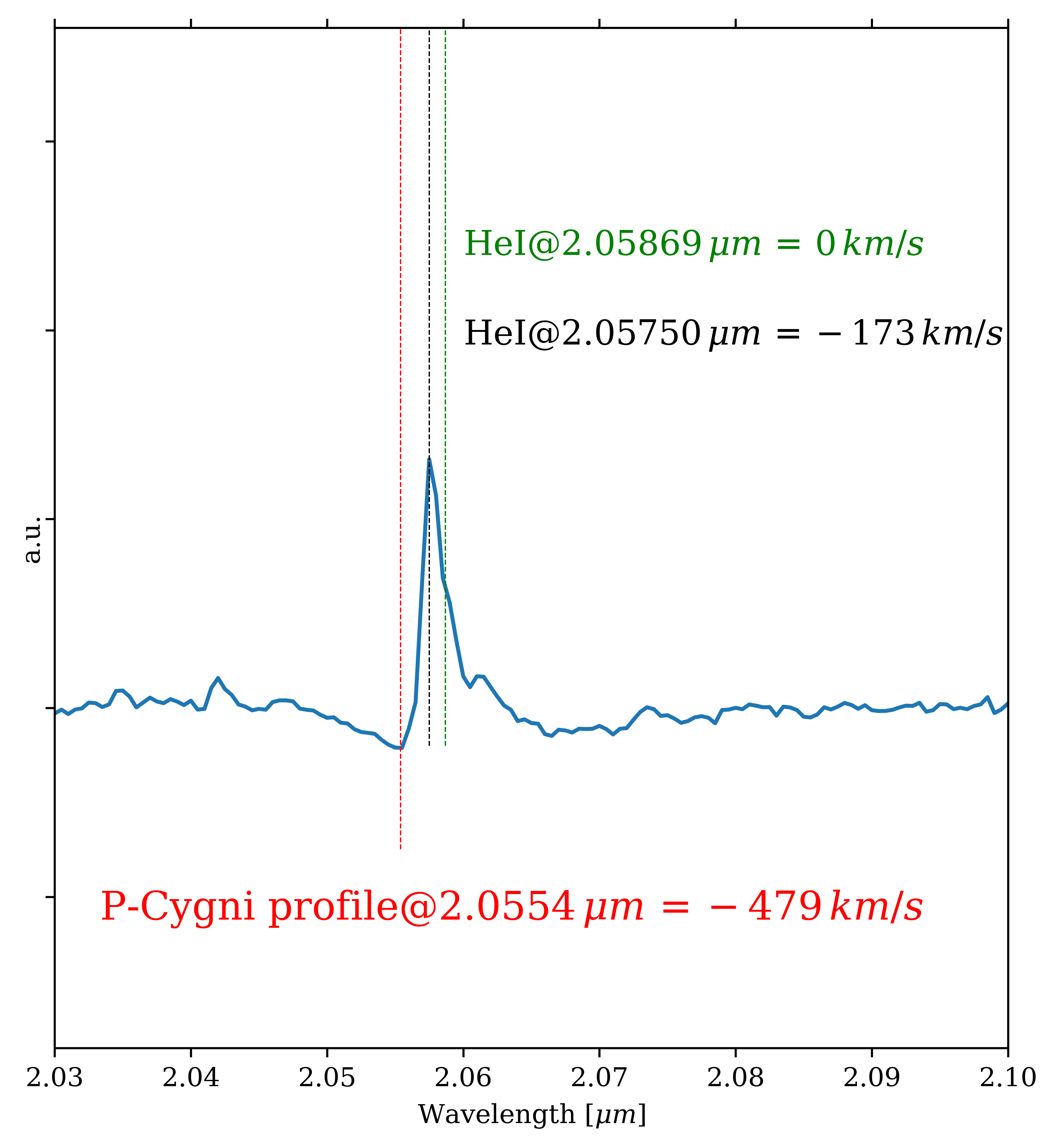}
	\caption{P-Cygni profile of X3a observed with SINFONI. The infrared HeI (transition $2p^1 P^0-2s^1S$) line with a rest wavelength of $2.0586\,\mu m$ is marked with a green dashed line. The spectrum of X3a exhibits a blue-shifted HeI line with a related line of sight velocity of -173 km/s. Furthermore, a P-Cygni profile at $2.0554\,\mu m$ with -479 km/s is clearly traceable.}
\label{fig:pcygni_sinfo}
\end{figure}

\begin{figure*}[htbp!]
	\centering
	\includegraphics[width=0.8\textwidth]{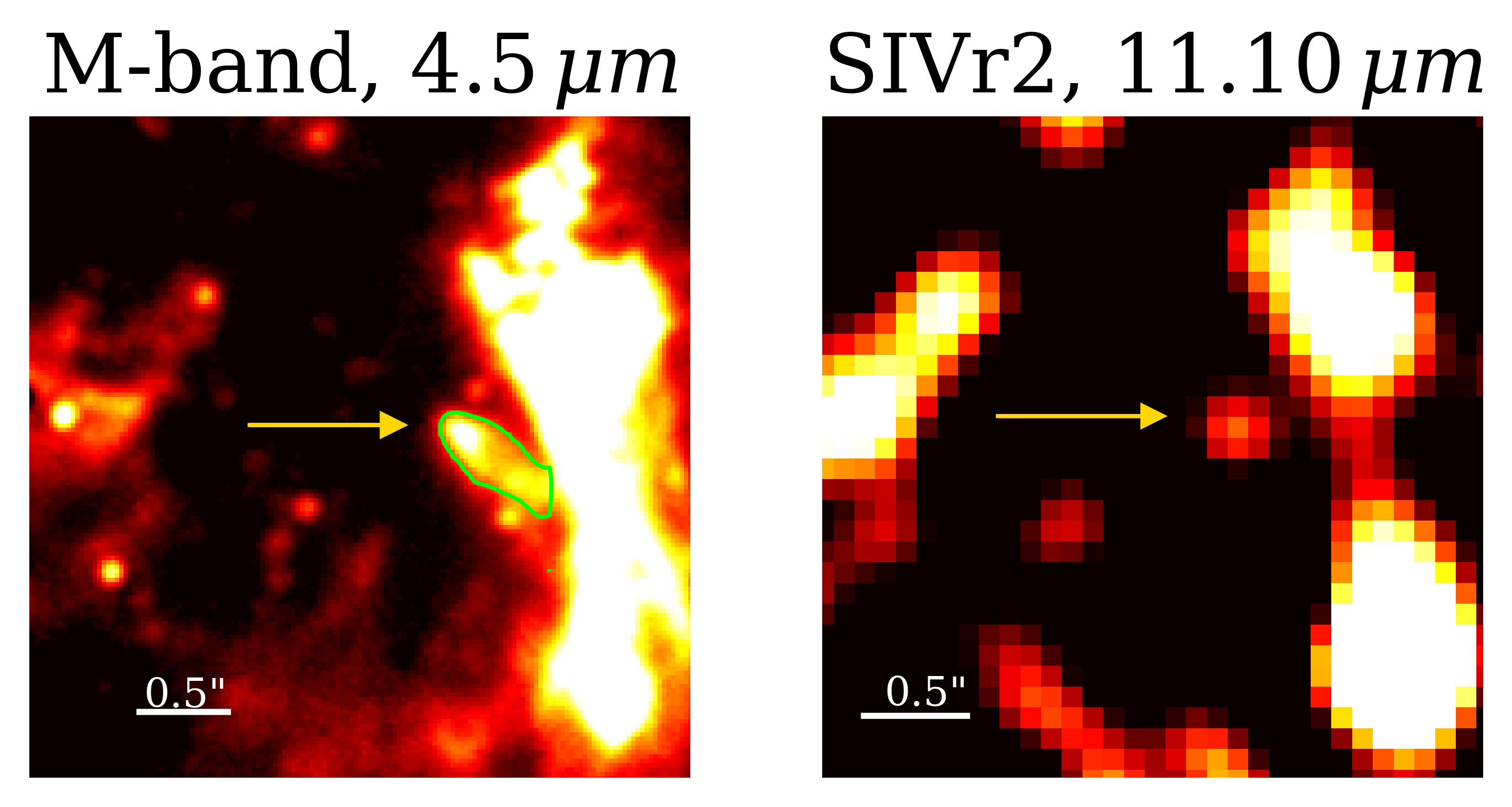}
	\caption{Comparison of the M-band continuum with the SIVr2 line emission of X3. The left image was observed with NACO in 2012, the right image shows observations carried out with VISIR in 2004. The golden colored arrow indicates the position of X3, the M-band observations exhibits lime colored contour lines of the bow shock source at 35$\%$ of the peak emission (see Table \ref{tab:mag_flux}). The length of the bow shock observed at 4.5$\mu m$ is about 0.8 arcsec measured from tip to tail, the FHWM of the SIVr2 line emission is about 0.3 arcsec. Here, North is up, East is to the left.}
\label{fig:sivr2_visir}
\end{figure*}

\begin{figure*}[htbp!]
	\centering
	\includegraphics[width=1.\textwidth]{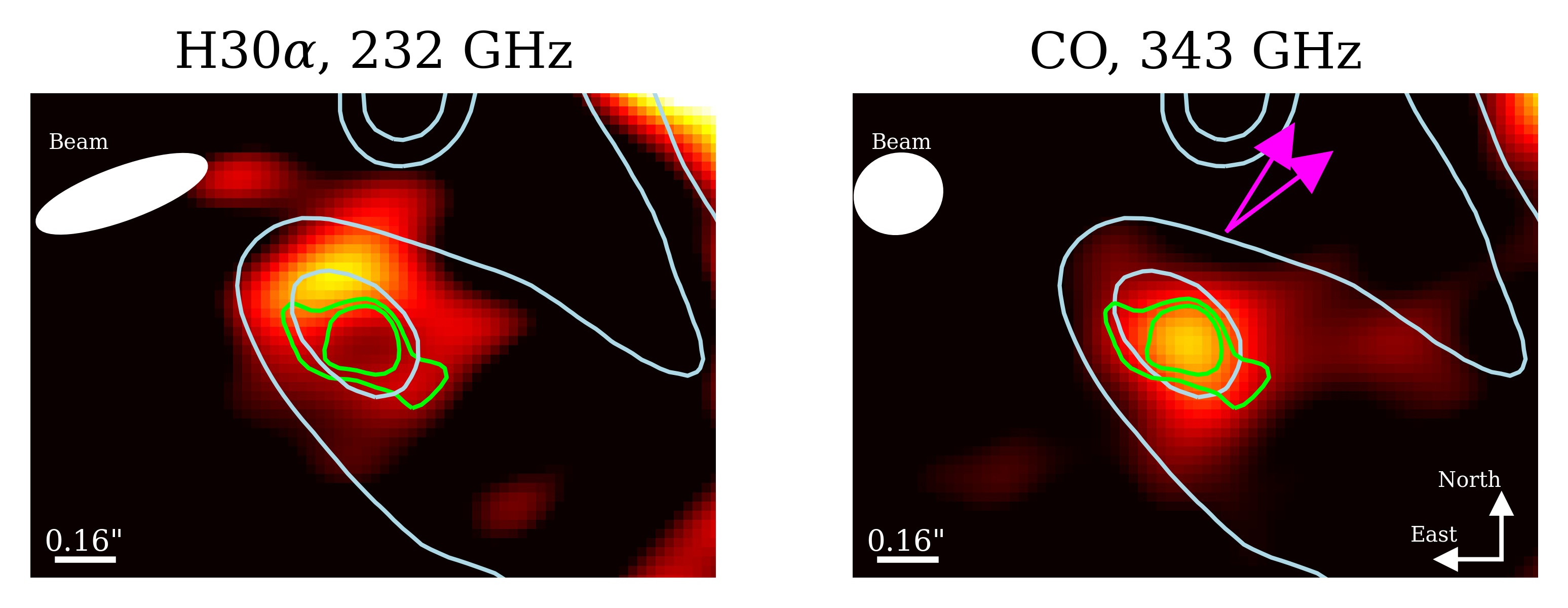}
	\caption{Gas emission observed with ALMA overlaid with K- and L-band contour lines of X3a and X3. With lime-colored contours, we indicate the K-band peak emission of 2.8 mJy at 20$\%$ and 40$\%$ for X3a as observed in 2015 (see Table \ref{tab:mag_flux}). {With light-blue contours}, the L-band dust emission of X3 of the same epoch is shown. The background image was observed with ALMA and shows H30$\alpha$ and CO radio/submm continuum emission at the position of X3a. {While the H30$\alpha$ line seems to be arranged in a ring-like structure around X3a following the morphology of the recently observed massive protostar G28.20-0.05, the CO line exhibits a compact density distribution. Please note that the K-band contours do not imply the true size of the stellar source X3a. In the right figure, we show proper motion vector with a velocity of about $244\pm 27$km/s (magenta colored, Fig. \ref{fig:proper_motion}).}}
\label{fig:x3_disk}
\end{figure*}

Inspecting SINFONI data observed in 2014 for spectroscopic analysis, we find several emission lines, such as Br10, Br$\delta$, Br$\gamma$, HeI, Pa$\gamma$, and the prominent [FeIII] multiplet that are related to X3a (see Fig. \ref{fig:x3_spec_sinfo}). For the telluric-corrected\footnote{We use the standard star Hip094122 for the telluric correction.} SINFONI spectrum, we applied a 3-pixel aperture to the position of X3a. {To avoid the contamination of the spectrum, we do not apply a background subtraction (see Appendix \ref{sec:appendix_spectral_analysis} for a further discussion).}
The observed Br$\gamma$ line exhibits a double-peak profile with a blue- and a red-shifted velocity of -152 km/s and +55 km/s, respectively. The isolation of the Doppler-shifted Br$\gamma$ lines and the subtraction of the underlying continuum reveal a compact gas emission at the position of X3a which is shown in Fig. \ref{fig:linemaps_sinfo_sinfo}. By selecting the individual Br$\gamma$ peaks, we create two related line maps that show the distribution of the ionized gas. We append a short movie that shows the offset of the line that we interpret as {photoionized outflows originating in a gaseous disk close to X3a.} 
{ Futhermore, we construct} a position-position-velocity (PPV) map based on the Doppler-shifted Br$\gamma$ emission and include it in Fig. \ref{fig:linemaps_sinfo_sinfo}. This PPV map shows a continuous velocity gradient with $\Delta v\,\simeq \,200$km/s. Analyzing the SINFONI spectrum, we find no tracers of CO band heads at 2.29$\mu m$, 2.32$\mu m$, 2.35$\mu m$ and 2.38$\mu m$ implying a young age of X3a. The dimensions of the Br$\gamma$-line distribution can be treated as an upper limit due to uncertainties imposed by the variable background and the marginally resolved emission.\newline 
Furthermore, we find a P-Cygni profile for the near-infrared HeI line observed with a blue-shifted velocity of about -480 km/s (Fig. \ref{fig:pcygni_sinfo}), which indicates a stellar outflow of $\sim 400$ km/s when corrected for the LOS velocity of the source, which is $\sim -50$ km/s as inferred from the average of red and blue peaks of Br$\gamma$ line. Interestingly, the blue-shifted Br$\delta$ velocity of -478 km/s matches the peak velocity of the P-Cygni profile, implying a correlation and the potential origin of the recombination line in the outflow. A detailed investigation of a possible connection between Br$\delta$ and the P-Cygni profile is beyond the scope of this work and should be analyzed separately. We note that \cite{Mizumoto2018} find [FeII] emission lines related to the investigated P-Cygni absorption profiles. From this work, it is plausible that emission and absorption features in the spectrum seem to be related.
{Table \ref{tab:emissionlines} lists all detected Dopplershifted NIR emission lines as observed with SINFONI. In addition to the H+K band, we investigate the observation campaign that covers the region around X3, which was carried out with VISIR in the N- and Q-band}.
\begin{table*}[hbt!]
    \centering
    \begin{tabular}{|c|ccc|}
         \hline %& flux [10$^{-16}$ erg/s/cm$^{2}$]
         \hline
           Spectral line ($@$rest wavelength [$\mu$m]) & transition & central wavelength [$\mu$m] & Velocity [km/s] \\
         \hline
          $\rm [FeII]@$1.6440 $\mu$m & $a^4D_{7/2} - a^4F_{9/2}$    & 1.6445 $\pm$ 0.0002 & 91 \\
          Br10 $@$1.7366 $\mu$m &         n=10-4         & 1.7360 $\pm$ 0.0002 & -104 \\
          Br$\delta$ $@$1.9450 $\mu$m &   n=8-4          & 1.9419 $\pm$ 0.0002 & -478 \\
          HeI $@$2.0586 $\mu$m &    $2p^1 P^0-2s^1S$     & 2.0554 $\pm$ 0.0002 & -479 \\
          HeI $@$2.0586 $\mu$m &    $2p^1 P^0-2s^1S$     & 2.0575 $\pm$ 0.0002 & -173 \\
          H$_2$ $@$2.1218 $\mu$m & v=1-0 S(1)            & 2.1232 $\pm$ 0.001  &  198 \\
          Br$\gamma$ $@$2.1661 $\mu$m & n=7-4            & 2.1650 $\pm$ 0.0002 &  -152 \\
          Br$\gamma$ $@$2.1661 $\mu$m & n=7-4            & 2.1665 $\pm$ 0.0002 &   55 \\
          $\rm [FeIII]@$2.1451 $\mu$m & $^3G_3\,-\,^3H_4$    & 2.1447 $\pm$ 0.0002 & -56 \\
          $\rm [FeIII]@$2.2178 $\mu$m & $^3G_5\,-\,^3H_6$    & 2.2175 $\pm$ 0.0002 & -41 \\
          $\rm [FeIII]@$2.2420 $\mu$m & $^3G_4\,-\,^3H_4$    & 2.2414 $\pm$ 0.0002 & -80 \\
          $\rm [FeIII]@$2.3479 $\mu$m & $^3G_5\,-\,^3H_5$    & 2.3475 $\pm$ 0.0002 & -51 \\
    \hline
    \end{tabular}
    \caption{Emission lines extracted from the SINFONI (NIR) data cube of 2014. We list the rest wavelength of the related emission line. Furthermore, we indicate the transition of the investigated species. The uncertainty of the derived Doppler-shifted velocity is about $\pm\,25$ km/s.}
    \label{tab:emissionlines}
\end{table*}
Based on these MIR observations, we trace NeII, PAH1, ArIII-, SIVr1-, SIVr2-, PAH2-, Q2-, and Q3-lines which are associated with YSOs \citep{Woitke2018}. In Fig. \ref{fig:sivr2_visir}, we present a SIVr2 continuum map with the indicated position of X3. All other lines are shown in Appendix \ref{sec:appendix_multiwavelength_detection}, Fig. \ref{fig:x3_all_visir}.    
Although the spatial pixel scale of the VISIR data is lower ($\sim\,0.1$arcsec) compared to the NACO data ($\sim\,0.027$arcsec), the elongation of the bow shock X3 should be detectable. We measure a rough size of about 0.8 arcsec in the M-band for the dusty emission associated with X3 (see Fig. \ref{fig:sivr2_visir}) whereas the SIVr2 line exhibits a FWHM of 0.3 arcsec. Hence, we expect an emission almost three times larger for the lines observed with VISIR if the MIR dust emission matches the distribution of the SIVr2 line. In contrast, the observations show that the MIR emission lines arise from a very dense and compact region which does not exhibit any elongation (please see also Appendix \ref{sec:appendix_multiwavelength_detection}, Fig. \ref{fig:x3_all_visir}).\newline
To cover longer wavelength regimes compared to the IR observations, we display the results of some of the ALMA observation campaigns targeting the vicinity of Sgr~A*. The presented radio data is also analyzed in \cite{Tsuboi2017}, \cite{tsuboi2019}, \cite{Tsuboi2020a}, and \cite{Tsuboi2020b} where the authors investigate, among other things, the IRS 13 cluster in detail.
In Fig.~\ref{fig:x3_disk}, we display the radio/submm emission at the position of X3a. In the same figure, we incorporate the contour lines of the stellar emission X3a (lime-colored) and the dust envelope (light blue-colored). The CO emission originates at the stellar position, { whereas} the H30$\alpha$ line is distributed around X3a {in a ring-like structure as it has been recently observed for the isolated massive YSO G28.20-0.05 by \cite{Law2022}}. Compared to the downstream bow-shock section, the H30$\alpha$ emission at the tip of the bow shock is enhanced by $\sim $20$\%$. This increased flux density at the position of the bow shock front implies heating caused by the interaction of X3 with the ambient medium.
Taking into account the ALMA observations presented in Fig. \ref{fig:x3_disk} and considering the SINFONI/VISIR line observations, it is evident that the ionized gas species seem to originate at different components inside the X3 system.

\subsection{Photometric analysis}

{ In this subsection}, we present the results of the photometric analysis. {Since \cite{Gautam2019} finds that more than 50$\%$ of their star sample (about 570) located in the NSC are variable, we are limited in our selecting of a proper reference star that is used to estimate the magnitudes and flux densities of the X3-system. Hence, variability but also fluctuating extinction values \citep[][]{Peissker2020c} confront the analysis with uncertainties that need to be targeted. Because the bright and close-by members E1, E2, and E3 of the IRS13 cluster show varying extinction in all investigated bands \citep[see Fig. \ref{fig:x3_system} and][]{muzic2008, Fritz2011}, we will address these fluctuations using an asymmetric uncertainty range for the estimated magnitude and flux densities (Table \ref{tab:mag_flux}). Furthermore, we used} the most observed object in the NSC, the S-cluster star S2 as a calibration source for the K-band whenever possible. Including the H-band, we note that the flux analysis of S2 in the literature shows inconsistencies. Therefore, we fit a blackbody SED to the emission of S2 with known individual numerical results (Table \ref{tab:mag_fux_reference_values}) for the K-\citep{Sabha2012}, L- and M-band \citep[both values are taken from][]{Viehmann2006}. For the SED model presented in Fig. \ref{fig:s2_sed} (Appendix \ref{sec:appendix_s2_sed}), we implement common B2V stellar-type parameters of $\rm R\,=\,8.5\,R_{\odot}$ and $\rm T_{eff}\,=\,22500$K \citep[see also][]{Hanson1996}. From the fit, we derive a flux density of $\rm 32.0\,\pm\,0.2\,mJy$ with the related magnitude of 16.0 mag \citep{Schoedel2010, Peissker2020b} for the S-cluster star S2 in the H-band. 
For mid-infrared analysis, we include the PAH1 and NeII emission of the close-by reference source IRS2L \citep[][]{Bhat2022} and the relation
\begin{equation}
    \rm mag_{\lambda}\,=\,-2.5\times log(f_{\lambda}/f_0)
    \label{eq:apparent_magnitude}
\end{equation}
where $\rm mag_{\lambda}$ is the magnitude of the related band and $f_0$ the zero flux adapted from \cite{Tokunaga2007}\footnote{\citet{Tokunaga2007} use Vega as a zero magnitude star.}.
\begin{table}[hbt!]
    \centering
    \setlength{\tabcolsep}{0.9pt}
    \begin{tabular}{|cccc|}
         \hline 
         \hline
         Band   & Magnitude & Flux$_{\lambda}$ & Reference  \\
            &  [mag] &  [Jy] &   \\

         \hline
         H-band  & 16.0 & 0.032  & S2, see text \\
         K-band  & 14.1 & 0.0147 & S2, \cite{Sabha2012} \\
         %L-band  & 11.1 & 0.0088  & S2, \cite{Hosseini2020} \\
         L-band  & 6.4 & 2.98  & IRS2L, \cite{Viehmann2006} \\
         M-band  & 5.5 & 3.98 & IRS2L, \cite{Viehmann2006} \\
         PAH1  & 0.9 & 21.4 & IRS2L, \cite{Bhat2022} \\
         NeII  & 0.4 & 18.9 & IRS2L, \cite{Bhat2022} \\
       \hline
    \end{tabular}
    \caption{Derredened reference values for S2 and IRS2L used in this work. The related references are listed \citep[see also][]{Viehmann2007}. For PAH1 and NeII, we use a zero flux of 50 Jy and 28.6 Jy, respectively \citep{Tokunaga2007}.}
    \label{tab:mag_fux_reference_values}
\end{table}
With the magnitude and flux information, we photometrically analyze almost 20 years of observations of the GC using
\begin{equation}
    \rm F_{\lambda}\,=\,F_{0}\,\times\,10^{(-0.4(mag_{2}-mag_{1}))}
    \label{eq:flux}
\end{equation}
where mag$_2$ refers to the { investigated object} and mag$_1$ to the reference source. The resulting magnitudes and fluxes are listed in Table \ref{tab:mag_flux}. The H-, K-, L-, and M-band data is observed with NACO, the narrow filter PAH1 and NeII with VISIR.
\begin{table*}[hbt!]%\subsection{Herbig-Haro object}

    \centering
    \setlength{\tabcolsep}{0.5pt}
    \begin{tabular}{|c|cccccccccccc|}
         \hline %& flux [10$^{-16}$ erg/s/cm$^{2}$]
           Year & \multicolumn{2}{c}{H-band} & \multicolumn{2}{c}{K-band} & \multicolumn{2}{c}{L-band} & \multicolumn{2}{c}{M-band} & \multicolumn{2}{c}{PAH1} & \multicolumn{2}{c|}{NeII}\\
           
          [YYYY] & [mag] & [mJy] & [mag] & [mJy] & [mag] & [Jy] & [mag] & [Jy] & [mag] & [Jy] & [mag] & [Jy] \\
        \hline
           2002 & $19.6^{+0.7}_{-0.3}$ & $1.1^{+0.4}_{-0.5}$ & $16.3^{+0.7}_{-0.3}$ & $1.9^{+0.6}_{-0.9}$ & $9.8^{+0.8}_{-0.3}$ & $0.13^{+0.03}_{-0.07}$  & - & - & - & - & - & - \\
           2003 & - & - & $15.8^{+0.7}_{-0.3}$ & $3.0^{+1.0}_{-1.4}$ & $9.9^{+0.8}_{-0.3}$  & $0.11^{+0.04}_{-0.06}$ & - & - & - & - & - & - \\
           2004 & $19.2^{+0.7}_{-0.3}$ & $1.6^{+0.6}_{-0.7}$ & $15.8^{+0.7}_{-0.3}$ & $3.0^{+1.0}_{-1.4}$ & $9.9^{+0.8}_{-0.3}$  & $0.11^{+0.04}_{-0.06}$ & - & - & $3.0^{+0.3}_{-0.3}$ & $3.1^{+0.9}_{-0.7}$ & $1.4^{+0.3}_{-0.3}$ & $7.5^{+2.4}_{-1.8}$ \\
           2005 & - & - & $15.5^{+0.7}_{-0.3}$ & $3.0^{+1.3}_{-0.9}$ & $9.7^{+0.8}_{-0.3}$  & $0.14^{+0.04}_{-0.07}$ & - & - & - & - & - & - \\
           2006 & - & - &         -            &        -            & $9.6^{+0.8}_{-0.3}$  & $0.15^{+0.05}_{-0.08}$ & - & - & - & - & - & - \\
           2007 & - & - & $15.9^{+0.7}_{-0.3}$ & $2.8^{+0.9}_{-1.4}$ & $9.5^{+0.8}_{-0.3}$  & $0.17^{+0.05}_{-0.09}$ & - & - & - & - & - & -  \\
           2008 & - & - & $16.0^{+0.7}_{-0.3}$ & $2.5^{+0.8}_{-1.2}$ & $9.4^{+0.8}_{-0.3}$  & $0.18^{+0.06}_{-0.09}$ & - & - & - & - & - & - \\
           2009 & - & - & $15.9^{+0.7}_{-0.3}$ & $2.8^{+0.9}_{-1.4}$ & $9.3^{+0.8}_{-0.3}$  & $0.20^{+0.07}_{-0.10}$ & - & - & - & - & - & - \\
           2010 & - & - & $15.7^{+0.7}_{-0.3}$ & $3.3^{+1.1}_{-1.5}$ & $8.9^{+0.8}_{-0.3}$  & $0.29^{+0.10}_{-0.15}$ & - & - & $2.4^{+0.3}_{-0.3}$ & $5.3^{+1.7}_{-1.3}$ & $1.5^{+0.3}_{-0.3}$ & $6.8^{+2.2}_{-1.6}$ \\
           2011 & - & - & $15.8^{+0.7}_{-0.3}$ & $3.0^{+1.0}_{-1.4}$ & $8.4^{+0.8}_{-0.3}$  & $0.47^{+0.15}_{-0.25}$ & - & - & - & - & - & - \\
           2012 & $19.2^{+0.7}_{-0.3}$ & $1.6^{+0.6}_{-0.7}$ & $15.8^{+0.7}_{-0.3}$ & $3.0^{+1.0}_{-1.4}$ & $8.4^{+0.8}_{-0.3}$  & $0.47^{+0.15}_{-0.25}$ & $7.9^{+0.7}_{-0.3}$  & $0.43^{+0.14}_{-0.24}$ & - & - & - & - \\
           2013 & - & - &         -            &        -            & $8.7^{+0.8}_{-0.3}$  & $0.35^{+0.12}_{-0.18}$ & - & - & - & - & - & - \\
           2014 & - & - &         -            &        -            & - & - & - & - & - & - & - & - \\
           2015 & - & - &         $15.9^{+0.7}_{-0.3}$ & $2.8^{+0.9}_{-1.4}$  & $8.5^{+0.8}_{-0.3}$  & $0.43^{+0.13}_{-0.23}$ & - & - & - & - & - & - \\
           %2016 & & & $16.6^{+0.7}_{-0.3}$ & $1.4^{+0.5}_{-0.4}$ & & & & & & & & \\
           2016 & - & - &         -            &            -        & $8.4^{+0.8}_{-0.3}$  & $0.47^{+0.15}_{-0.25}$ & - & - &$2.0^{+0.3}_{-0.3}$ & $7.7^{+2.3}_{-1.8}$ & $1.3^{+0.3}_{-0.3}$ & $8.2^{+1.8}_{-2.0}$ \\
           2017 & - & - &         -            &            -        & $8.4^{+0.8}_{-0.3}$ & $0.47^{+0.15}_{-0.25}$ & - & - & - & - & - & - \\
           2018 & - & - & $15.9^{+0.7}_{-0.3}$ & $2.8^{+0.9}_{-1.4}$ & $8.2^{+0.8}_{-0.3}$  & $0.56^{+0.18}_{-0.29}$ & - & - & $2.0^{+0.3}_{-0.3}$ & $7.7^{+2.3}_{-1.8}$ & $1.3^{+0.3}_{-0.3}$ & $8.2^{+1.8}_{-2.0}$ \\
           2019 & $19.6^{+0.7}_{-0.3}$ & $1.1^{+0.4}_{-0.5}$ & $16.1^{+0.7}_{-0.3}$ & $2.3^{+0.7}_{-1.0}$ & -  & - & - & - & - & - & - & - \\
           2020 & - & - & $16.0^{+0.7}_{-0.3}$ & $2.5^{+0.8}_{-1.2}$ & - & - & - & - & - & - & - & - \\
           \hline
           Averaged & $19.4\,\pm\,0.2$ & \multicolumn{1}{|l|}{$1.4\,\pm\,0.3$} & $15.9\,\pm\,0.2$ & \multicolumn{1}{|l|}{$2.8\,\pm\,0.4$} & $9.0\,\pm\,0.6$ & \multicolumn{1}{|l|}{$0.29\,\pm\,0.15$} & $7.9\,\pm\,0.5$  & \multicolumn{1}{|l|}{$0.43\,\pm\,0.1$} & $2.3\,\pm\,0.4$ & \multicolumn{1}{|l|}{$5.9\,\pm\,1.9$} & $1.3\,\pm\,0.1$  & \multicolumn{1}{|l|}{$7.7\,\pm\,0.6$} \\
       \hline
    \end{tabular}
    \caption{Results of the photometric multiwavelength analysis of X3. Please note that all flux densities are given in Jy except for the H and K band. Lower magnitude values equal brighter and higher source emission.}
    \label{tab:mag_flux}
\end{table*}
We adapt the listed averaged L-band magnitude of 9.0 mag from Table \ref{tab:mag_flux} to estimate the L-band magnitude of X3c.
With Eq. \ref{eq:apparent_magnitude}, we calculate an averaged L-band magnitude of $10.6\,\pm\,0.6$ for X3c. The uncertainties are adapted from the averaged L-band magnitude of X3 and represent the standard deviation of the individual measurements per epoch. %In the following subsection, we will use the derived flux density values to create  input spectrum for HYPERION.

\subsection{Spectral Energy Distribution}
\label{sec:results_sed}

\begin{figure}[htbp!]
	\centering
	\includegraphics[width=.45\textwidth]{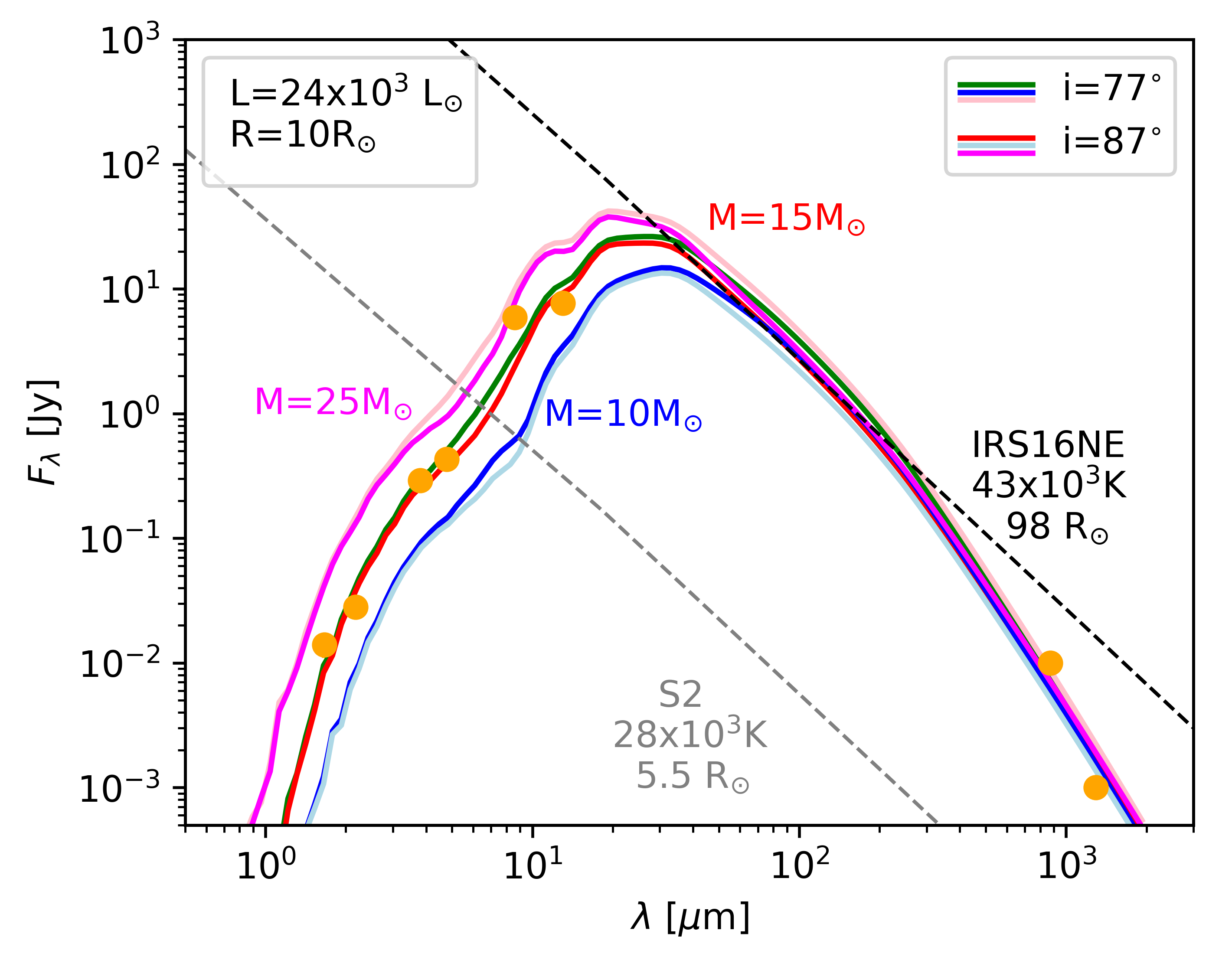}
	\caption{Spectral energy distribution of the X3 system in comparison with early-type stars. We show the flux density in [Jy] as a function of wavelength in [$\mu m$] with a logarithmic axis scaling. The inclination angles of 77.5$^{\circ}$ and 87.5$^{\circ}$ with a related stellar mass of 15M$_{\odot}$ are marked with green and red, respectively. The (light)blue lines indicate the lower stellar mass limit of $10 M_{\odot}$. In contrast, magenta/pink shows the upper mass limit of $25 M_{\odot}$ resulting in a final stellar mass estimate of $15^{+10}_{-5} M_{\odot}$. The estimated flux densities are represented by orange dots whereas the size correlates with the uncertainties. The grey and black dashed lines represent the blackbody SED fits of S2 and IRS16NE, respectively \citep{Krabbe1995, Habibi2017}.}
\label{fig:x3_sed}
\end{figure}

\begin{figure}[htbp!]
	\centering
	\includegraphics[width=.45\textwidth]{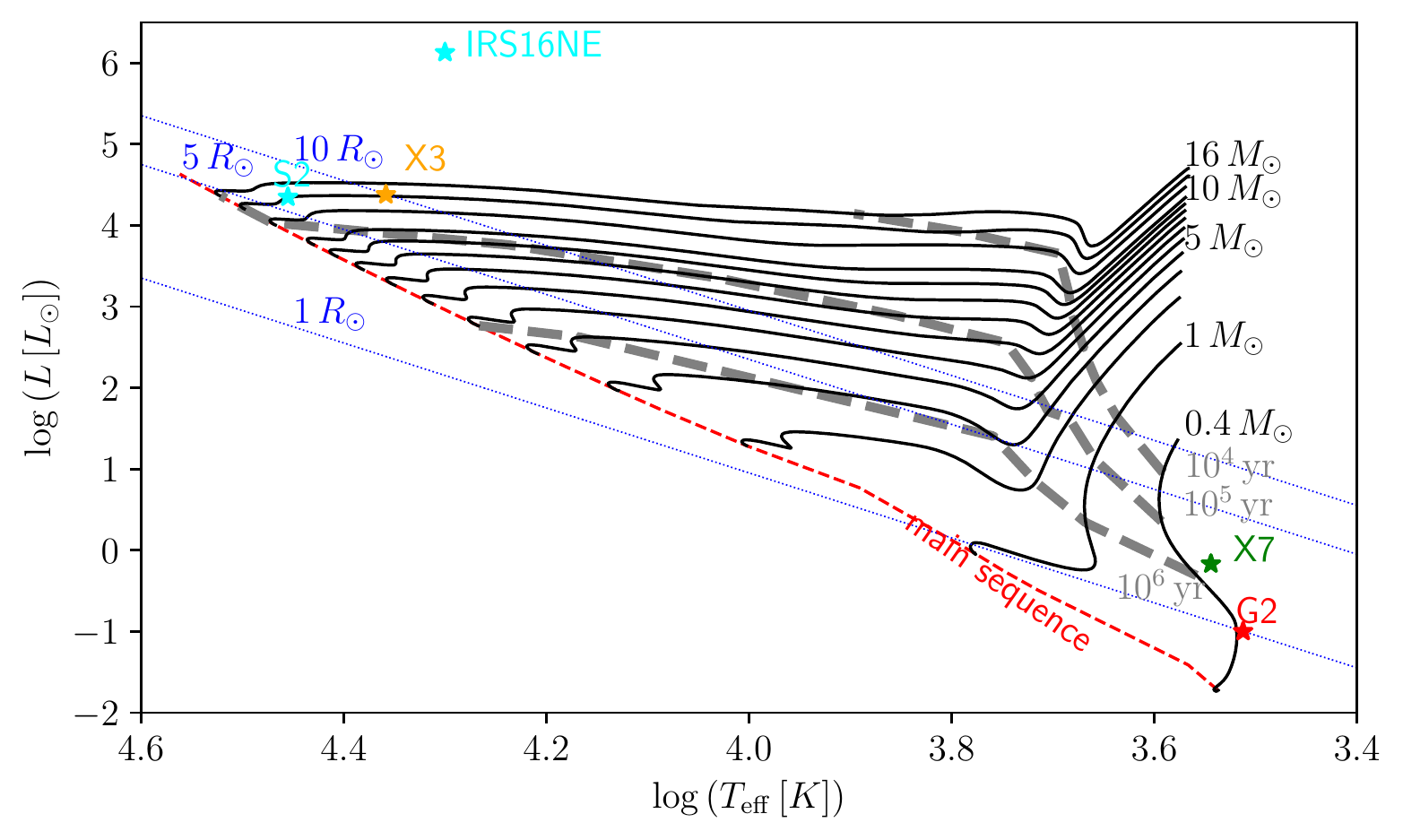}
	\caption{Position of the X3 in the Hertzsprung-Russell diagram. The X3 position (orange asterisk) in the Hertzsprung-Russell diagram is compared with the pre-main-sequence star evolutionary tracks according to the PARSEC code \cite{2012MNRAS.427..127B}, calculated for the near-Solar metallicity of $Z=0.01$. Evolutionary paths of stars in the mass range 1-10\,$M_{\odot}$ are shown (with 1\,$M_{\odot}$ increments) as well as $0.4$, $12$, $14$, and $16\,M_{\odot}$ stars are depicted. In addition, we plot $10^4$, $10^5$, $10^6\,{\rm yrs}$ isochrones, the { zero-age} main sequence, and the lines standing for $1\,R_{\odot}$, $5\,R_{\odot}$, and $10\,R_{\odot}$ (blue dotted lines). The inferred temperature and the luminosity of X3 are consistent with being a pre-main-sequence massive star ($\sim 14-15\,M_{\odot}$ having an age of $\sim 40\,000$ years). For comparison, we also plot the positions of X7 and G2 objects according to the broad-band SED fits of \cite{peissker2021} and \cite{peissker2021c}, respectively. In contrast to X3, they follow the evolutionary track of a low-mass star of $\sim 0.4\,M_{\odot}$ with the age of $\gtrsim 10^6$ years. The denoted OB stars S2 and IRS16NE are also clearly offset from the X3 star.}
\label{fig:hr_x3}
\end{figure}

Due to the robust H- and K-band detection of X3a between 1995 and 2020 (see Fig. \ref{fig:detection_x3a_x3b}, Fig. \ref{fig:x3_all_k}, and Fig. \ref{fig:osiris_2020}), the rich line emission in various bands, and the strong outflow {that might emerge from the star or a protoplanetary disk} (Fig. \ref{fig:pcygni_sinfo}), the question about the stellar nature of the object arises. At this point, we can exclude a late type stellar nature due to the missing infrared CO band heads at 2.29$\mu m$, 2.32$\mu m$, 2.35$\mu m$ and 2.38$\mu m$. The absence of these absorption feature clearly separates X3a from late-type stars \citep[][]{Buchholz2009}. We address the question about the nature of X3a with the support of the 3D Markov Chain Monte Carlo radiative transfer code HYPERION to model the composite Spectral Energy Distribution (SED) of the X3 system \citep{Robitaille2011}; { see also \citet{Zajacek2017} for a similar set-up}. For HYPERION, we assume that X3a is a YSO motivated by the above arguments. The 3D dust continuum radiative transfer code calculates individual spectra and allows for a variety of input parameters. From the observed magnitudes and flux densities of X3a, we arrange the input spectrum for HYPERION as the flux density (in [Jy]) as a function of wavelength (in [$\mu m$]). See Table \ref{tab:mag_flux} for the numerical values of the related input spectrum.
This spectrum is then combined with the input values of the model listed in Table~\ref{tab:sed_values}. We choose 10$^6$ photons and 10$^4$ ray-tracing sources, resulting in an average computation time of 12-24 hours per simulation run. In total, we used more than 100 different setups to derive suitable best-fit parameters for the emission of the X3-system. For example, we checked different stellar mass assumptions between 0.5-25 M$_{\odot}$ in steps of 1 M$_{\odot}$.
\begin{table}[hbt!]
    \centering
    \begin{tabular}{|cc|}
         \hline 
         \hline
           Properties & Setting\\
           \hline
           Radius [R$_{\odot}$]     & 10 \\     
           Luminosity [L$_{\odot}$] & 24$\times\,10^3$ \\     
           Mass [M$_{\odot}$] & 15 \\  
           Disk mass [M$_{\odot}$] & 0.01 \\
           Disk radius (min) [R$_{\odot}$] & 50 \\     
           Disk radius (max) [AU] & 700 \\  
           Disk radius height[R$_{\odot}$] & 0.01 \\  
           Accretion [M$_{\odot}$/yr] & $10^{-6}$ \\     
           Ulrich env. (min) [R$_{\odot}$]  & 50 \\     
           Ulrich env. (max) [AU]  & 1000 \\     
           Number of Photons & 10$^6$ \\
           Ray-tracing sources & 10$^4$ \\
           Number of Iterations & 10 \\
           \hline
           
           \hline
    \end{tabular}
    \caption{Final input parameters for HYPERION. In addition to these settings, we define a input spectrum related to the derived fluxes listed in Table \ref{tab:mag_flux}.} 
    \label{tab:sed_values}
\end{table}
The results of the model are shown in Fig. \ref{fig:x3_sed}. Among the best-fit parameters are a stellar mass of $15M_{\odot}$ with a related effective temperature of $24\times 10^3$ K for X3a. Considering the values listed in Table \ref{tab:mag_flux}, we transfer the uncertainties of the estimated magnitudes and flux densities of X3 to the comparison plot. Due to the logarithmic scaling of Fig. \ref{fig:x3_sed}, these uncertainties are about the size of the used symbols. Since the uncertainties do not entirely reflect the data distribution, we constrain the mass of X3a to $15^{+10}_{-5} M_{\odot}$.  
Consistent with the previous study by \cite{muzic2010}, we independently constrain an inclination angle between 77.5$^{\circ}$ and 87.5$^{\circ}$ for the X3 system. When comparing the mass and luminosity of the X3 system with the GAIA YSO survey, a source with comparable characteristics is identified as MWC 297 \citep{Wichittanakom2020}. We note that the YSO MWC 297 is classified as a Herbig Ae/Be star which will be adapted for the X3 system.
{ Using} the derived mass and the effective temperature of the star, we use the PARSEC code { output files} \citep{2012MNRAS.427..127B} to elaborate the Hertzsprung-Russell diagram (Fig. \ref{fig:hr_x3}). Taking into account evolutionary paths of pre-main-sequence stars of different masses, we can estimate an age of X3 to be $\sim 0.04$ Myr.
The projected stagnation radius of $R_{\rm BS}\sim 0.4''\sim 3300\,{\rm AU}$, which is close to the deprojected value due to the high inclination, allows us to estimate the mass-loss rate of X3. For this, we use the ram-pressure equilibrium relation \citep[e.g.,][]{Wilkin1996} 
\begin{equation}
    R_{\rm BS}=(\dot{m}_{\rm w}v_{\rm w}/[4\pi \mu m_{\rm H} n_{\rm a} v_{\rm rel}^2])^{1/2}
\end{equation}
where the ambient number density $n_{\rm a}\sim 26\,{\rm cm^{-3}}$ is close to the Bondi radius \citep{2003ApJ...591..891B}. The terminal stellar wind velocity $v_{\rm w}\sim 400\,{\rm km\,s^{-1}}$ is adapted from the HeI P Cygni profile (Fig. \ref{fig:pcygni_sinfo}), and the relative velocity with respect to the ambient medium can be approximated by $v_{\rm rel}\gg v_{\star}\gtrsim 1000$ km/s due to the nearly perpendicular orientation of the bow shock with respect to the proper motion vector (Fig. \ref{fig:x3_disk}). From the above numerical approach we obtain $\dot{m}_{\rm w}\sim 2.65\times 10^{-6}-1.06\times 10^{-5}\,{\rm M_{\odot}\,yr^{-1}}$. This result is at the higher end of typical mass-loss rates for Hebig Ae/Be stars \citep{1995A&A...302..169N} and is comparable to the input accretion rate for the HYPERION SED calculation. We note that the estimate is strongly dependent on the exact value of the relative velocity value and on the bow-shock stagnation radius.

Furthermore, the stellar orbital velocity of X3a can be estimated simply as $v_{\star}\,\simeq \,(v^2_{\rm LOS}+v^2_{\rm PROP})^{1/2}$ where $v_{\rm LOS}$ defines the LOS velocity and $v_{\rm PROP}$ is the velocity related to the proper motion. With $v_{\rm PROP}\,=\,244\pm 27$km/s from the fit shown in Fig. \ref{fig:proper_motion} and the Br$\gamma$-based line center LOS velocity of $v_{\rm LOS}\,\sim (v_{\rm blue}+v_{\rm red})/2\,\sim -48.5 $km/s, we obtain the estimate of the stellar orbital velocity of X3a, $v_{\star}\,=\,249$km/s. The mean orbital distance then is $d_{\rm X3}\simeq GM_{\rm SgrA*}/v_{\star}^2\simeq 0.28\,{\rm pc}$, which is slightly larger than the projected distance of $\sim 0.1\,{\rm pc}$ of the X3 system from Sgr A*. This is expected due to the inclined stellar orbit with a non-zero LOS velocity traced in the emission lines.

For the modeled SED, we assumed the presence of a massive {protoplanetary disk with a related radius of $\lesssim 700$ AU around} the star. If the double-peak Br$\gamma$ line originates in the {inner disk which gets spatially extended by outflows}, the characteristic Keplerian velocity is estimated as $v_{\rm Br\gamma, K} \gtrsim 103.5$km/s, which serves as a lower limit for the Keplerian velocity in the disk due to an uncertain inclination \citep{Kraus2012}. The mean radius for the {origin of the} Br$\gamma$ emitting material is $r_{\rm disk,Br\gamma}\lesssim Gm_{\rm X3a}/v_{\rm Br\gamma, K}^2\sim 1.2\,{\rm AU}$. This is in agreement with the orbiting bound ionized material around a YSO that could be either inflowing or outflowing, that is, an accretion or a decretion disk \citep{2014A&A...561A...2A}.
This provides a consistency test of the overall kinematics of the system (stellar orbital motion around Sgr~A* as well as the bound gas motion around the X3a star) which is in agreement with the proposed computational model (Fig. \ref{fig:x3_sed}). {We note that the observed Br$\gamma$ line is spatially extended as it is expected for Herbig Ae/Be stars \citep{Tatulli2007, Davis2011}. We will elaborate this relation in the following section.}

\section{Discussion} \label{sec:discuss}
In the following, we will discuss the various findings of the { presented} analysis. First, we will summarize the different components related to the X3 system. Second, we will motivate the stellar nature of the bow shock source. Third, we will propose a speculative scenario to explain the formation process of the young star.

\subsection{The X3 system}

Analyzing NIR data observed with the VLT and KECK, we trace X3a along with X3 between 2002 and 2020 which is reflected by a matching proper motion of 244$\pm$27 km/s. In the case of two non-related sources, we would expect a significant difference for the proper motion but also the trajectory. Considering the statistical analysis of \cite{Sabha2012} for a random fly-by event, the 3d motion of X3 and X3a is a six-parameter problem ($x$, $y$, $v_{{pm}_x}$, $v_{{pm}_y}$, $v_{{los}_x}$, $v_{{pm}_y}$) where it is unlikely that at least 4 parameters are arranged in a way that $x_{X3}=x_{X3a}$ and $y_{X3}=y_{X3a}$. However, several studies \citep[see, for example,][]{Gallego-Cano2018} investigate the stellar density as a function of distance { from} Sgr~A*. Hence, we expect by definition that at larger distances the chance for a random encounter is { significantly} decreased. Considering the KLF for IRS 13E taken from \cite{Paumard2006}, we find for the position of the X3 system in total {$\sim $8 sources/(1 arcsec)$^2$}. Taking into account the spatial pixel scale of NACO, we { obtain the mean value of 0.14 sources per resolution element (i.e., pixel)}. For the {almost consecutive} NACO observations between 2002 and 2018, we find a random source at a random position of about {2 $\%$. The authors \cite{Sabha2012} and \cite{Eckart2013} estimate a similar value for the higher crowded S-cluster. Considering that a random fly-by encounter of two stars dissolves after three years, it becomes negligible that such an event can explain} the co-movement of the H-, K-, and L-band detection of X3 and X3a.\newline
Regarding X3b, it is reasonable to assume that the thermal blob was created (or confused) between 1995 and 2002 and disappeared around 2012. While the ejection of a blob from the central stellar source X3a is rather speculative, the formation of dense material within the bow shock due to instabilities is supported by simulations of bubbles around the bow-shock tip \citep{Gardner2016}. {We can assume that unstable bubbles suffer from magnitude and flux variations because of their nature. Investigating the $K$-band magnitude of X3b in 2002 revealed a magnitude of about 16 mag which was higher compared to X3a (see Table \ref{tab:mag_flux}). Since no emission above the noise of X3b is detected in 2012, we compare this magnitude to the NACO detection limit of 18.5 mag and find therefore a decrease of $\Delta_{mag}\,=\,18.5-16.0\,=\,2.5$.}
Therefore, the analysis implies that X3b suffered a decrease in thermal energy between 2002 and 2012 due to radiative or adiabatic cooling, which justifies the classification as a thermal blob. Furthermore, the proper motion of X3a observed in the $H$- and $K$-band is in reasonable agreement with X3, which suggests that both sources belong to the same system. This argumentation is also valid for X3c although the nature of the compact blob is challenging to uncover. The X3c component could be related to the collimated bow-shock downstream flow, such as within the Bondi–Hoyle–Lyttleton (BHL) accretion flow model \cite{Matsuda2015}. This mechanism describes the shock formation downstream at a specific stagnation point. Taking into account the thermal pressure of the ambient hot plasma as discussed in \cite{Christie2016}, it is also plausible that the surrounding medium could create enough pressure to make the bow-shock shell closed and X3c would be the shock formed where the streams meet. As a consequence, the blob is expected to be variable and the material could be accreted over time. Although we did not find indications of an accretion process between 2002 and 2018 (see Fig. \ref{fig:distance_blobs}), it could be subject to future observations. However, the idea of high thermal pressure to explain the nature of X3c seems { appealing} since the cigar shape of X7 \citep[][]{peissker2021} could also be created by a dominant { thermal pressure of the} ambient medium.

\subsection{The nature of X3a}

\begin{figure}[htbp!]
	\centering
	\includegraphics[width=.45\textwidth]{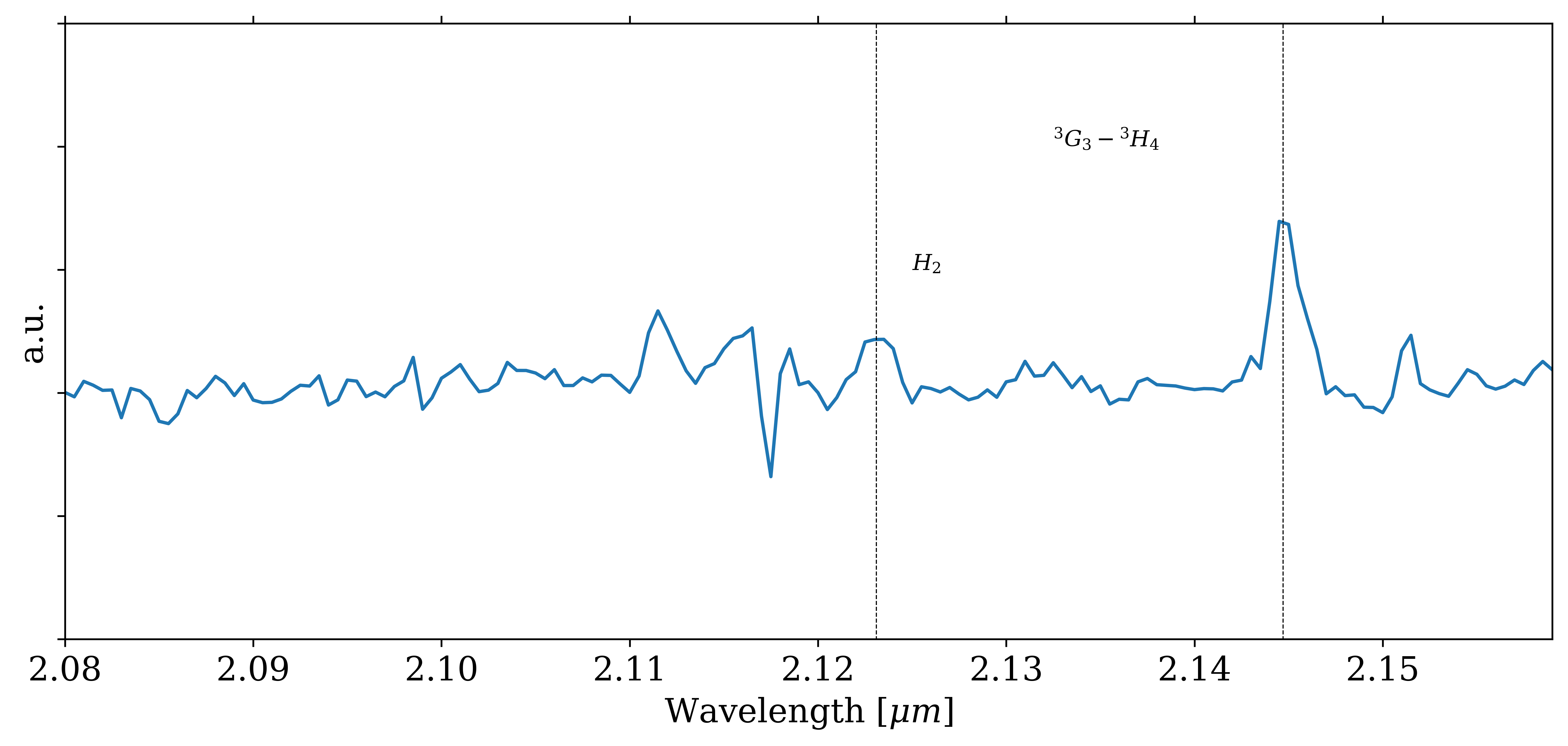}
	\caption{Zoomed-in view to the SINFONI extracted spectrum in the spectral region between 2.08$\mu m$ and 2.155$\mu m$ to show the Doppler-shifted H$_2$ line. Besides the H$_2$ line at 2.1232$\mu m$ (rest wavelength at 2.1218$\mu m$), one forbidden iron transition at 2.1447$\mu m$ is indicated. The presence of the Doppler-shifted emission line H$_2$ with v$_{H_2}\,=\,198$ km/s observed with SINFONI is an indicator of the presence of a protoplanetary disk \citep[][]{Glassgold2004}. {For the related line map of the H$_2$ detection, please consult Fig. \ref{fig:x3_h2_linemap}, Appendix \ref{sec:appendix_spectral_analysis}.}}
\label{fig:x3_spec_sinfo_h2}
\end{figure}

\begin{figure}[htbp!]
	\centering
	\includegraphics[width=.45\textwidth]{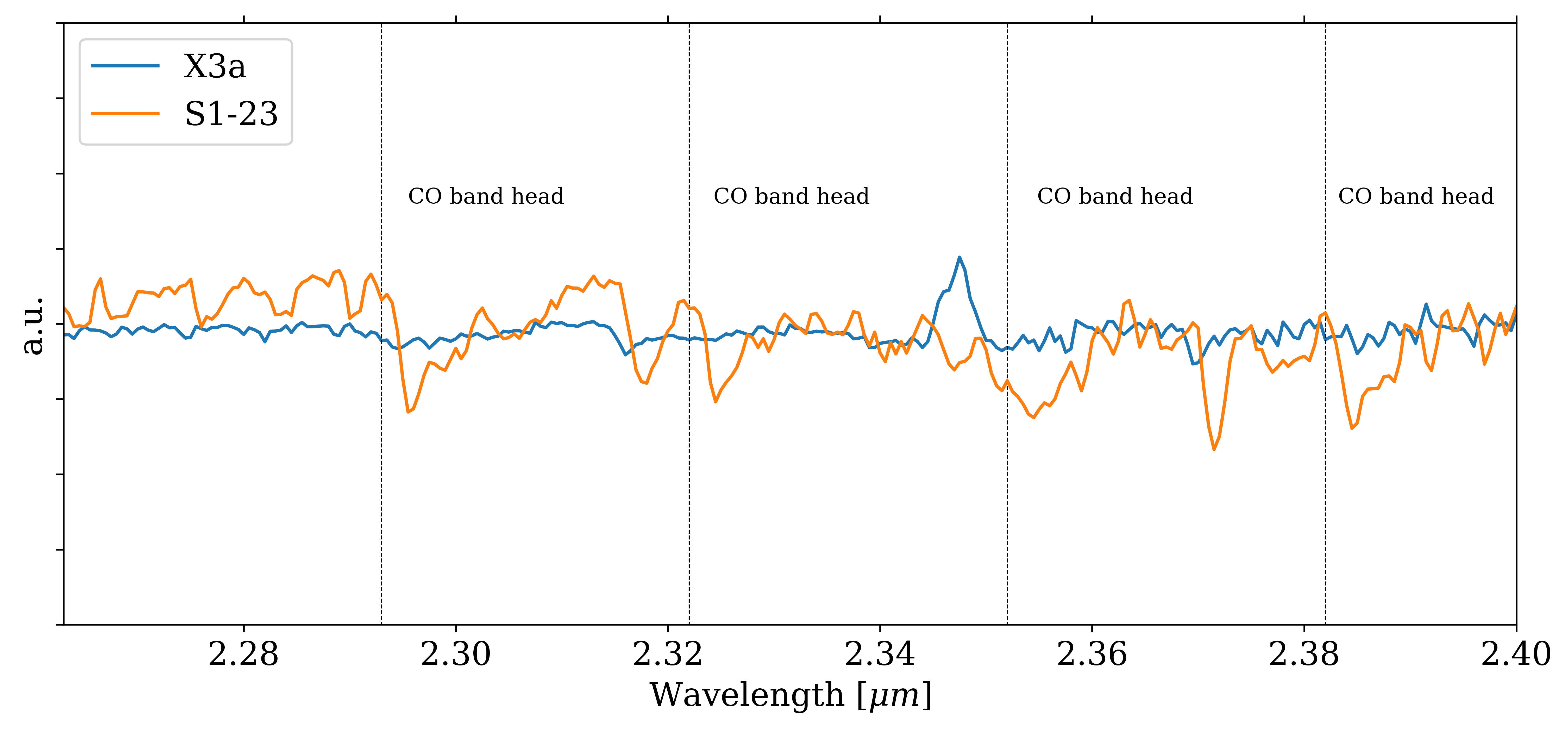}
	\caption{Search for IR CO band-heads for X3a and comparison with the late-type analog S1-23. For the identification of the late-type star S1-23, we use the spectral analysis of \cite{Gautam2019}.}
\label{fig:x3_spec_co}
\end{figure}

\begin{figure}[htbp!]
	\centering
	\includegraphics[width=0.45\textwidth]{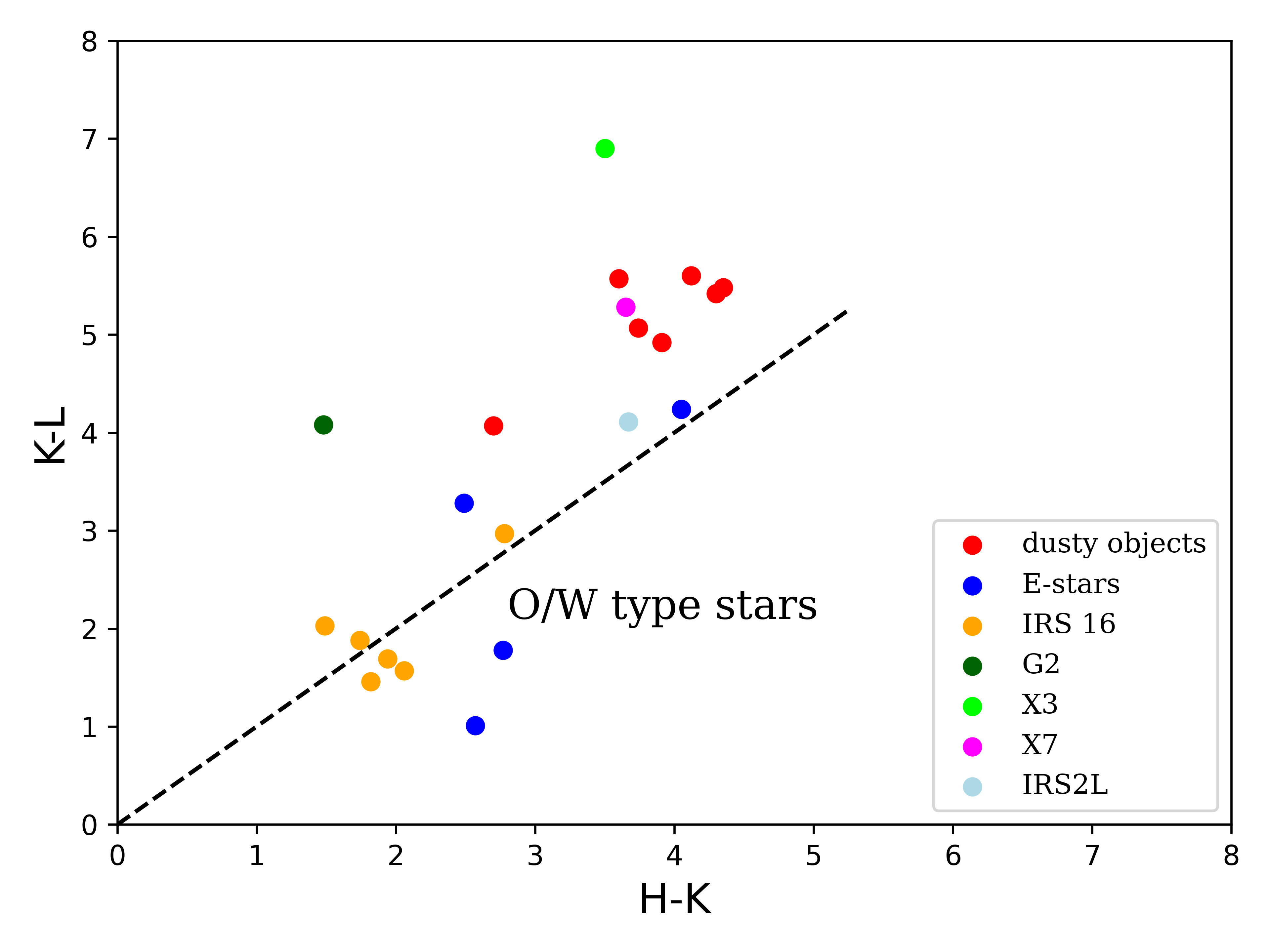}
	\caption{Color-color diagram of close-by sources. We use published H-K and K-L colors of close-by sources \citep{Eckart2004, Viehmann2006} and incorporate the estimated magnitudes for X3 (Table \ref{tab:mag_flux}), X7 \citep{peissker2021}, and G2 \citep{peissker2021c}. The dashed line represents an emitting blackbody at different temperatures. The presumably YSOs exhibit a higher K-L color compared to the O/W type stars in our sample.}
\label{fig:color_color_diagramm}
\end{figure}

\begin{figure*}[htbp!]
	\centering
	\includegraphics[width=1.0\textwidth]{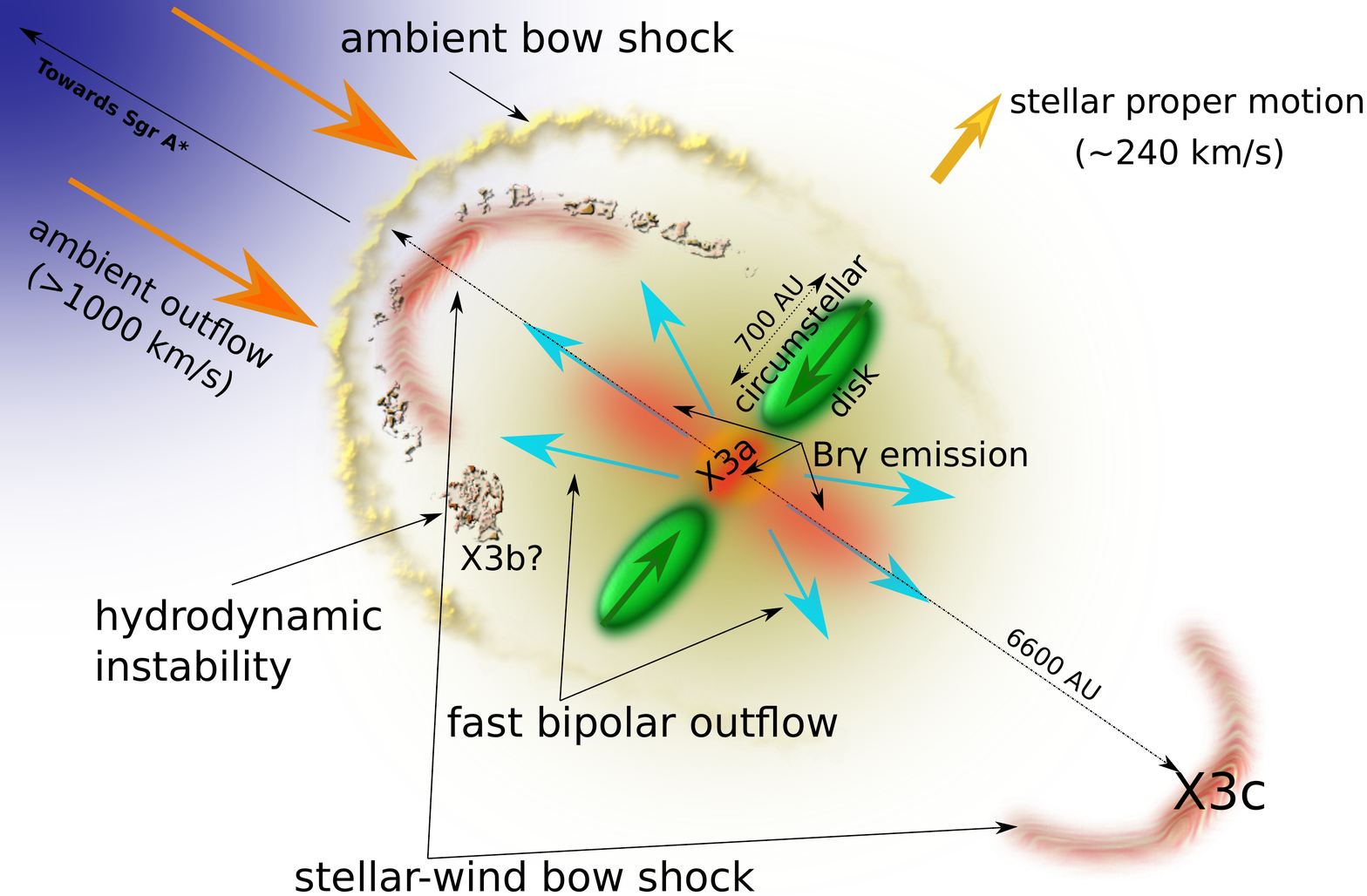}
	\vspace{-1.5cm}
	\caption{{ Sketch of the X3 system. The figure depicts the central massive star X3a that is surrounded by the circumstellar disk (green), whose inner ionized part and the associated disk outflows (blue) emit Br$\gamma$ emission (reddish regions). The stellar and disk outflows get shocked within the stellar-wind bow shocks. The downstream bow shock may be associated with the stable infrared continuum component X3c. Since the star moves supersonically with respect to the ambient medium, a bow shock forms in the ambient medium as well (yellow). The inner (stellar wind) and outer (ambient medium) bow shock shells mix together and hydrodynamic instabilities form downstream. The transient infrared component X3b might be associated with a larger instability formed downstream. Yellow and orange arrows indicate the stellar proper motion and the ambient outflow directions, respectively, including their inferred velocities. Drawn not to the scale.}  }
\label{fig:sketch}
\end{figure*}

Examining the different Doppler-shifted velocities for X3 as observed with SINFONI, it is implied that various components of the system are responsible for the emission. For example, the observed Br$\gamma$ {emission (Fig. \ref{fig:linemaps_sinfo_sinfo}) exhibits} a double-peak {line that shows a velocity gradient along the source implying the presence of a disk or strong outflows \citep{Davis2011}. We will discuss this particular feature in Sec. \ref{Sec:discussion_brgamma}. Furthermore, the} HeI line might be connected to outflows from a possible jet or strong stellar winds (see the P-cygni profile shown in Fig. \ref{fig:pcygni_sinfo}). Helium pumping can be considered to be the responsible mechanism for the forbidden iron multiplet \citep{peissker2021}. We note the detection of a Doppler-shifted H$_2$ line (Fig. \ref{fig:x3_spec_sinfo_h2}) {which might serve as an additional tracer for photoinized outflows originating close to the massive protostar X3a \citep{Kumar2002, Tanaka2016}. The detection of the NIR H$_2$ line is accompanied by the radio/submm CO emission presented in Fig. \ref{fig:x3_disk}. Both lines are common indicators for the presence of a protoplanetary disk and outflows of Herbig Ae/Be stars \citep{Thi2001, Davis2011}. Although the spatial resolution of SINFONI forbids a detailed determination of the exact origin of the H$_2$ emission line, the arrangement of the ionized CO and H30$\alpha$ line observed with ALMA and displayed in Fig. \ref{fig:x3_disk} suggest different origins of the detected gas species related to the X3 system.}
%Furthermore, the H$_2$ emission line also serves as a tracer for outflows and disks around massive YSOs \citep{Kumar2002}. 
{To exclude the possibility of a stellar classification as late-type, we inspect the SINFONI spectrum for CO absorption lines which are typical for evolved stars. Consequently, young stars do not show extended infrared CO band heads between $2.3-2.4\mu m$.} To visualize the difference of the CO band heads for different stellar ages, we compare the SINFONI spectrum of X3a with the late-type star S1-23 which is located about 2.36 arcsec east of IRS 13 \citep{Gautam2019}. In Fig. \ref{fig:x3_spec_co}, the two spectra of both sources are incorporated and the CO band head locations are marked. 
Due to the absence of the CO band head features in the X3a spectra, the stellar temperature must be higher compared to S1-23. From the normalized spectrum presented in Fig. \ref{fig:x3_spec_co}, we evaluated the depth of the CO band (CBD) using the feature at 2.36$\mu m$. We follow the analysis of \cite{Buchholz2009} and derive a CBD for X3a of -0.1, for S1-23 a value 0.4. From a critical point of view, the CBD might be biased by telluric correction, background, or noise level. However, the CBD is only considered to provide a rough classification that separates late-type stars from young stellar sources and serves as an independent parameter that underlines our findings for X3a.\newline 
This classification is supported by the broad-band spectral energy distribution with a prominent NIR and MIR excess, the double-peak profile of the Br$\gamma$ line with the velocity gradient of $\sim 200\,{\rm km/s}$ indicating a rotating structure {or outflows}, and the prominent P-Cygni profile of HeI emission line. { Therefore, we conclude} that the X3 system can be described as a YSO with a {protoplanetary} disk embedded in a bow-shock dust envelope. With the support of stellar evolutionary tracks shown in Fig. \ref{fig:hr_x3}, we propose that X3a is a young Herbig Ae/Be star with a mass of $15^{+10}_{-5} M_{\odot}$ and { an estimated age of a few} $10^4$yr.
Estimating the $H$-$K$ and $K$-$L$ colors from Table~\ref{tab:mag_flux} and comparing it with the surrounding stars and dusty objects of the nearby cluster IRS 13, we find further support for the pre-main-sequence nature of X3a (see Fig. \ref{fig:color_color_diagramm}).
%From the above discussion, it is evident that the ionized gas species observed with SINFONI, VISIR, and ALMA seem to be related to different components within the X3 system. Considering the continuous velocity gradient of the disk or the stellar outflow indicated by the observed P-Cygni profile, we expect different line-of-sight (LOS) velocities as presented in Table \ref{tab:components}
To get a better understanding of the setup of the X3 system, we display a possible arrangement for the different components (proto-star, protoplanetary disk, dust envelope, bow shocks) in the sketch shown in Fig. \ref{fig:sketch}.
%\begin{figure*}[htbp!]
%	\centering%	\includegraphics[width=1.0\textwidth]{2022-11-03_1137_labels.pdf}
%	\caption{Sketch of the X3 system inspired by \cite{SiciliaAguilar2016} and \cite{Haworth2016}. The blue envelope represents the dust emission associated with the YSO. With the submm/radio observations carried out with ALMA, we observe the optical thick CO emission and the embedded H30$\alpha$ line. With increasing temperature, ionized gas such as Br$\gamma$ is deposited in the accretion disk (magenta colored) close to the central stellar source colored here in red. The figure is not to scale and shows a static representation of a possible setup. Since the X3 system suffers from the interaction with the ambient medium, the setup of the young star may be more complex compared to this naive representation.}
%\label{fig:sketch}
%\end{figure*}

{ We note that this sketch is not to the scale, but it captures the main components and dynamics of the X3 system. In particular, the interaction of the stellar and disk outflows (Fig. \ref{fig:pcygni_sinfo}) leads to the formation of two bow shocks, one of which (downstream one) may be associated with the stable infrared continuum component X3c. Furthermore, due to the supersonic motion of X3a with respect to the ambient medium, a bow shock forms in the ambient medium as well and the shocked gasesous-dusty material flows downstream, giving rise to the elongated shape of the X3 system visible in the L-band. Due to several molecular and recombination-line tracers, the X3a star is surrounded by a circumstellar disk ($\sim 700\,{\rm AU}$) in radius, whose inner part is ionized and is the source of the double-peaked Br$\gamma$ line. The Br$\gamma$ emission is further extended above and below the disk plane due to bipolar disk winds. The whole model may be even more complex, e.g. due to the presence of the so-called transition disk that connects the dust envelope with the accretion disk \citep[][]{Haworth2016a}.} Such a transition area also increases the size of {an accretion} disk compared to the sketch shown in Fig. \ref{fig:sketch}. Recent observations of (massive) YSOs suggest the existence of this feature which extends possible disk sizes \citep[][]{Frost2019, GravityCollaborationYSO2020}. While it can be argued that additional background noise may impose uncertainties that increase the confusion in order to estimate the exact disk size (Fig. \ref{fig:linemaps_sinfo_sinfo}), observations of massive YSOs in the Carina Nebula revealed even larger disk dimensions compared to the X3 system \citep{Preibisch2011}. {Using the settings for the disk radius of the model from Table \ref{tab:sed_values}, we estimate an expected size of almost 100 mas, which is in reasonable agreement with the dimensions displayed in Fig. \ref{fig:linemaps_sinfo_sinfo}. In the next section, we will provide a detailed approach of the composition of the X3 system.}
%It is interesting to note that complex molecules such as PAH1 and ArIII are expected to be localized in disks around YSOs \citep{Petit2021}. Since the VISIR observations (Fig. \ref{fig:sivr2_visir} and Fig. \ref{fig:x3_all_visir}, Appendix \ref{sec:appendix_multiwavelength_detection}) exhibit the same dimensions compared to the Br$\gamma$ line (Fig. \ref{fig:linemaps_sinfo_sinfo}), we use 

\subsection{The disk structure of the X3 system}
\label{Sec:discussion_brgamma}

{In Fig. \ref{fig:linemaps_sinfo_sinfo}, we show the detection of ionized Doppler-shifted Br$\gamma$ with an FWHM of about 0.15" ($\sim$ 1200 AU) which is about the size of the SINFONI PSF ($\sim$ 0.25") for the corresponding spatial plate scale. The diffraction-limited detection of the integrated Br$\gamma$ line shown in Fig. \ref{fig:linemaps_sinfo_sinfo} is confined by the resolution of the telescope and indicates that the ionized hydrogen originates in a smaller area than 0.15". In contrast, the single blue- and red-shifted linemaps (see the PPV diagram in Fig. \ref{fig:linemaps_sinfo_sinfo} and the attached movie) exhibit a slightly extended size of about 0.25" ($\sim$ 2000 AU) that matches the FWHM of the SINFONI PSF.\newline
Although the Br$\gamma$ line is often used as a tracer for accretion disks, the quantities of the X3 system exceed typical disk sizes of a few ten to about 100 AU \citep[][]{Beck2010, Beck2019}. However, \cite{Davis2011} and \cite{Ward2017} show spatially extended Br$\gamma$ lines of massive YSOs that exceed dimensions of several hundred up to 1000 AU. This apparent inconsistency can be explained by photoionized disk-wind outflows and, more generally, stellar winds that increase the spatial distribution of the Br$\gamma$-line. For example, \cite{Kraus2012} performed VLTI/AMBER observations of Br$\gamma$ gas distributions and showed that the Doppler-shifted line exhibits a photocenter offset compared to the central emission. Although \citet{Kraus2012} investigate classical Be stars, \citet{Tanaka2016} simulate density distributions n$_H$ of hydrogen around massive protostars. \citet{Tanaka2016} report strongly ionized outflows of high mass protostars with associated temperatures of about 5000-10000 K on 100-2000 AU scales. Since Br$\gamma$ can be detected at temperatures around 8000 K \citep[][]{Wojtczak2022}, the results of \cite{Tanaka2016} are in agreement with the observations carried out by \cite{Davis2011}, \cite{Ward2017} and this work. Due to the high mass of the X3 system (Fig. \ref{fig:x3_sed}), it is expected that ionized stellar winds distribute gas species on scales of several hundred AU \citep[][]{Tanaka2016}. This argument follows the analysis of \cite{Davis2011} who interpret the spatially extended Br$\gamma$ emission as a tracer for outflows which was independently shown by VLTI/AMBER observations carried out by \cite{Tatulli2007}.\newline
Although we cannot spatially resolve the accretion disk of the X3 system, we can safely assume that we observe a superposition of the warm disk material with strongly ionized outflows that are traced by the detected P-Cygni profile. Another mechanism might impose constraints on the detection of the size of the Br$\gamma$ emission area. As we show in Fig. \ref{fig:x3_disk}, the H30$\alpha$ line is arranged in the ring-like structure with an approximate diameter of about 0.25" or 2000 AU exhibiting a comparable morphology as the massive protostar G28.200.05 observed by \cite{Law2022}. The size of the ionized hydrogen matches the Br$\gamma$ emission but also the MIR lines detected with VISIR (see Fig. \ref{fig:sivr2_visir} and Fig. \ref{fig:x3_all_visir}, Appendix \ref{sec:appendix_further_detections}) and ISAAC (Fig. \ref{fig:linemap_isaac}, Appendix \ref{sec:appendix_further_detections}) implying that all detected atomic and molecular species originate in the same region. The enhanced intensity of the H30$\alpha$ line at the tip of the bow shock of about 20$\%$ might be related to the interaction of the ambient medium with the X3 system which is in agreement with the 3D magnetohydrodynamic models of $\zeta$ Ophiuchi carried out by \cite{Green2022}. The increased temperature in the bow-shock tip caused by the interaction of the supersonic X3 system with the ambient medium could also be responsible for the increased Br$\gamma$ emission size. The authors of \cite{Scoville2013}, for example, shows a spatially increased Br$\gamma$ emission created by the interaction of a young T Tauri star with the ambient medium around Sgr~A*. This contributes to the high temperature within the disk material in combination with the photoionized disk-wind outflows. Because there might be an interplay between the bow-shock induced temperatures and strong winds, gas flows through the disk cannot be ruled out. A probe for these gas flows are polycyclic aromatic hydrocarbons (PAH, see Fig. \ref{fig:x3_all_visir}). We use the SED representing IRS 3 (Fig. \ref{fig:x3_system}) derived by \cite{Pott2008} to normalize the PAH1 and PAH2 detection (Fig. \ref{fig:x3_all_visir}, Appendix \ref{sec:appendix_multiwavelength_detection}) of the X3 system and infer the intensity ratio $\rm I_{PAH1}/I_{PAH2}$ of $\sim\,6$. Following the analysis of Herbig Ae/Be stars shown in \cite{Maaskant2014}, we can conclude that the PAH total emission can be distributed in a radius around the protostar between a few AU up to 1000 AU. The PAH intensity ratio implies gas flows between optical thick disks close and further away from the protostar.\newline 
For the X3 system, we can therefore conclude that the presence of a thick accretion disk and an optically thin region of gas flow \citep[][]{Maaskant2014} or transition disk \citep[][]{Haworth2016a} embedded in an optically thick dust envelope is the most plausible explanation for the observed lines. Strongly ionized outflows are responsible for the spatially extended Br$\gamma$ emission. The size of the Br$\gamma$ confined area matches all other lines detected in this work in agreement with the hypothesis of domination photoionized winds. For a visualization of the different components and interplays, please see the sketch displayed in Fig. \ref{fig:sketch}.}

\subsection{The gravitational stable X3 system}

Although the gravitational footprint of Sgr~A* can be traced at a distance of $\sim 0.1$ pc, the high stellar mass of X3a effectively shields the inner region of the system on the scale of $\sim 100\,{\rm AU}$. To provide a quantitative estimate of the gravitational imprint of Sgr~A* on the X3 system, we estimate the tidal radius with
\begin{equation}
    r_{\rm t}\,\sim \,R_{\rm X3} (2M_{\rm SgrA*}/m_{\rm X3a})^{1/3}
\end{equation}
where $R_{\rm X3}\,\sim \,0.005 $pc is the approximate size of the X3 bow shock, $M_{\rm SgrA*}\,=\,4\times 10^{6} M_{\odot}$ is the mass of Sgr~A* \citep{Peissker2022}, and $m_{\rm X3a}$ is the mass of X3a ($\sim 15\,M_{\odot}$). 
Using the above relation, we estimate the tidal radius of $r_{\rm t}\sim 0.41 $pc, which is greater than or comparable to the X3 distance from Sgr~A*, $d_{\rm X3}\sim 0.3\,{\rm pc}$ (see Sec. \ref{sec:results_sed}). Therefore, the outermost bow-shock can be affected by the ambient tidal field of Sgr~A*. However, the Hill radius of the X3a is $r_{\rm Hill}\sim d_{\rm X3}[m_{\rm X3a}/(3M_{\rm SgrA*})]^{1/3}\sim 700$ AU, which implies that the stellar core and the surrounding circumstellar material are bound and not significantly affected by the gravitational field of Sgr~A*. The extended sphere of influence of X3a has likely also determined the fate of the thermal warm blob X3b, which disappeared when it reached a comparable distance from X3a. The remnant of X3b will approach the star on the free-fall timescale of $t_{\rm ff}\sim \pi (r_{\rm Hill}/2)^{3/2}(Gm_{\rm X3a})^{-1/2}\sim 850$ years, see also Fig.~\ref{fig:distance_blobs}. The disappearance of X3b may be related to faster hydrodynamical processes, specifically the interaction with the bound material around the star or the stellar outflow and subsequent shocks. 

\subsection{Dust temperature and sublimation radius}

{ Given the large bolometric luminosity of the X3a star, $L_{\rm bol}\simeq 24\times 10^3\,L_{\odot}$, and the stellar radius, $R_{\star}\simeq 10\,R_{\odot}$, see Table~\ref{tab:sed_values}, we obtain the effective temperature of $T_{\star}\simeq 22700\,{\rm K}$. This implies that the X3a star emits most of its radiation in the UV domain ($\lambda_{\rm max}\sim 128\,{\rm nm}$) and due to the large luminosity, the circumstellar dust evaporates at the sublimation radius $r_{\rm sub}$ at a certain distance from the star where it reaches the temperature of $T_{\rm sub}\sim 1500\,{\rm K}$. To estimate $r_{\rm sub}$, we use the model of \citet{1987ApJ...320..537B} for the temperature of the graphite grains at the distance $r_{\star}$ from the star given the UV luminosity $L_{\rm UV}$ of the central source and the optical depth $\tau_{\rm UV}$ of the material, in which grains are located. For $L_{\rm UV}\sim L_{\rm bol}$, we obtain the dust temperature scaled approximately to the sublimation temperature,}
\begin{equation}
    T_{\rm dust}\simeq 1501.3 \left[\left(\frac{L_{\rm UV}}{24\times 10^3\,L_{\odot}} \right) \left(\frac{r_{\star}}{23\,{\rm AU}} \right)^{-2} e^{-\tau_{\rm UV}}\right]^\frac{1}{5.6}\,{\rm K}\,,
    \label{eq_dust_temp1}
\end{equation}
{ hence, if the dust is not shielded, it evaporates at $r_{\rm sub}\sim 23\,{\rm AU}$. At larger distances, colder dust contributes to the mid- and far-infrared thermal emission of the X3 system. The peak in the SED close to 20-30\,${\rm \mu m}$, see Fig.~\ref{fig:x3_sed}, can be interpreted by the thermal emission of dust with the temperature of $\sim 100\,{\rm K}$, which must be shielded from the star by the optically thick circumstellar disk and outflows. If the optical depth reaches values close to $\tau_{\rm UV}\sim 10$, dust with $T_{\rm dust}\sim 100\,{\rm K}$ is located at $r_{\star}\sim 300\,{\rm AU}$ according to Eq.~\ref{eq_dust_temp1}.}  

{ A comparable dust sublimation radius is obtained from the numerically determined relation by \citet{2004ApJ...617.1177W},
\begin{equation}
    r_{\rm sub2}=R_{\star}\left(\frac{T_{\rm sub}}{T_{\star}} \right)^{-2.085}\,,
    \label{eq_dust_temp2}
\end{equation}
which gives $r_{\rm sub2}\sim 289R_{\star}\sim 13.4\,{\rm AU}$. Eq.~\ref{eq_dust_temp2} takes into account shielding by an optically thick inner disk wall, hence the dust can then exist still closer in comparison with the optically thin limit ($\tau_{\rm UV}=0$) evaluated in Eq.~\ref{eq_dust_temp1}. 

\begin{figure}
    \centering
    \includegraphics[width=\columnwidth]{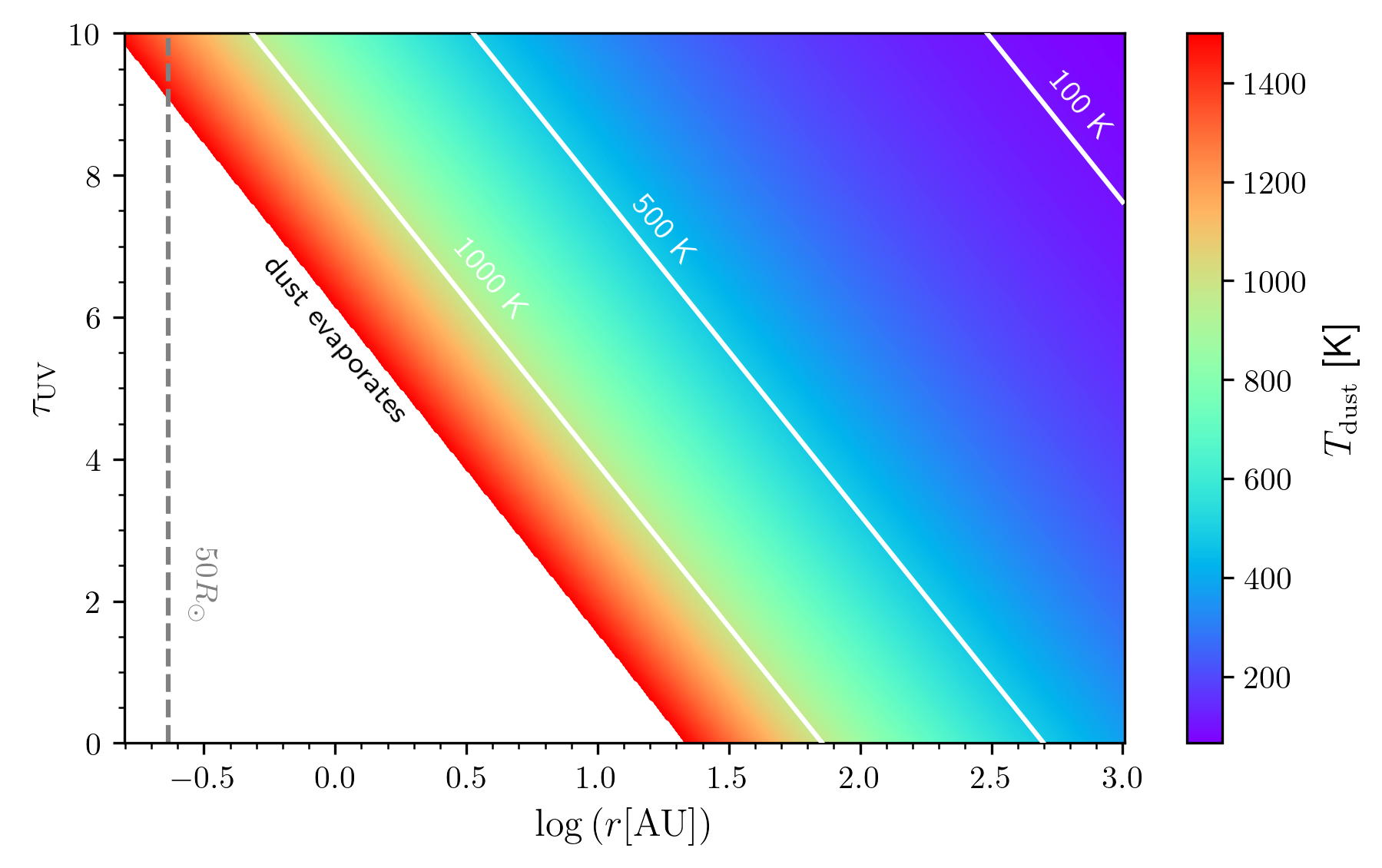}
    \caption{{ Dependence of the dust temperature $T_{\rm dust}$ on the distance from the X3a star ($\log{r}$) and the optical depth $\tau_{\rm UV}$ of the circumstellar medium at a given distance. The vertical dashed gray line marks the distance of 50$R_{\odot}$ for orientation. The contours for $100$, $500$, and $1000\,{\rm K}$ are depicted as white solid lines. The white region in the left corner stands for the parameter space where $T_{\rm dust}>1500\,{\rm K}$.}}
    \label{fig_dust_temp_X3}
\end{figure}

In Fig.~\ref{fig_dust_temp_X3}, we show the distribution of the dust temperature around the X3a star as a function of the distance from the star (in AU) and the optical depth, using the simplified model of \citet{1987ApJ...320..537B} for a single point source with the UV luminosity of $L_{\rm UV}=24000\,L_{\odot}$. The plot shows a strong dependence of the dust temperature at a given distance on the optical depth, and hence on the shielding by the opaque circumstellar disk and outflows.

}

\subsection{Formation history of X3}

As a consequence of this analysis, the presence of a high-mass YSO requires { to assess} the question of the origin and the formation process of the X3 system. We will therefore discuss three different possible scenarios for the formation of the X3 system.\newline
\begin{enumerate}
    \item For the first scenario, we consider the proposed top-heavy initial mass function for young stars in the GC proposed by \cite{Lu2013}. The authors assume a single starburst about 6-7 Myr ago that explains the presence of investigated young stars. Due to the age of the X3 system of $\sim 0.04$ Myr, this formation scenario can be excluded.
    \item The second scenario is proposed by, among others, the authors of \cite{Jalali2014}. This scenario of in-spiralling molecular clouds with masses of about 100 M$_{\odot}$ might be appealing since \cite{Yusef-Zadeh2013} already found indications of high-mass star formation, which was underlined by the observations of \cite{Moser2017} and \cite{Hsieh2021}. However, there should be at least some tracers of a recent massive cloud infall, such as elongated gaseous-dust features or trail stars \citep[for further discussion, see also][]{Paumard2006}.
    \item The third and the last discussed scenario is a continuous star formation process in the GC. In this scenario, it is believed that ongoing star formation processes take place in the so-called (counter-)clockwise disks as proposed by \cite{Paumard2006}. Due to high gas and dust temperature \citep{Cotera1999}, the turbulent environment \citep{Genzel2000}, and strong magnetic field, this scenario seems at least questionable. On the other hand, it can explain the variety of stellar members and regions in the inner parsec, such as IRS13, IRS 16, and early/late-type stars \citep[][]{Krabbe1995, Habibi2017, Gautam2019}. Overdensities in the two disks could have created cluster structures like the mentioned IRS 13 and IRS 16 regions. 
\end{enumerate}
We consider a combination of the second and third scenario as the most plausible explanation for the origin of the X3 system {without excluding the starburst scenario as a general concept}. Specifically, a cluster of young stars could have started to form outside the inner parsec, e.g. in the region of the current Circum-nuclear disk, and at the same time, it was migrating towards Sgr~A*. {Due to cloud-cloud interactions, the inital birthplace of the X3 sytem could have lost angular momentum. The dynamical instability timescales that are required for the formation of the X3 system are in agreement with Kelvin-Helmholtz timescales as we have shown for a young T-Tauri star \citep[][]{peissker2021c}.} As suggested by \cite{Maillard2004}, IRS 13 might be an example of an evaporating cluster with ongoing star formation. \cite{Maillard2004} propose that IRS 13 is the remnant core of a massive cluster. As a result, IRS 13 could be the birth site of some of the high-mass stars in the GC as discussed by the authors. \cite{Portegies-Zwart2003} favor a similar explanation for the IRS 16 stars as they also propose an in-spiral of the cluster caused by dynamical friction.\newline
Taking into account the direction and proper motion of the studied dusty objects of IRS 13 \citep{Eckart2004}, our estimated value of v$_{\rm PROP}\,=\,244\pm 27$km/s for the X3 system coincides with the these presumably YSOs of the cluster \citep[][]{muzic2008, muzic2010}. Furthermore, v$_{\rm PROP}$ of X3 matches the proper motion of the evolved WR stars E2 and E4 \citep[][]{Zhu2020}. Therefore, we cannot exclude the possibility that the X3 system might be a cluster member of IRS 13. Considering the young age, X3a must have formed {\it in situ} due to the migration timescales \citep{Morris1993}. This implies that IRS 13 could have initially served as the birthplace of the X3 system in agreement with the discussion about the origin of some high mass stars by \cite{Maillard2004}. During the infall of IRS 13, the X3 system might have been separated from the cluster due to tidal disintegration and the initial velocity dispersion of forming stellar cores. This could also explain the spatial offset of about 1 arcsec between the X3 system and IRS 13. Consequently, the in-spiralling and evaporating cluster should have lost more sources similar to X3a \citep{Paumard2006}. Tracers of this event might be the {low- and high-mass YSOs observed by \cite{Yusef-Zadeh2013, Yusef-Zadeh2015, Yusef-Zadeh2017} in the inner two parsec of the GC. If IRS 13 is, however,} not classified as a cluster but rather as a fragmented disk structure, the theoretical scenarios proposed by \citet{Bonnel2008}, \citet{Hobbs2009}, and \citet{Jalali2014} that describe the in-spiral of massive molecular clouds should be considered to explain the formation of the X3 system. In particular, the simulations by \citet{Bonnel2008} favor the formation of high-mass stars, which could serve as a plausible explanation for the high-mass YSO described here. {While this formation scenario seems appealing, it does not explain the discovery of 11 low-mass bipolar outflow sources by \cite{Yusef-Zadeh2017}. These bipolar outflow sources are located in the inner parsec, but also in the S-cluster as we suggest in \cite{Peissker2019}. Recently, \cite{Owen2022} proposed a new formation path for X8 \citep{Peissker2019} and similar sources \citep[][]{Peissker2020b, Ciurlo2020} that requires the presence of giant planets in the related disk of the young protostar. Although all of the above scenarios aim to explain a specific stellar type or group in the NSC, no approach is able to address the rich presence of various stars of different ages. For example, the young stars in the S-cluster exhibit an age range of $\sim\,3-15$ Myr clearly inconsistent with the starburst 6 Myr ago \citep[][]{Lu2013} or an infalling molecular cloud with conjunt star formation processes \citep[][]{Jalali2014}.}
In the mid-term perspective, upcoming GC observations with the James Webb Space Telescope utilizing MIRI {IFU data} could add valuable insights into {these scenarios}.

%\subsection{Alternative models}
%
%Although we investigated various different aspects that favor the classification of X3a as a YSO, we want to discuss %some alternative explanations for the observed results. From the analysis presented, we can already exclude a %late-type stellar nature of X3a. This limits the exploration of another explanation for the observations to younger %ages of the system.
%\begin{enumerate}
%    \item{Binary system:} Since the observations of X3a are limited to the current resolution, it is possible that %the system contains more than one stellar component. For example, the IRS 13 cluster was historically considered %to be one single source. With increasing resolution power, it was found that the IRS 13 "source" consist of %different components. Hence, the observed velocity gradient along the source could be interpenetrated as two %sources that are gravitiationally bound.
%    \item{Central stars of planetary nebulae:}
%\end{enumerate}

\section{Conclusion} \label{sec:conclusion}
In this work, we have presented the observations of the first { candidate} high-mass YSO close to Sgr~A* using a data baseline of almost 30 years with about four different telescopes in various wavelength regimes. In the following, we will outline our key findings and the related interpretation as discussed above.
\begin{enumerate}
    \item In the NIR/MIR, we have identified several components related to the X3 system in the H-, K-, L-, and M-band,
    \item Because of the broad wavelength coverage, a coreless gas/dust feature can be excluded,
    \item Based on the extensive data baseline covering two decades of observations, the components of the system (X3, X3a, X3b, X3c) move with a comparable proper motion towards the IRS 13 cluster,
    \item {The H- and K-band detection of X3a between 1995 and 2020 imply a stellar classification of this component of the X3 system. It is therefore plausible to classify X3a as the embedded stellar source of the dusty envelope X3,}
    \item Due to the missing NIR CO absorption lines and the depth of the 2.36$\mu m$ spectral feature, X3a { is consistent with} a young (proto)star { rather than an evolved, late-type star,}
    \item The hot blob X3b with a decreasing distance towards the central stellar source X3a is below the detection limit in 2012 with no traceable emission in the following epochs. { The independently calculated theoretical Hill radius matches the NIR observations and implies that X3b was likely accreted in 2011/2012,}
    \item For the hot L-band blob X3c, we find a constant distance towards X3a which suggests that it is created due to the { thermal} pressure of the ambient medium,
    \item The spectroscopic footprint reveals a rich abundance of NIR emission lines,
    \item We { detect} P-Cygni profile of the HeI line which indicates the presence of a wind with over -400 km/s { terminal velocity}. { Such a high wind velocity is a common property of young stars},
    \item A detailed analysis of the Br$\gamma$ line reveals a continuous velocity gradient { that coincides with the position of X3a implying a physical connection. This emission line is most likely connected to photoionized outflows/winds that might originate close to the protostar or a gaseous accretion disk,} 
    \item This interpretation of the Br$\gamma$ emission correlates with the MIR lines that originate in a dense and compact region,
    \item The {spatial distribution and dimensions of the NIR and MIR emission matches the size of the radio/submm observations of ionized H30$\alpha$ line implying that the ionization process might be produced by the same mechanism, namely, photoionized outflows},
    \item {The H30$\alpha$ line is arranged in a ring-like structure with a diameter of about 2000 AU and most likely shielded in an optical thick envelope in agreement with recent independent observations of massive YSOs, i.e., G28.200.05},
    \item In general, the organic and complex molecules observed in the {NIR and MIR are associated tracers for the presence of a YSO},
    \item Based on our { 3d MCMC radiative transfer calculations}, we { infer the stellar mass of $15^{+10}_{-5} M_{\odot}$ and an age of a few $10^4$ years for the X3 system,}
    \item Considering the 3d distance and proper motion of the X3 system, it may have been a former member of the IRS 13 cluster.
\end{enumerate}
In terms of the { future} perspective, we expect more insights into the X3 system with ERIS (VLT), MIRI (JWST), GRAVITY (VLTI), and METIS (ELT).

\begin{acknowledgments}
This work was supported in part by the
Deutsche Forschungsgemeinschaft (DFG) via the Cologne
Bonn Graduate School (BCGS), the Max Planck Society
through the International Max Planck Research School
(IMPRS) for Astronomy and Astrophysics as well as special
funds through the University of Cologne. We acknowledge support for the Article Processing Charge from the DFG (German Research Foundation, 491454339). MZ acknowledges the financial support by the National Science Center, Poland, grant No. 2017/26/A/ST9/00756 (Maestro 9) and the NAWA financial support under the agreement PPN/WYM/2019/1/00064 to perform a three-month exchange stay at the Charles University in Prague and the Astronomical Institute of the Czech Academy of Sciences. MZ also acknowledges the GA\v{C}R EXPRO grant 21-13491X (``Exploring the Hot Universe and Understanding Cosmic Feedback") for financial support. Part of this
work was supported by fruitful discussions with members of
the European Union funded COST Action MP0905: Black
Holes in a Violent Universe and the Prague--Cologne Exchange Program for university students. VK thanks the Czech Science Foundation
(No.\ 21-11268S). AP, JC, SE, and GB contributed useful points to the discussion. We also would like to 
thank the members of the SINFONI/NACO/VISIR and ESO's Paranal/Chile team for their support and collaboration.
\end{acknowledgments}
%Conditions and Impact of Star Formation is carried out within the Collaborative Research Centre 956, sub-project [A02], funded by the Deutsche Forschungsgemeinschaft (DFG) – project ID 184018867.

This research has made use of the Keck Observatory Archive (KOA), which is operated by the W. M. Keck Observatory and the NASA Exoplanet Science Institute (NExScI), under contract with the National Aeronautics and Space Administration.

\vspace{5mm}
\facilities{HST (NICMOS), VLT (SINFONI, VISIR, ISAAC, and NACO), ALMA, KECK (NIRCAM2).}

\software{astropy \citep{2013A&A...558A..33A,2018AJ....156..123A},  
          Hyperion \citep{Robitaille2011},
          DPuser \citep{Ott2013}.
          }

\bibliography{bib}{}
\bibliographystyle{aasjournal}

\appendix

This Appendix provides additional detections of the X3 system and lists the data used in this work. In addition, we present a preliminary Keplerian solution for the orbit of the X3 system.

\section{Data}
\label{ref:data_appendix}

Most of the data used in this work are freely available from the ESO, ALMA, and Keck archives and have already been analyzed in several related publications, such as \cite{Witzel2012}, \cite{Shahzamanian2017}, \cite{Peissker2020b}, \cite{Peissker2020c}, \cite{Peissker2020d}, \cite{peissker2021}, and \cite{peissker2021c}. 

\begin{table*}[h!]
\centering
\begin{tabular}{ccc}
\hline
\hline
\multicolumn{3}{c}{NACO H-band}\\
\hline
Date  & Observation ID & \multicolumn{1}{p{1.5cm}}{\centering number \\ of exposures } \\
\hline
2002.08.30 & 60.A-9026(A)  & 25  \\
2004.03.29 & 072.B-0285(B) & 92 \\
2006.10.15 & 078.B-0136(A)  & 48   \\
2007.04.04 & 179.B-0261(A)  & 192   \\
2007.07.21 & 179.B-0261(D)  & 192   \\
2010.05.09 & 183.B-0100(T)  & 21   \\
2012.07.21 & 088.B-0308(B)  & 28   \\
2018.07.14 & 0101.B-0570(A)  & 29   \\
2019.05.03 & 5102.B-0086(D)  & 29   \\

\hline  
\end{tabular}
\caption{H-band data observed with NACO between 2002 and 2019.}
\label{tab:naco_data0}
\end{table*}

\begin{table*}[h!]
\centering
\begin{tabular}{ccc}
\hline
\hline
\multicolumn{3}{c}{NACO K-band}\\
\hline
Date  & Observation ID & \multicolumn{1}{p{1.5cm}}{\centering number \\ of exposures } \\
\hline
2002.07.31 & 60.A-9026(A)  & 61   \\
2003.06.13 & 713-0078(A)   & 253  \\
2004.07.06 & 073.B-0775(A) & 344  \\
2004.07.08 & 073.B-0775(A) & 285  \\
2005.07.25 & 271.B-5019(A) & 330  \\
2005.07.27 & 075.B-0093(C) & 158  \\
2005.07.29 & 075.B-0093(C) & 101  \\
2005.07.30 & 075.B-0093(C) & 187  \\
2005.07.30 & 075.B-0093(C) & 266  \\
2005.08.02 & 075.B-0093(C) & 80   \\
2006.08.02 & 077.B-0014(D) & 48   \\
2006.09.23 & 077.B-0014(F) & 48   \\
2006.09.24 & 077.B-0014(F) & 53   \\
2006.10.03 & 077.B-0014(F) & 48   \\
2006.10.20 & 078.B-0136(A) & 47   \\
2007.03.04 & 078.B-0136(B) & 48   \\
2007.03.20 & 078.B-0136(B) & 96   \\
2007.04.04 & 179.B-0261(A) & 63   \\ 
2007.05.15 & 079.B-0018(A) & 116  \\ 
2008.02.23 & 179.B-0261(L) & 72   \\ 
2008.03.13 & 179.B-0261(L) & 96   \\ 
2008.04.08 & 179.B-0261(M) & 96   \\ 
2009.04.21 & 178.B-0261(W) & 96   \\ 
2009.05.03 & 183.B-0100(G) & 144  \\ 
2009.05.16 & 183.B-0100(G) & 78   \\ 
2009.07.03 & 183.B-0100(D) & 80   \\ 
2009.07.04 & 183.B-0100(D) & 80   \\ 
2009.07.05 & 183.B-0100(D) & 139  \\ 
2009.07.05 & 183.B-0100(D) & 224  \\ 
2009.07.06 & 183.B-0100(D) & 56   \\ 
2009.07.06 & 183.B-0100(D) & 104  \\ 
2009.08.10 & 183.B-0100(I) & 62   \\ 
2009.08.12 & 183.B-0100(I) & 101  \\
2010.03.29 & 183.B-0100(L) & 96   \\ 
2010.05.09 & 183.B-0100(T) & 12   \\ 
2010.05.09 & 183.B-0100(T) & 24   \\ 
2010.06.12 & 183.B-0100(T) & 24   \\ 
2010.06.16 & 183.B-0100(U) & 48   \\
2011.05.27 & 087.B-0017(A) & 305  \\
2012.05.17 & 089.B-0145(A) & 169  \\
2013.06.28 & 091.B-0183(A) & 112  \\
2017.06.16 & 598.B-0043(L) & 36   \\
2018.04.24 & 101.B-0052(B) & 120  \\
\hline  
\end{tabular}
\caption{K-band data observed with NACO between 2002 and 2018.}
\label{tab:naco_data1}
\end{table*}

\begin{table*}[h!]
\centering
\begin{tabular}{ccc}
\hline
\hline
\multicolumn{3}{c}{NACO L-band}\\
\hline
Date  & Observation ID & \multicolumn{1}{p{1.5cm}}{\centering number \\ of exposures }   \\
\hline
2002.08.30 &  060.A-9026(A) & 80 \\  % 16 & $L'$ \\
2003.05.10 &  071.B-0077(A) & 56 \\  % 16 & $L'$ \\
2004.07.06 &  073.B-0775(A) & 217\\  % 43.3 & $L'$ \\
2005.05.13 &  073.B-0085(E) & 108\\  % 21.6 & $L'$ \\
2005.06.20 &  073.B-0085(F) & 100\\  % 20   & $L'$ \\
2006.05.28 &  077.B-0552(A) &  46\\  %  9.2 & $L'$ \\
2006.06.01 &  077.B-0552(A) & 244\\  % 48.8 & $L'$ \\
2007.03.17 &  078.B-0136(B) &  78\\  % 15.6 & $L'$ \\
2007.04.01 &  179.B-0261(A) &  96\\  % 19.2 & $L'$ \\
2007.04.02 &  179.B-0261(A) & 150\\  % 30   & $L'$ \\
2007.04.02 &  179.B-0261(A) &  72\\  % 14.4 & $L'$ \\
2007.04.06 &  179.B-0261(A) & 175\\  % 35   & $L'$ \\
2007.06.09 &  179.B-0261(H) &  40\\  %  8   & $L'$ \\
2008.05.28 &  081.B-0648(A) &  58\\  % 11.6 & $L'$ \\
2008.08.05 &  179.B-0261(N) &  64\\  % 12.8 & $L'$ \\
2008.09.14 &  179.B-0261(U) &  49\\  %  9.8 & $L'$ \\
2009.03.29 &  179.B-0261(X) &  32\\  %  6.4 & $L'$ \\
2009.03.31 &  179.B-0261(X) &  32\\  %  6.4 & $L'$ \\
2009.04.03 &  082.B-0952(A) &  42\\  %  8.4 & $L'$ \\
2009.04.05 &  082.B-0952(A) &  12\\  %  2.4 & $L'$ \\
2009.09.19 &  183.B-0100(J) & 132\\  % 26.4 & $L'$ \\
2009.09.20 &  183.B-0100(J) &  80\\  % 16   & $L'$ \\
2010.07.02 &  183.B-0100(Q) & 485\\  % 97   & $L'$ \\
2011.05.25 &  087.B-0017(A) & 29 \\  %  5.8 & $L'$ \\
2012.05.16 &  089.B-0145(A) & 30 \\  %  6   & $L'$ \\
2013.05.09 &  091.C-0159(A) & 30 \\  %  6   & $L'$ \\
2015.09.21 &  594.B-0498(G) & 420   \\
2016.03.23 &  096.B-0174(A) & 60 \\  %  12  & $L'$ \\
2017.03.23 &  098.B-0214(B) & 30 \\  %   6  & $L'$ \\
2018.04.22 & 0101.B-0065(A) & 68 \\  % 13.6 & $L'$ \\
2018.04.24 & 0101.B-0065(A) & 50 \\  % 10   & $L'$ \\
\hline  
\end{tabular}
\caption{L-band data observed with NACO between 2002 and 2018.}
\label{tab:naco_data2}
\end{table*}

\begin{table*}[h!]
\centering
\begin{tabular}{|ccccc|}
\hline
\hline
%\multicolumn{5}{c}{NACO M-band}\\
%\hline
        Date & Observation ID  & Exp. Time & Band & Instrument/Telescope  \\  
        (YYYY:MM:DD) &  & (s) &  &   \\ \hline\hline %table heading\hline
         2012:03:17  & 088.B-1038(A) & 2780 & M & NACO/VLT\\  % 16 & $L'$ \\
\hline  
\end{tabular}
\caption{M-band data observed with NACO in 2012.}
\label{tab:naco_data3}
\end{table*}

\begin{table*}[h!]
\centering
\begin{tabular}{|ccccc|}
\hline
\hline
%\multicolumn{5}{c}{VISIR N/Q-band}\\
%\hline
        Date & Observation ID  & Exp. Time & Filter/Band & Instrument/Telescope  \\  
        (YYYY:MM:DD) &  & (s) &  &   \\ \hline\hline %table heading\hline
          2004:05:09 & 60.A-9234(A)  & 35 & PAHr1 & VISIR/VLT\\  % 16 & $L'$ \\
          2004:05:08 & 60.A-9234(A)  & 32 & PAH   & VISIR/VLT\\  % 16 & $L'$ \\
          2004:05:09 & 60.A-9234(A)  & 130 & ArIII & VISIR/VLT\\  % 16 & $L'$ \\
          2004:05:09 & 60.A-9234(A)  & 12 & SIVr1 & VISIR/VLT\\  % 16 & $L'$ \\
          2004:05:09 & 60.A-9234(A)  & 30 & SIVr2 & VISIR/VLT\\  % 16 & $L'$ \\
          2004:05:07 & 60.A-9234(A)  & 950 & PAH2  & VISIR/VLT\\  % 16 & $L'$ \\
          2004:05:09 & 60.A-9234(A)  & 35 & NeIIr1 & VISIR/VLT\\  % 16 & $L'$ \\
          2004:05:08 & 60.A-9234(A)  & 171 & NeII & VISIR/VLT\\  % 16 & $L'$ \\
          2004:05:09 & 60.A-9234(A)  & 104 & NeIIr2 & VISIR/VLT\\  % 16 & $L'$ \\
          2004:05:09 & 60.A-9234(A)  & 62 & Q2     & VISIR/VLT\\  % 16 & $L'$ \\
          2004:05:09 & 60.A-9234(A)  & 340 & Q3     & VISIR/VLT\\  % 16 & $L'$ \\
\hline  
\end{tabular}
\caption{VISIR MIR data covering the N- and Q-band carried out in 2004.}
\label{tab:visir_data}
\end{table*}

\begin{table*}[h!]
\centering
\begin{tabular}{|ccccc|}
\hline
\hline
%\multicolumn{5}{c}{ISAAC L-band}\\
%\hline
        Date & Observation ID  & Exp. Time & Band & Instrument/Telescope  \\  
        (YYYY:MM:DD) &  & (s) &  &   \\ \hline\hline %table heading\hline
         2003:07:11  & 71.C-0192(A)  & 266 & SL & ISAAC/VLT \\  % 16 & $L'$ \\
\hline  
\end{tabular}
\caption{ISAAC data observed in 2003 \citep[][]{Moultaka2005, moultaka2015}. The SL filter corresponds to 2.55-4.2$\mu m$ (LWS3).}
\label{tab:isaac_data}
\end{table*}

\begin{table*}[htbp!]
        \centering
        \begin{tabular}{|ccccc|}
        \hline\hline
             Date & Observation ID  & Exp. Time & Band & Instrument/Telescope  \\  
        (YYYY:MM:DD) &  & (s) &  &   \\ \hline\hline %table heading
        
        2014.08.30 & 093.B-0218(B) &  2700  &  H+K   &  SINFONI/VLT    \\
        \hline 
        \end{tabular}   
\caption{SINFONI data used for the spectroscopic analysis in this work. The final mosaic is created from nine single data cubes with an exposure time of 300s/each. Adaptive Optics is enabled for every observation.}
\label{tab:data_sinfo}
\end{table*}

\begin{table*}[htbp!]
        \centering
        \begin{tabular}{|ccccc|}
        \hline\hline
             Date & Observation ID  & Exp. Time & Band & Instrument/Telescope  \\  
        (YYYY:MM:DD) &  & (s) &  &   \\ \hline\hline %table heading
        
        2019.08.14 & K311 &  1110  &  L   &  NIRCAM2/Keck    \\
        2019.08.14 & K311 &  952  &  K   &  NIRCAM2/Keck    \\
        2019.08.14 & K311 &  89  &  H   &  NIRCAM2/Keck    \\
        2020.08.03 & E337 &  59  &  K   &  OSIRIS/Keck    \\
        \hline 
        \end{tabular}   
\caption{Keck data observed with NIRCAM2 and OSIRIS. The L-band data is a combination of 37 single observations with an exposure time of 30s/each. The K-band data observed with NIRCAM2 consist of 34 single observations and an exposure time of 28s/each. The observations in 2020 are carried out with the OSIRIS science cam. For the stacked final mosaic, we used four single observations with an exposure time of 14.7s/each. The three single H-band observations exhibit an exposure time of about 30s/each.} 
\label{tab:data_keck}
\end{table*}

\begin{table*}[htbp!]
        \centering
        \begin{tabular}{|ccccc|}
        \hline\hline
             Date & Observation ID  & Exp. Time & Band & Instrument/Telescope  \\  
        (YYYY:MM:DD) &  & (s) &  &   \\ \hline\hline %table heading
        2016:08:30 & 2015.1.01080.S & 11250  &   7 (CO)  &  ALMA    \\
        2017:05:09 & 2015.1.01080.S  & 27155   &  5 (H30$\alpha$)   &  ALMA    \\
        \hline 
        \end{tabular}   
\caption{ALMA data used in this work. The CO data is previously analyzed in, e.g., \cite{Tsuboi2017}.} 
\label{tab:data_alma}
\end{table*}

\section{Multi-wavelength detection of X3}
\label{sec:appendix_multiwavelength_detection}

\begin{figure*}[htbp!]
	\centering
	\includegraphics[width=0.7\textwidth]{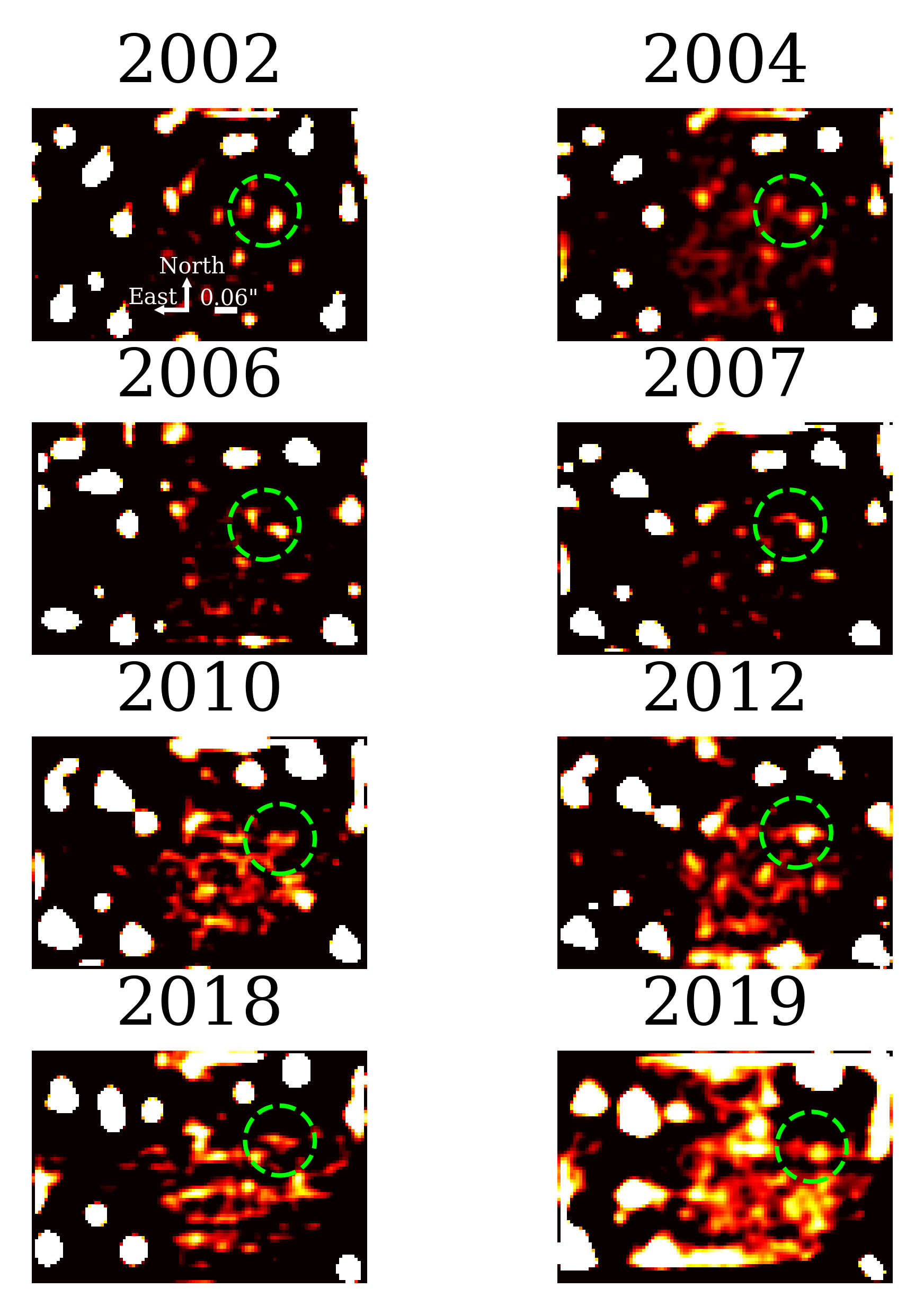}
	\caption{Observation of the X3-system in the H-band with NACO (VLT) and NIRC2 (KECK) between 2002 and 2019. We indicate the position of X3a with a lime-colored dashed circle. Until 2010, we observe X3b which seems to coincide with X3a in 2012. After 2012, X3b is below the detection limit. In contrast, X3a can be detected throughout the available data at the expected position.}
\label{fig:x3_all_h}
\end{figure*}

\begin{figure*}[htbp!]
	\centering
	\includegraphics[width=0.9\textwidth]{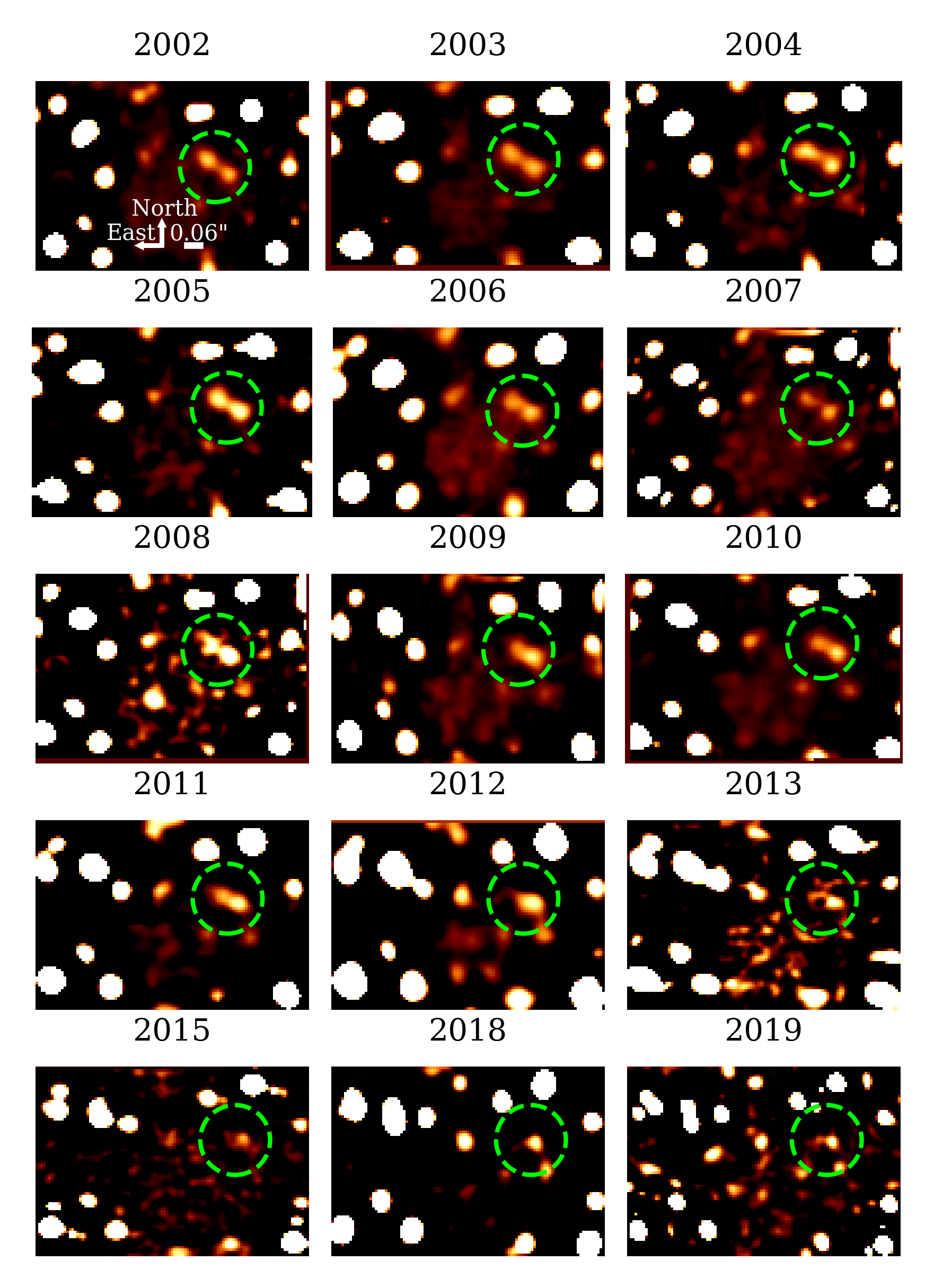}
	\caption{Observation of the X3-system in the K-band with NACO (VLT) and NIRC2 (KECK) between 2002 and 2019. The data between 2002 and 2018 was observed with NACO, the data in 2019 with NIRC2. We indicate the position of X3a and X3b with a lime-colored dashed circle. Likewise in the H-band, we do not find any emission above the noise for X3b after 2012. For a continuum observation of X3a in 2020, please see Fig. \ref{fig:osiris_2020}.}
\label{fig:x3_all_k}
\end{figure*}

\begin{figure*}[htbp!]
	\centering
	\includegraphics[width=0.7\textwidth]{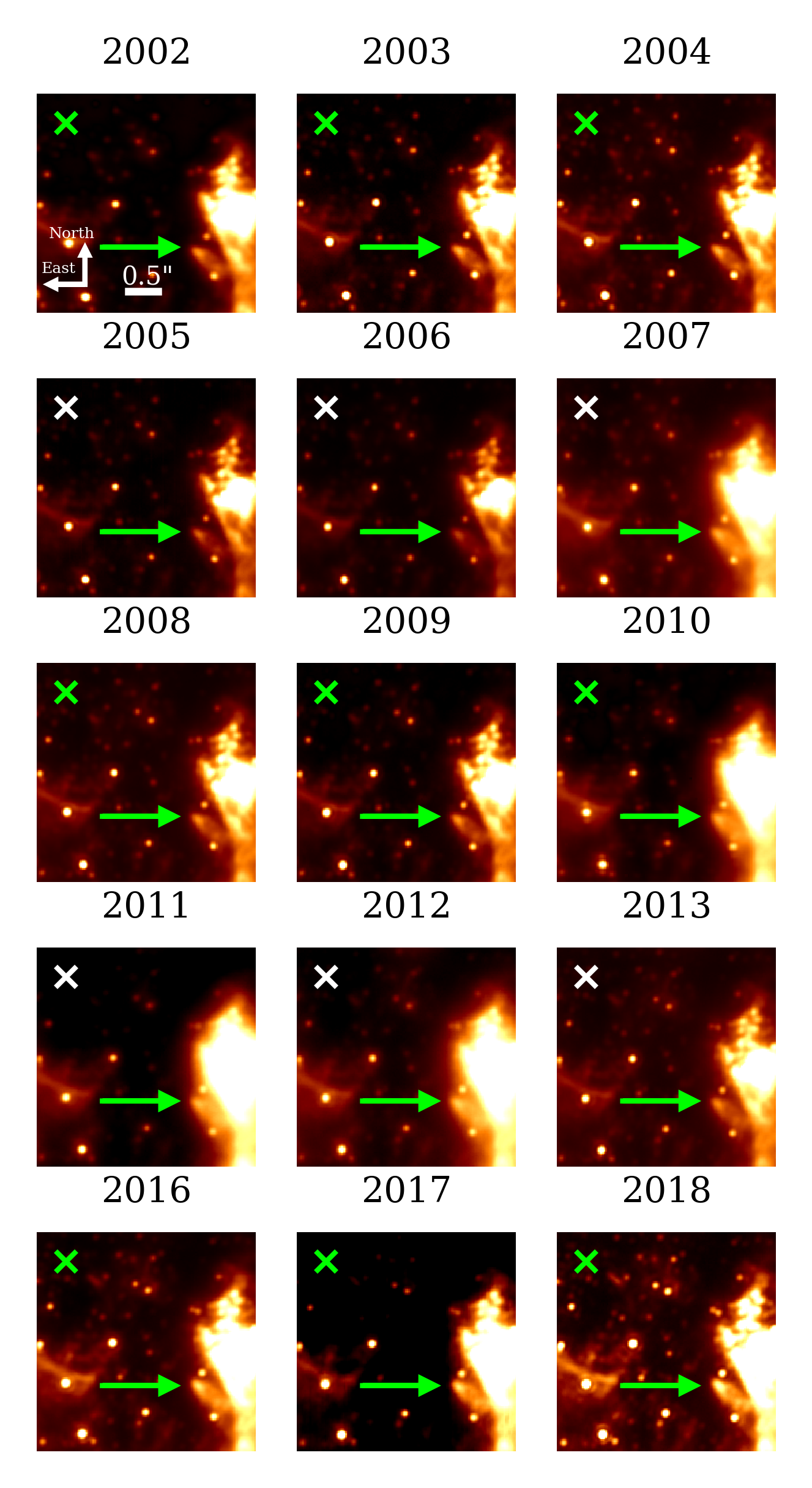}
	\caption{Observation of the X3-system in the L-band with NACO between 2002 and 2018. The position of the $\times$ indicates the position of Sgr~A*. The lime-colored arrow is pointing to the position of the bow shock X3.}
\label{fig:x3_all_l}
\end{figure*}

In this section, we present the multi-wavelength evolution of the X3 system covering almost two decades of NIR and MIR observations in the GC. This rich data set allows us to determine the properties of the X3 system. Although the H-band suffers from increased background noise (Fig. \ref{fig:x3_all_h}), we find, in agreement with the K-band data, two objects (X3a and X3b) at the position of the green dashed circle (see Fig. \ref{fig:x3_all_k}). This green circle also represents the rough position of the dusty L-band envelope shown in Fig. \ref{fig:x3_all_l}. Based on the proper motion analysis shown in Fig. \ref{fig:proper_motion}, we find that the components move along toward the north-west (for the projected direction, see also Fig. \ref{fig:x3_disk}).
In the H- and K-band, we find two emission blobs that we denote as X3a and X3b. Although X3a can be observed throughout the data (see also Fig. \ref{fig:osiris_2020}), X3b seems to fade out between 2002 and 2011. We also noticed a decreasing distance between the two sources. We note that X3b was brighter in 2002 compared to X3a.
In addition to the NACO data presented, we use MIR narrow-line VISIR { filters} to investigate the X3 system for complex molecules. As is expected for a YSO, we find { prominent} { emission lines in the MIR that originate within a compact area} \citep[][]{Berrilli1992, Hartmann1993, Brooke1993}.

\begin{figure*}[htbp!]
	\centering
	\includegraphics[width=0.7\textwidth]{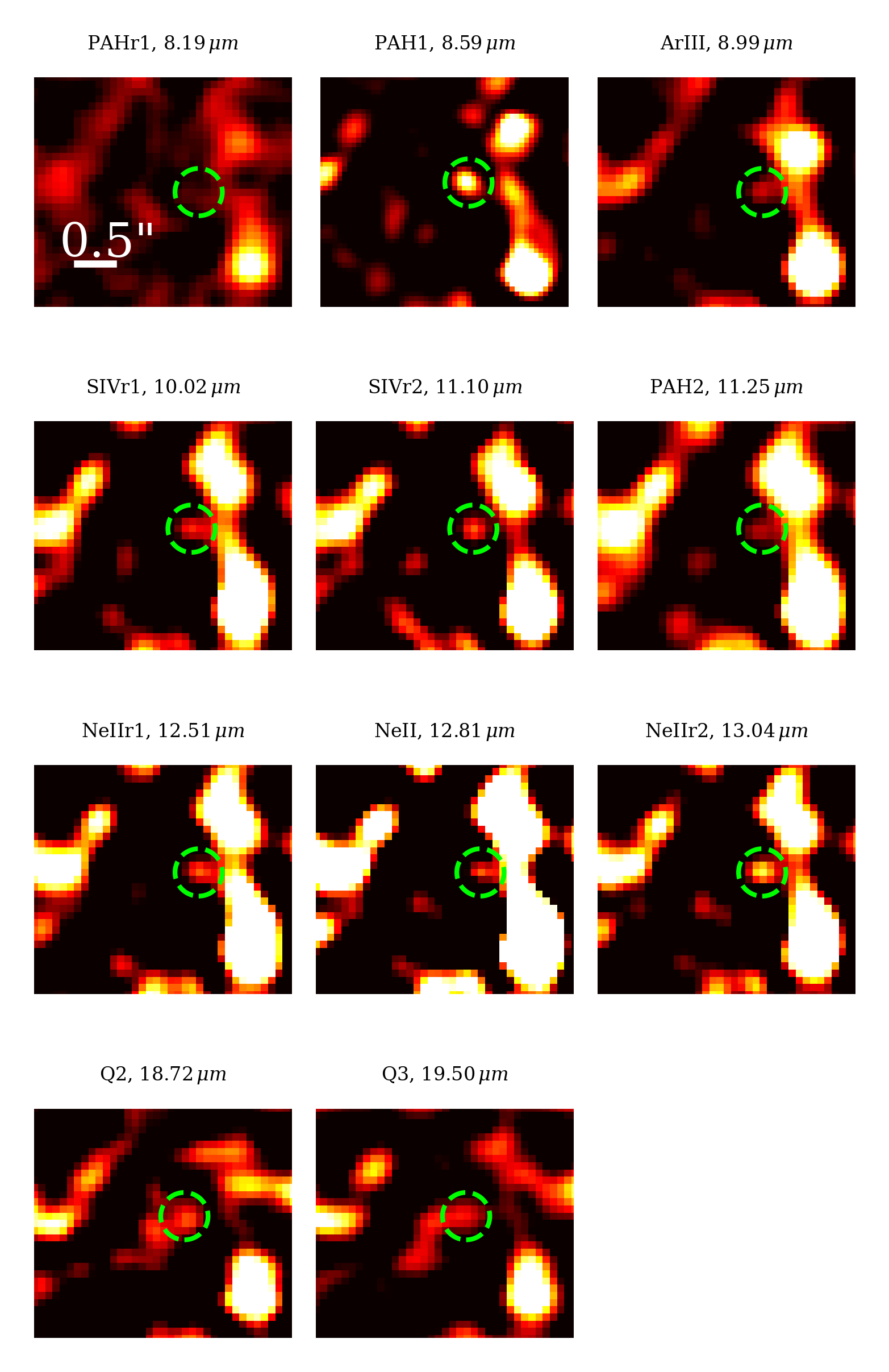}
	\caption{Observation of the X3-system with VISIR {of 2004}. The titles of these sub-panels refer to the { corresponding} filters of VISIR. Every (complex) observed molecule originates in a compact region indicated by a lime-coloured dashed circle. {In every box, North is up and East is to the left.}}
\label{fig:x3_all_visir}
\end{figure*}

\section{The Keplerian orbit of the X3 system}

Due to the bright emission of X3 in the L-band in combination with the long data baseline, we investigated its evolution to derive a Keplerian solution of the orbit. We are aware of the challenges due to the elongated shape of X3. A more sophisticated approach should incorporate all bands in combination with a precise measurement of the line-of-sight velocity of X3a to reduce the uncertainties. { Currently,} the Keplerian orbit can be treated as a quantitative approach. A more qualitative treatment of the orbit will be subject to future publications using 3d IFU data of MIRI/JWST.
\begin{table*}[]
\centering
\begin{tabular}{|ccccccc|}
\toprule
ID     & $a$ (mas)&	$e$  &	$i$($^o$)&$\omega$($^o$)&	$\Omega$($^o$)&	$t_{\rm peri}$(yr)\\
\hline \hline
X3   & 440.26$\pm$50 & 0.708$\pm$0.21 & 78.49$\pm$7.9 & 16.32$\pm$9.3 & 226.31$\pm$6.7 & 2090.05  \\
\hline
\end{tabular} %inc $\pm$8.83
\caption{Orbital elements of X3 using L-band NACO data between 2002 and 2018.}
\label{tab:kepler}
\end{table*}

%440.2693857677963 , 0.7081075676381773 , 1.3788729900954406 , 0.2858707657047062 , 3.9502755342896334 , 2090.0525458496436

\section{Further detections of the X3 system}
\label{sec:appendix_further_detections}

\begin{figure}[htbp!]
	\centering
	\includegraphics[width=0.5\textwidth]{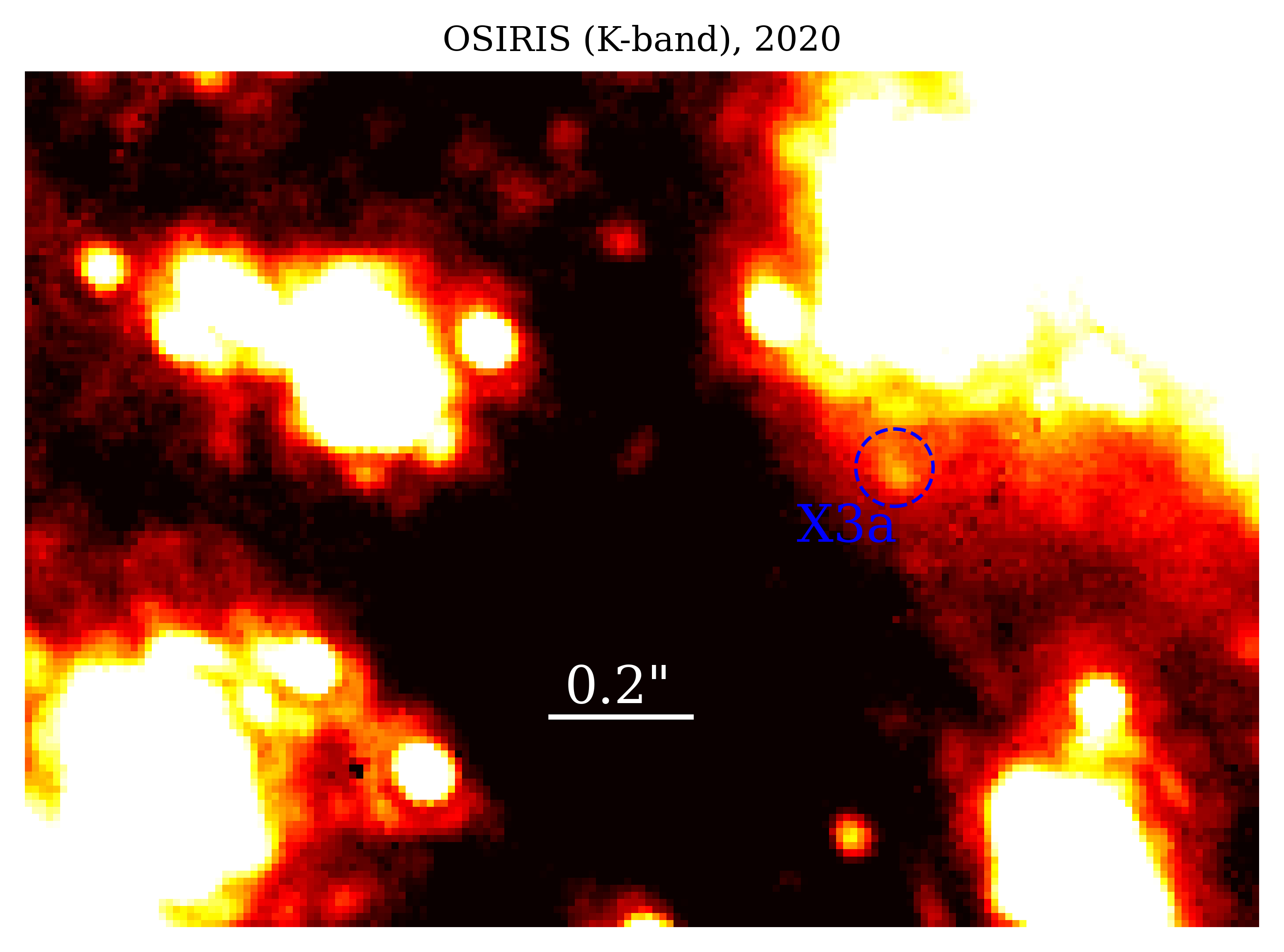}
	\caption{Continuum detection of X3a observed with OSIRIS in 2020. No filter is applied to the K-band data observed with KECK. Here, the dominant emission of the close-by cluster IRS 13 but also the nearby stars is eminent. Especially the PSF wings of S3-374 are suppressing the emission of X3a which { motivates} the use of a high-pass filter. The compact emission of X3a is located at the expected position based on the proper motion derived in this work (Fig. \ref{fig:proper_motion}). Consistent with the NACO K-band data, we derive a magnitude of $mag_{K-band}\,=\,16.0^{+0.7} _{-0.3}$ mag.}
\label{fig:osiris_2020}
\end{figure}

\begin{figure}[htbp!]
	\centering
	\includegraphics[width=0.5\textwidth]{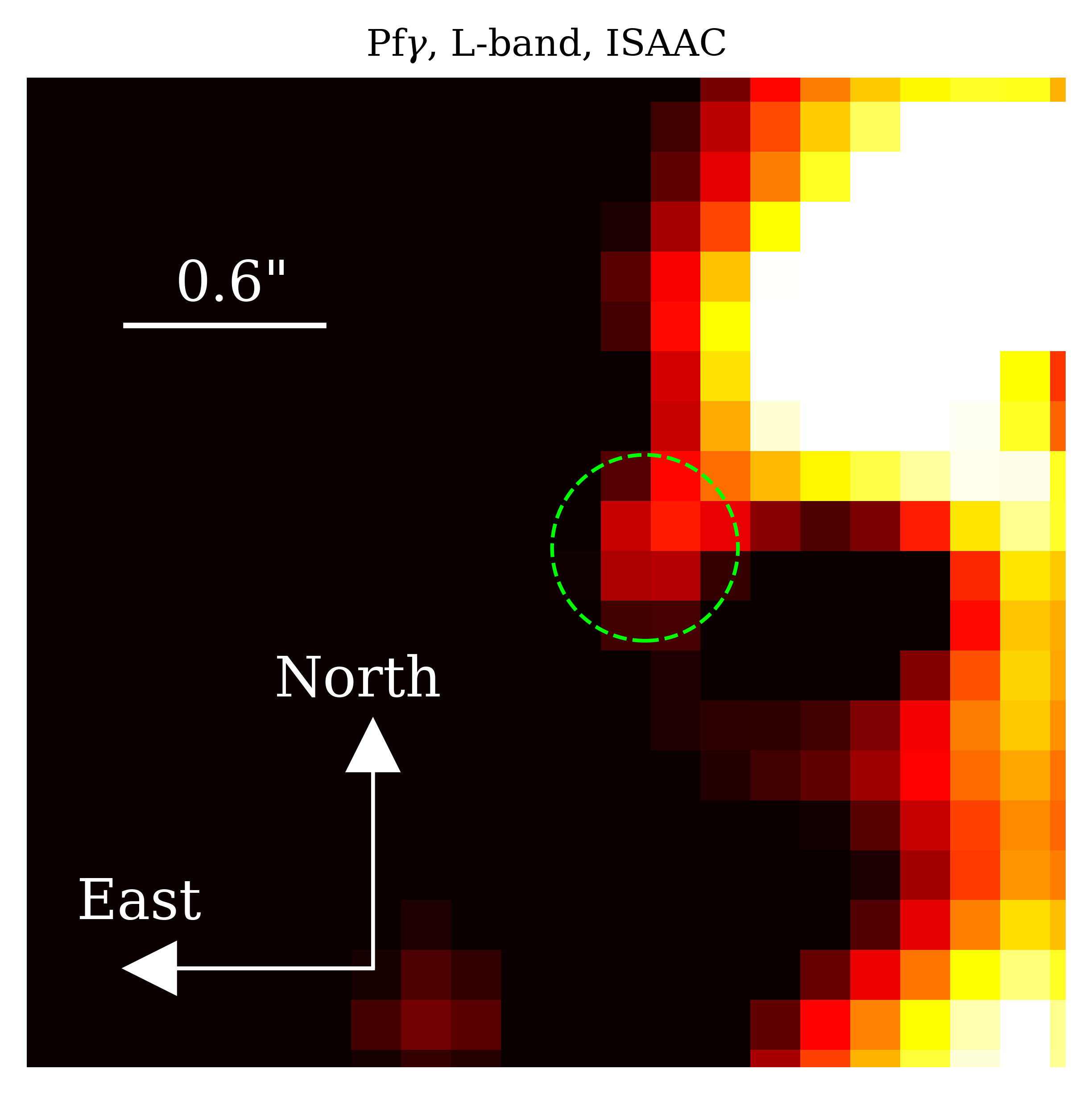}
	\caption{Combined continuum and Pf$\gamma$ line emission observed with ISAAC. The background subtracted line map exhibits the X3 system (green dashed circle) close to the bright IRS 13 cluster. The used ISAAC data cube was previously analyzed in \cite{Moultaka2005} and \cite{moultaka2015}.}
\label{fig:linemap_isaac}
\end{figure}

In this section, we will show additional detections of the X3 system that support the analysis in this work. In Fig. \ref{fig:osiris_2020}, we present data observed with the OH-Suppressing Infrared Imaging Spectrograph (OSIRIS) in 2020. The NIR instrument is mounted at the Keck telescope and works in a way comparable to that of SINFONI (VLT) with a slightly lower spatial pixel scale of 10 mas. We downloaded the reduced data from the Keck Online Archive and applied no further correction to the data. In agreement with the proper motion analysis presented in this work, we find X3a at the expected position (blue dashed circle in Fig. \ref{fig:osiris_2020}).
In addition to the NIR detection with OSIRIS, we find the prominent MIR emission line Pf$\gamma$ (rest wavelength $@\,3.7405\mu m$) in the ISAAC data cube that is analyzed in \cite{Moultaka2005}. In Fig. \ref{fig:linemap_isaac}, we show a line map of the related line. Due to the spatial and spectral resolution of the ISAAC data, the detection suffers from increased noise and artefacts.
However, we find well above the confusion limit an indisputable emission of the X3 system showing hot dust, which is related to the envelope. The detection of a Pf$\gamma$ line is in agreement with the analysis of protoplanetary disks around YSOs \citep{McClure2020}.

\section{Spectral analysis of the X3 system}
\label{sec:appendix_spectral_analysis}

\begin{figure}[htbp!]
	\centering
	\includegraphics[width=.5\textwidth]{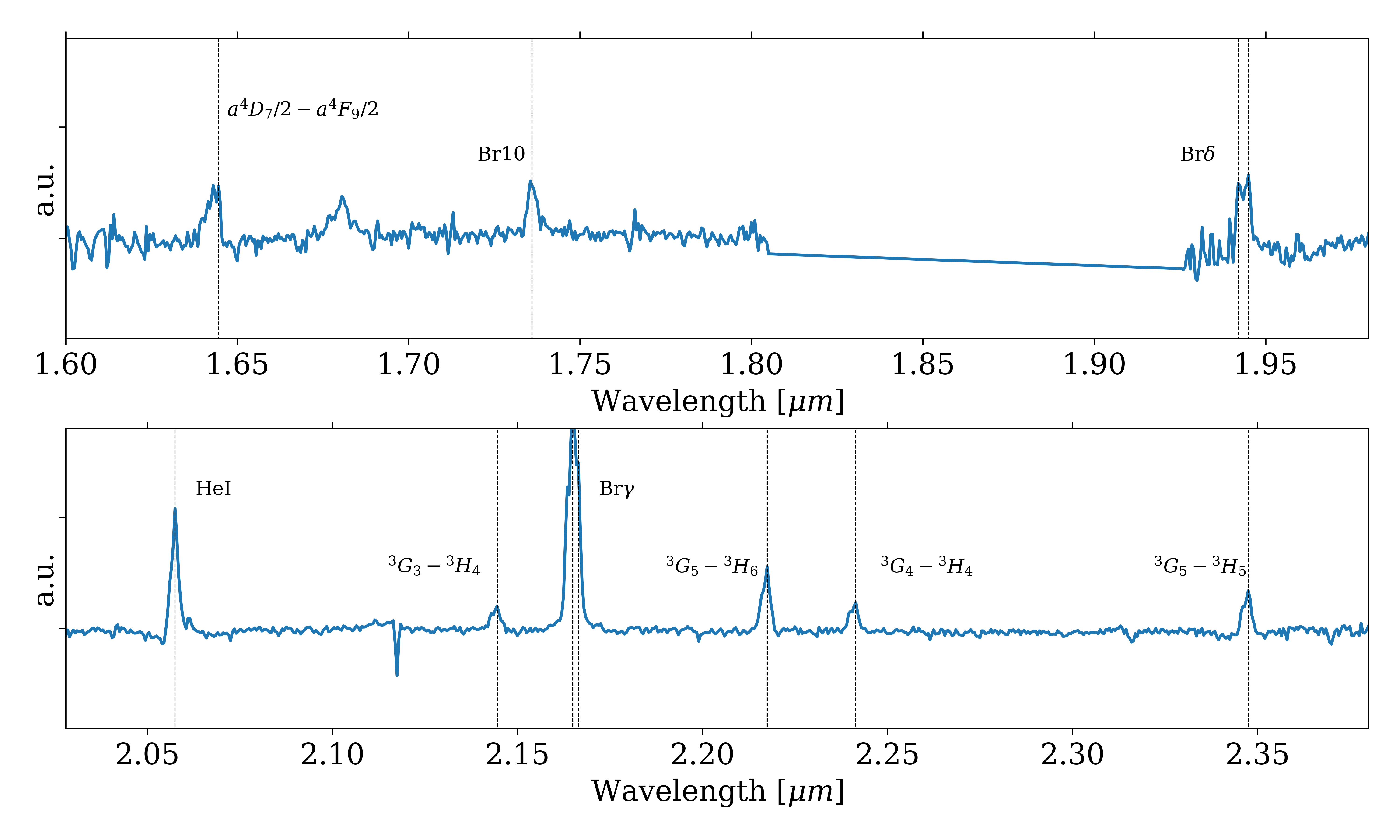}
	\caption{Near-infrared spectrum of X3a. We use a PSF sized aperture with and subtract a background which is located 1.5" north of X3a.}
\label{fig:x3_spec_sinfo_bs_2}
\end{figure}

\begin{figure}[htbp!]
	\centering
	\includegraphics[width=.5\textwidth]{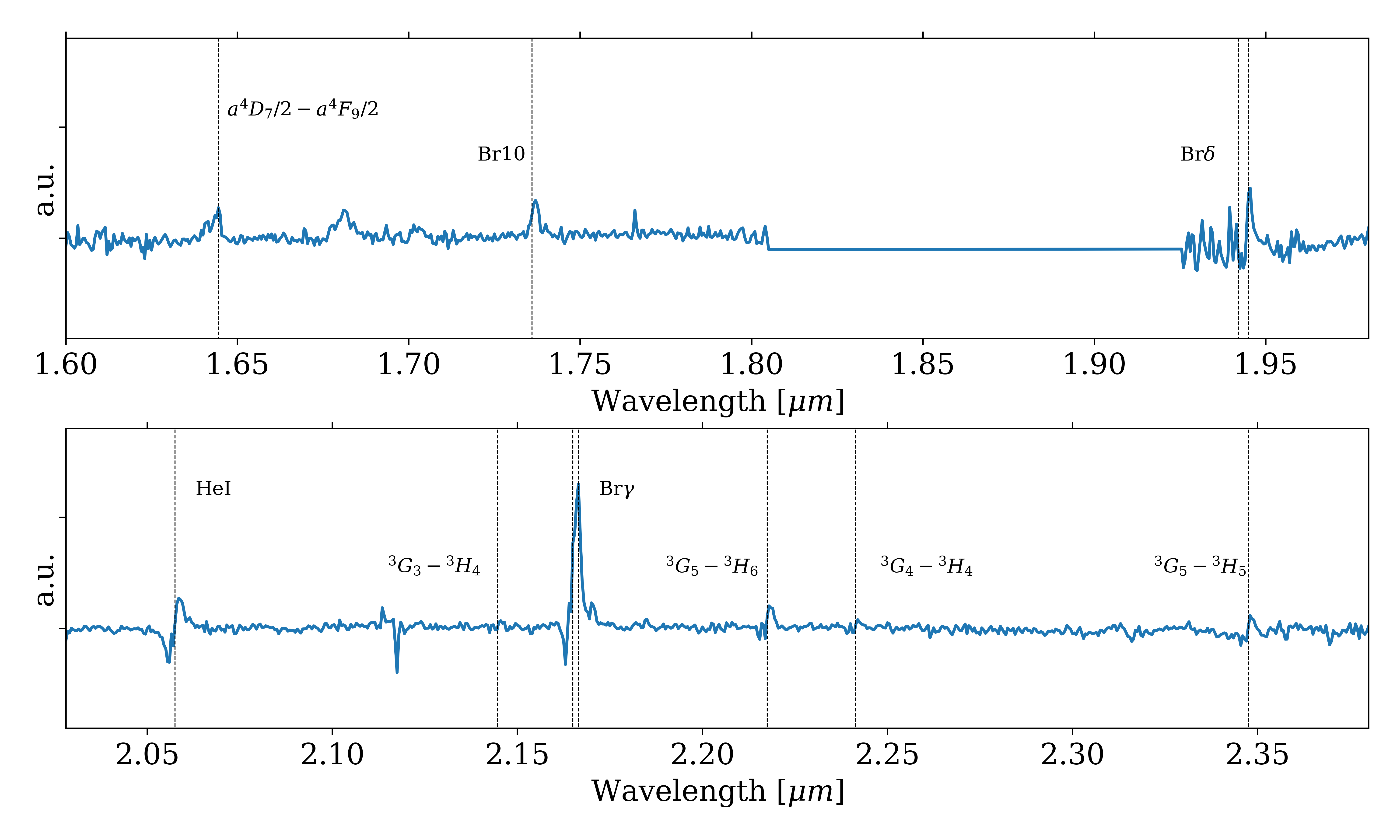}
	\caption{Same figure as Fig. \ref{fig:x3_spec_sinfo_bs_2} but with a background that is located 0.55" south of X3a}
\label{fig:x3_spec_sinfo_bs_4}
\end{figure}

\begin{figure}[htbp!]
	\centering
	\includegraphics[width=.5\textwidth]{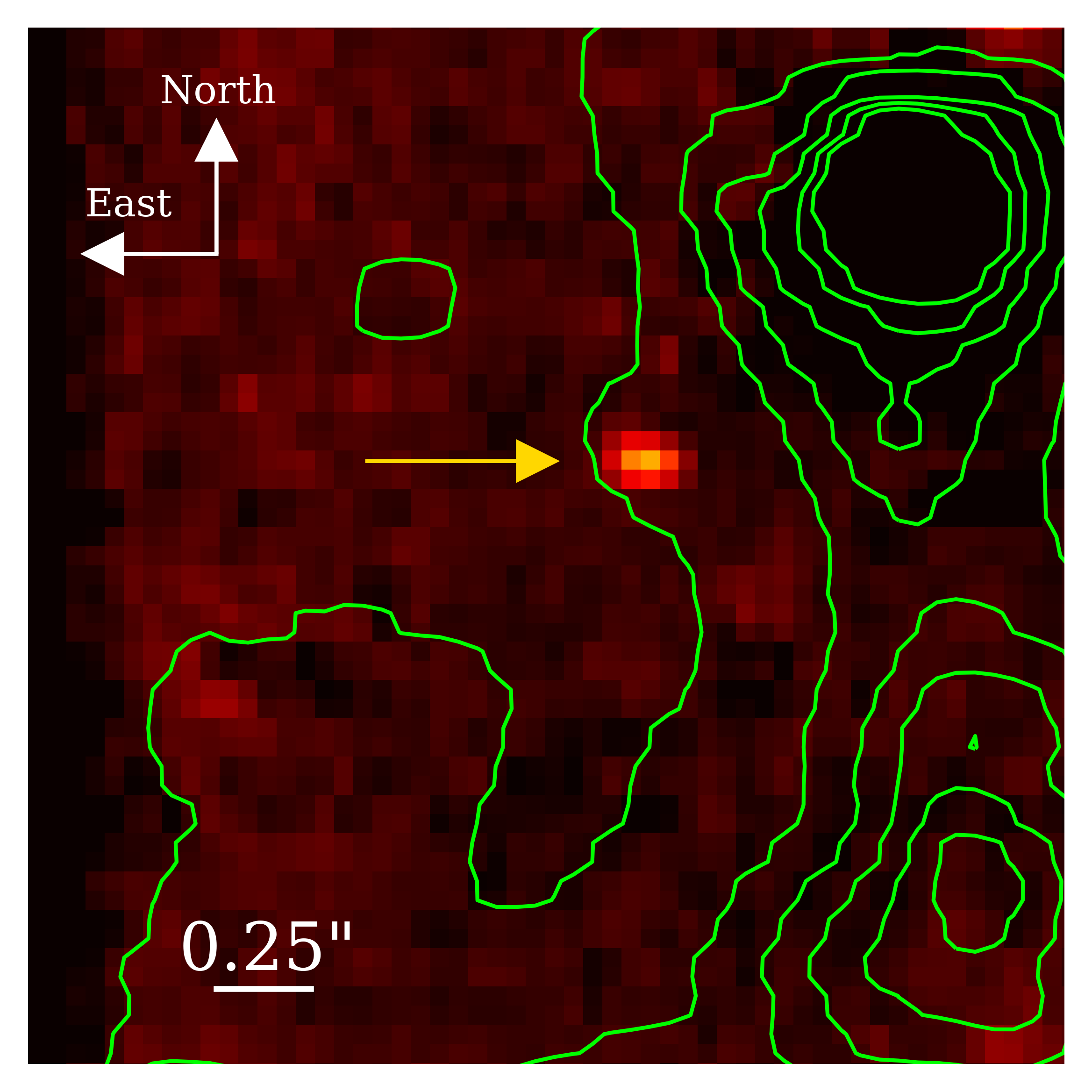}
	\caption{Line map of the H$_2$ emission line observed with SINFONI in 2014. The same FOV is shown in Fig. \ref{fig:linemaps_sinfo_sinfo}, the lime colored contour lines are adapted from the same figure showing the continuum emission of the environment of the X3 system. With a golden arrow, we indicate the position of X3a and the H$_2$ emission.}
\label{fig:x3_h2_linemap}
\end{figure}

For the spectrum shown in Fig. \ref{fig:x3_spec_sinfo}, we did not apply a background subtraction. While this might be a reasonable approach for many objects and sources outside the Galactic center, we have shown in \cite{Valencia-S.2015} the contamination that emerges from a variable background. In Fig. \ref{fig:x3_spec_sinfo_bs_2} and Fig. \ref{fig:x3_spec_sinfo_bs_4}, we show the same source spectrum as in Fig. \ref{fig:x3_spec_sinfo} but with a nearby background subtraction. Due to the limitation of the field of view with $3.2"\,\times\,3.2"$ and the nearby IRS 13 cluster (see Fig. \ref{fig:x3_system} and Fig. \ref{fig:linemaps_sinfo_sinfo}), we limit the selection to a northern region around 1.5" away from X3a. We also inspect the impact of a background about 0.55" south of the source. As it is evident from both spectra, the overall detection of all features and lines is maintained.
This is not completely unexpected since we investigate Doppler-shifted emission lines that are rather {\it unique} due to the limited FOV. In \cite{peissker2021c}, we have shown that SINFONI data suffers observed in the GC might suffer from increased background noise. However, the isolation of Doppler-shifted lines and the related channel maps remains unquestionable and are a reliable tool, to exhibit a physical connection of the investigated object and its spectrum.   
We demonstrate this relation by creating a line map for the Doppler-shifted H$_2$ line as presented in Fig. \ref{fig:x3_spec_sinfo_h2}. We integrate over the FWHM of the hydrogen line and subtract the closest channels which define the background. For a better comparison, we incorporate the contours from Fig. \ref{fig:linemaps_sinfo_sinfo}, left plot. The resulting line map is shown in Fig. \ref{fig:x3_h2_linemap} which marks an unambiguous detection of the Doppler shifted H$_2$ line.

%Spectrum (H+K band) of the X3-system extracted from the SINFONI data observed in 2014. Because of different line strengths, we subdivide the spectrum into two subplots (top and bottom). While the emission shows tracers as a double-peaked Br$\gamma$ line around 2.1661$\mu m$, we note a P-Cygni profile close to the blueshifted HeI line at 2.0575$\mu m$. We find no NIR CO tracers which excludes a late-type nature of X3a. We will discuss the missing CO absorption lines in detail in Sec. \ref{sec:discuss}.

\section{Spectral Energy Distribution of the S-cluster star S2}
\label{sec:appendix_s2_sed}

\begin{figure}[htbp!]
	\centering
	\includegraphics[width=0.5\textwidth]{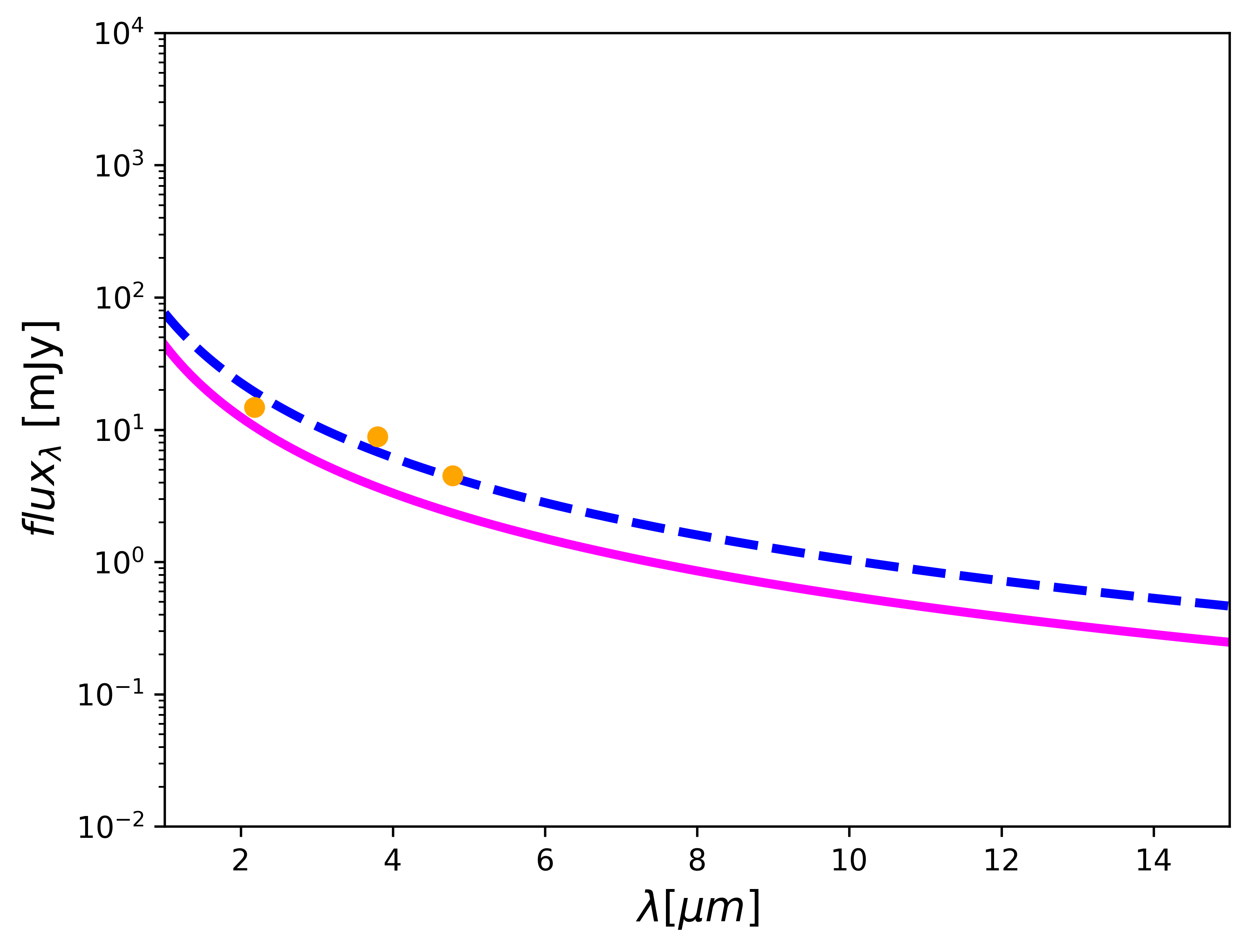}
	\caption{Spectral Energy Distribution of the B2V star S2 based on a { single-temperature} black-body fit. With this fit represented with a {blue dashed line}, we { estimate} the H-band flux density of S2 that can be used as a zero-flux point for the photometric analysis. We use $T_{\rm eff}=22500\,K$ and $R=8.5R_{\odot}$. The {magenta solid} line represents $T_{\rm eff}=28500K$ and $R=5.5R_{\odot}$ adopted from \cite{Habibi2017}.}
\label{fig:s2_sed}
\end{figure}

Due to the lack of a flux density analysis of the S-cluster star S2 \citep[][]{Eckart1996Natur, Parsa2017,2020A&A...644A.105H} in the H-band, we implement literature values derived for the K- \citep[][]{Sabha2012}, L- \citep[][]{Viehmann2006}, and M- bands \citep[][]{Viehmann2006}, see Fig. \ref{fig:s2_sed}.
We used a { single-temperature} black-body fit to estimate the H-band flux of about $32.0\pm 0.2$mJy. Based on the spectral atlas of \cite{Hanson1996}, we use common B2V star properties to reproduce the stellar energy distribution.

\end{document}